\documentclass[twocolumn]{aastex631}
\usepackage{upgreek}
\usepackage{xspace}
\usepackage{amsmath,amssymb}
\usepackage{rviewport}
\usepackage{bm}
\usepackage{nicefrac}
\usepackage{booktabs}
\usepackage{multirow}
\usepackage{mathtools}
\usepackage{placeins}
\usepackage{afterpage}

\defcitealias{2024ApJ...970...95U}{UF23}
\defcitealias{2012ApJ...761L..11J}{JF12}

\input{style}
\newcommand{\bdisk}{\ensuremath{b_\mathrm{d}}}          
\newcommand{\bann}{\ensuremath{b_\mathrm{a}}}           
\newcommand{\Vdisk}{\ensuremath{V_\mathrm{d}}}          
\newcommand{\Sdisk}{\ensuremath{S_\mathrm{d}}}          
\newcommand{\Edisk}{\ensuremath{E_\mathrm{d}}}          
\newcommand{\smallodot}{{\scriptscriptstyle\odot}}
\newcommand{\bdisksun}{\ensuremath{b_{\mathrm{d},\mkern-1.5mu{\smallodot}}}}
\newcommand{\lrad}{\ensuremath{l_r}}                    
\newcommand{\rcusp}{\ensuremath{r_c}}                   
\newcommand{\Rmax}{\ensuremath{R_\mathrm{max}}}         
\newcommand{\wmax}{\ensuremath{w_\mathrm{max}}}         
\newcommand{\zdisk}{\ensuremath{z_\mathrm{d}}}          
\newcommand{\bannpeak}{\ensuremath{b_{\mathrm{a},0}}}   
\newcommand{\rann}{\ensuremath{r_\mathrm{a}}}           
\newcommand{\wann}{\ensuremath{w_\mathrm{a}}}           
\newcommand{\zann}{\ensuremath{z_\mathrm{a}}}           
\newcommand{\Rsun}{\ensuremath{R_\smallodot}} 
\newcommand{\ncre}{\ensuremath{n_\mathrm{cre}}}         

\newcommand{\nRegions}{20} 
\newcommand{\nFitRegions}{16} 

\newcommand{\muG}{\ensuremath{\upmu}\textrm{G}}
\newcommand{\muK}{\ensuremath{\upmu}\textrm{K}}

\newcommand{\I}{\ensuremath{I}}
\newcommand{\Q}{\ensuremath{Q}}
\newcommand{\U}{\ensuremath{U}}

\newcommand{\PI}{\ensuremath{\text{PI}}}

\newcommand{\MINUIT}{{\scshape Minuit}\xspace}

\newcommand{\DRAGON}{{\scshape Dragon}\xspace}
\newcommand{\HEALPix}{{\scshape HEALPix}\xspace}
\newcommand{\GALPROP}{{\scshape Galprop}\xspace}
\newcommand{\ROOT}{{\scshape Root}\xspace}

\newcommand{\nLOS}{\ensuremath{\hat{\bm{n}}}}
\newcommand{\ddip}{\ensuremath{\bm{d}}}
\newcommand{\Planck}{\textit{Planck}\xspace}
\newcommand{\WMAP}{\textit{WMAP}\xspace}

\definecolor{rossoCP3}{cmyk}{0,.88,.77,.40}
\definecolor{darkBlue}{rgb}{0, 0, 0.8}
\definecolor{darkGreen}{rgb}{0, 0.5, 0}

\newcommand{\hDiff}{\ensuremath{h_{D}}}

\newcommand{\ospurNorm}{\ensuremath{\kappa_\text{os} \, \Delta s\, b_\text{os}^2}}
\newcommand{\ospurL}{\ensuremath{\ell_\text{os}}}
\newcommand{\ospurDeltaL}{\ensuremath{\Delta \ell_\text{os}}}
\newcommand{\ospurDeltaB}{\ensuremath{\Delta b_\text{os}}}

\newcommand{\modelNe}{\textnormal{\texttt{neCL}}\xspace}
\newcommand{\modelXr}{\textnormal{\texttt{expX}}\xspace}
\newcommand{\modelBase}{\textnormal{\texttt{base}}\xspace}
\newcommand{\modelSpur}{\textnormal{\texttt{spur}}\xspace}

\newcommand{\modelTwist}{\textnormal{\texttt{twistX}}\xspace}
\newcommand{\modelKappa}{\textnormal{\texttt{nebCor}}\xspace}
\newcommand{\modelCreTen}{\textnormal{\texttt{cre10}}\xspace}
\newcommand{\modelCreSix}{\textnormal{\texttt{cre06}}\xspace}

\newcommand{\modelCreOS}{\textnormal{\texttt{creOS}}\xspace}

\newcommand{\modelBubBest}{\textnormal{\texttt{bubBest}}\xspace}
\newcommand{\modelBubDust}{\textnormal{\texttt{bubDust}}\xspace}

\newcommand{\modelDisk}{\textnormal{\texttt{expDisk}}\xspace}
\newcommand{\modelRing}{\textnormal{\texttt{ringDisk}}\xspace}

\def\Offline{\mbox{$\overline{\textrm%
{Off}}$\hspace{.05em}\protect\raisebox{.4ex}%
{$\protect\underline{\textrm{line}}$}}\xspace}

\newcommand{\UFCoh}{\citetalias{2024ApJ...970...95U}\xspace}

\newcommand{\ModelD}{Disk\xspace}
\newcommand{\ModelDR}{Disk+Ring\xspace}
\newcommand{\ModelDRF}{Disk+Ring+Spur\xspace}
\newcommand{\ModelDF}{Disk+Spur\xspace}
\newcommand{\ModelDRS}{Disk+Ring+Spiral\xspace}
\newcommand{\brms}{\ensuremath{b_\mathrm{rms}}}

\makeatletter
\newcommand{\aasragged}{%
  \def\raggedcolumn@sw##1##2{##1}}
\newcommand{\aasflush}{%
  \def\raggedcolumn@sw##1##2{##2}}
\makeatother

\begin{document}
\global\let\tabular\savetabular

\title{The Random Magnetic Field of the Milky Way}

\author[0000-0002-7651-0272]{Michael Unger}
\email{michael.unger@kit.edu}
\affiliation{Institut f\"ur Astroteilchenphysik, Karlsruher Institut f\"ur Technologie, Karlsruhe 76344, Germany}
\affiliation{Institutt for fysikk, Norwegian University of Science and Technology (NTNU), Trondheim, Norway}
\author[0000-0003-2417-5975]{Glennys R. Farrar}
\email{gf25@nyu.edu}
\affiliation{Center for Cosmology and Particle Physics, Department of Physics, New York University, New York, NY 10003, USA
}

\begin{abstract}
\noindent
We constrain the large-scale isotropic random component of the
Galactic magnetic field using the \Planck reconstruction of the
408~MHz synchrotron sky. Our analysis simultaneously fits the
random-field structure, the synchrotron contributions of the coherent
field and local foregrounds, and isotropic and dipolar offsets. A new
element of the analysis is the use of excess polarized emission at
30~GHz to model local foregrounds, allowing us to retain 97\% of the
sky without absorbing these structures into Galaxy-wide features of
the magnetic field. Across variations in the coherent-field model,
cosmic-ray electron distribution, sky mask, and field profile, we
consistently find that the dominant random-field component is a
vertically compact disk with a local rms strength of about
$4~\upmu\mathrm{G}$ and a $1/e$ scale height of approximately $1$~kpc,
substantially thinner than in most previous models. An annular
enhancement in the inner Galaxy improves the fit to the data, whereas we
find no evidence for either a large-scale spiral pattern or a thick
random-field disk. The fits also yield an intensity monopole of
4--7~K, whose possible origin we discuss. We quantify the implications
of the inferred random field for the angular smearing of
ultrahigh-energy cosmic rays. Compared with previous models, the
sky-median smearing angle is smaller by up to a factor of 1.7, and by
up to 2.4 in individual directions.
\end{abstract}

\section{Introduction}
The energy density of the magnetic field of the Galaxy is comparable
to that of cosmic rays and of the turbulent motions of the
interstellar gas, making the magnetic field a dynamically important
constituent of the interstellar
medium~\citep{2001RvMP...73.1031F,2013pss5.book..641B}. The Galactic
Magnetic Field (GMF) also governs the propagation of charged particles
through the Galaxy, from the diffusion of Galactic cosmic rays to the
deflection of ultra-high-energy particles arriving from extragalactic
sources. The GMF is commonly decomposed into a large-scale coherent
component and a random component that fluctuates on scales of a few to
a hundred parsecs.  The global structure of the coherent field is
mainly constrained by Faraday rotation measures and polarized
synchrotron emission, whereas the random field is traced by the
unpolarized part of the synchrotron intensity, by fluctuations of the
rotation measures, and by depolarization; its global properties are
correspondingly less well known.  See \citet{2026arXiv260602149H} for
a recent review.

A variety of approaches have been used to model the Galactic random
field \citep{2019Galax...7...52J}, with most previous determinations
of its large-scale strength and spatial distribution relying primarily
on the total synchrotron intensity of the Galaxy measured by
\citet{1981A&A...100..209H,1982A&AS...47....1H}. \citet{2008A&A...477..573S},
following the early parametric investigation of the global structure
of the Galactic synchrotron emissivity by \citet{1985A&A...153...17B},
found the overall scale of an assumed spatially homogeneous random
field to be $\sim$3~\muG{} by visual comparison of the implied
synchrotron intensity to the 408~MHz data.
\citet{2010MNRAS.401.1013J} gave estimates of the coherent, ordered
random, and isotropic random field components in the Galactic plane.
A major advance toward a full three-dimensional description of the
random GMF was made by \citet{2012ApJ...761L..11J}, who fitted a
13-parameter model, having a smooth halo component and independent
random-field amplitudes in the spiral arms, to the \textit{WMAP}7
22~GHz total intensity~\citep{WMAP7}.  \citet{2013MNRAS.436.2127O}
constrained the random field strength together with the cosmic-ray
electron density using multi-frequency radio surveys. The
\citet{2016A&A...596A.103P} proposed modifications to several of these
models through qualitative comparison to the \Planck synchrotron total
and polarized intensity as well as the observed dust polarization.

Other observables, in particular fluctuations of the Faraday rotation
of pulsars and extragalactic sources, provide complementary
constraints on the strength and structure of the random field, e.g.,
\citet{1993MNRAS.262..953O,
  1996ApJ...458..194M,2004ApJ...610..820H,2008ApJ...680..362H,2013MNRAS.436.2326P}.
Because rotation measure (RM) fluctuations can originate from
variations in both the line-of-sight magnetic field and the
thermal-electron density, their interpretation requires assumptions
about both; we do not use them in this work.

Here we revisit the global modeling of the random field and improve on
previous work in several respects. First, we use the total synchrotron
intensity reconstructed by the \Planck component-separation analysis
\citep{2016A&A...594A..10P}, which in particular separated the
contribution of anomalous microwave emission that had contaminated the
\textit{WMAP}7 22~GHz total intensity estimate to which the model of
\citet{2012ApJ...761L..11J} had been tuned. Second, we account for
prominent local structures as foreground emission rather than
attributing them to Galaxy-wide features of the random field.  Third,
all parameters of our model are determined simultaneously in a fit to
the data, instead of being set by visual comparison or partial
optimization as in some of the previous studies. Finally, to account
for the uncertainty in the contribution of the coherent field to the
total intensity, we derive the random field for each model in the
ensemble of coherent-field models of \citet{2024ApJ...970...95U} (hereafter
\UFCoh), and propagate the uncertainty of the coherent field into the
determination of the random field. Since the \UFCoh models account for
the anisotropic (``striated'') random field as a rescaling of the
coherent field, the random component determined in the present work is
the isotropic part of the random field.

We model the total synchrotron intensity reconstructed by \Planck (see
Sec.\,\ref{sec:data}) as
\begin{equation}
I_\text{tot} = I_\text{coh} + I_\text{rand} + I_\text{fg} + I_\text{off}.
\label{eq:total}
\end{equation}
The synchrotron contributions $I_\text{coh}$ and $I_\text{rand}$ are
produced by cosmic-ray electrons, modeled as described in
Sec.\,\ref{sec:ncre}. The calculation of the synchrotron intensity for
a given magnetic-field configuration is detailed in
Sec.\,\ref{sec:synchro}.
The contribution $I_\text{coh}$ from the
coherent field, including its possible striated enhancement, is
presented in Sec.\,\ref{sec:coh}. The term $I_\text{fg}$ accounts for
localized foreground features not captured by the large-scale field
model (Sec.\,\ref{sec:fg}), while $I_\text{off}$ represents a
large-scale intensity offset modeled as the sum of isotropic and
dipolar contributions (Sec.\,\ref{sec:offsets}). The isotropic
random-field model underlying $I_\text{rand}$, the primary target of
our analysis, is presented in Sec.\,\ref{sec:rand}. Analytic
approximations that identify which features of the sky constrain the
field parameters are derived in Sec.\,\ref{sec:synpar}, and the
procedure used to fit the model parameters to the \Planck intensity
map is given in Sec.\,\ref{sec:analysis}. The impact of the modeling
choices is assessed in Sec.\,\ref{sec:modelingChoices}, and the final
results are given in Sec.\,\ref{sec:results}. Their implications are
discussed in Sec.\,\ref{sec:discussion}, before we conclude in
Sec.\,\ref{sec:conclusions}.

\section{Data}
\label{sec:data}

\begin{figure}[t]
\def\labelpad{0.8cm}
\gridline{\fig{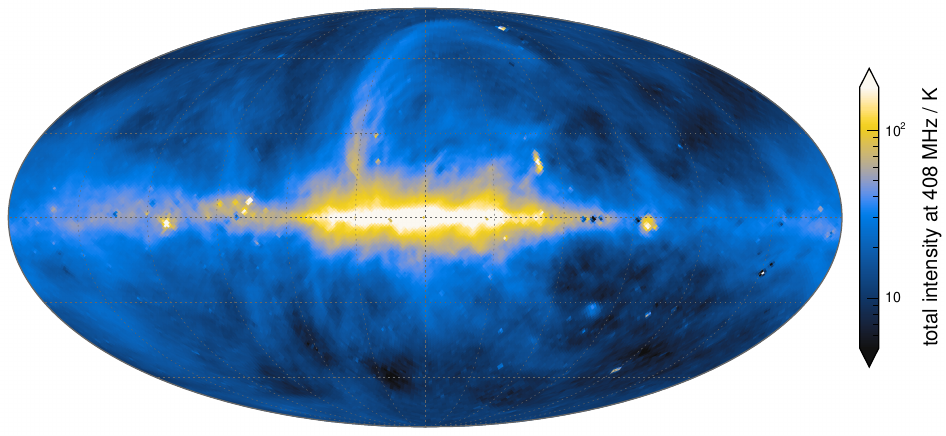}{\columnwidth}{(a) Input map.\hspace*{\labelpad}}}
\gridline{\fig{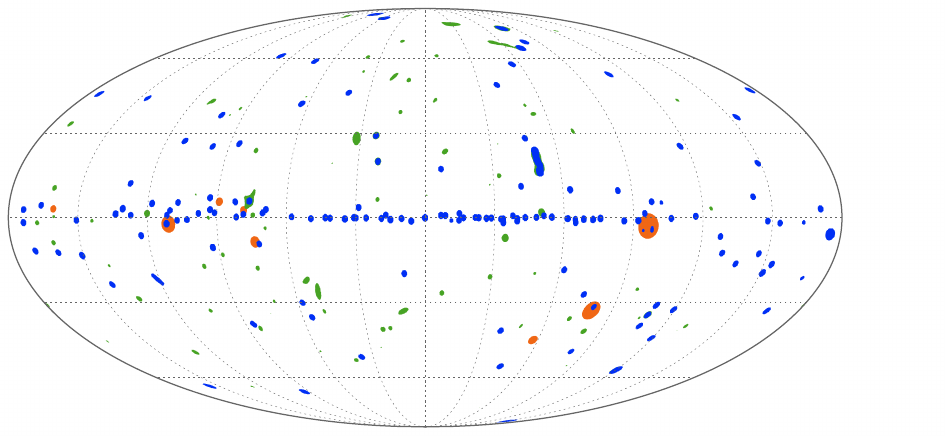}{\columnwidth}{(b) Source mask.\hspace*{\labelpad}}}
\gridline{\fig{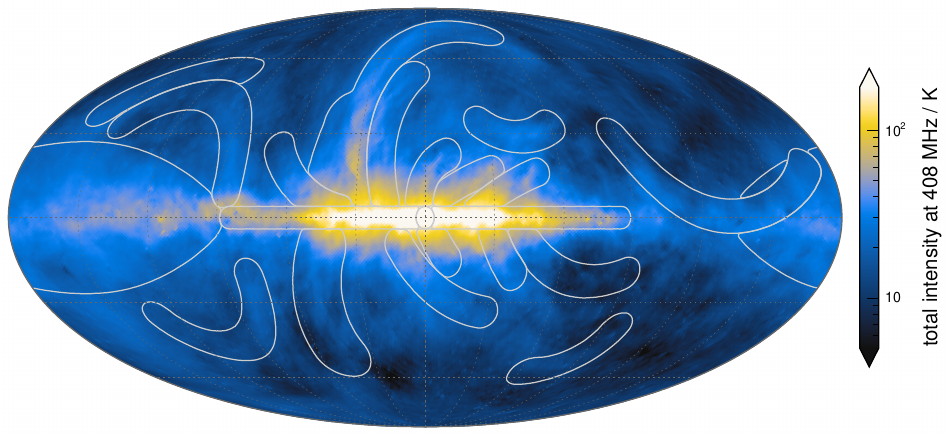}{\columnwidth}{(c) De-sourced map with foreground region contours.\hspace*{\labelpad}}}
\caption{Synchrotron intensity at 408~MHz. Panel (a) shows the
  synchrotron sky map from the \Planck component-separation analysis
  \citep{2016A&A...594A..10P}, degraded to $N_{\rm side}=64$. Panel
  (b) shows the source mask used in this analysis, with polarized
  sources in blue, extragalactic sources in green, and further masked
  regions in orange. Panel (c) shows the de-sourced map after
  inpainting the masked pixels for display purposes, with the contours of the foreground regions
from Fig.\,\ref{fig:polintensity}\,(b) superimposed. Intensities in
  (a) and (c) are clipped to the range 5--180~K.}
\label{fig:intensity}
\end{figure}

We use the 408~MHz synchrotron total-intensity map from the \Planck
2015 foreground analysis \citep{2016A&A...594A..10P}. At this
frequency, the diffuse Galactic radio sky is dominated by synchrotron
emission, with free-free emission as the main Galactic contaminant. In
the \Planck analysis, the free-free contribution was estimated
simultaneously with the synchrotron emission in a multi-frequency
parametric foreground fit. The map is based on the all-sky 408~MHz
survey of \citet{1981A&A...100..209H,1982A&AS...47....1H}, using the
destriped reprocessing of \citet{2015MNRAS.451.4311R}. Its absolute
zero level and residual dipole were fixed to the values derived by
\citet{2017A&A...597A.131W}. It is shown in Fig.\,\ref{fig:intensity}\,(a).

\begin{figure}[t]
\def\labelpad{0.8cm}
\gridline{\fig{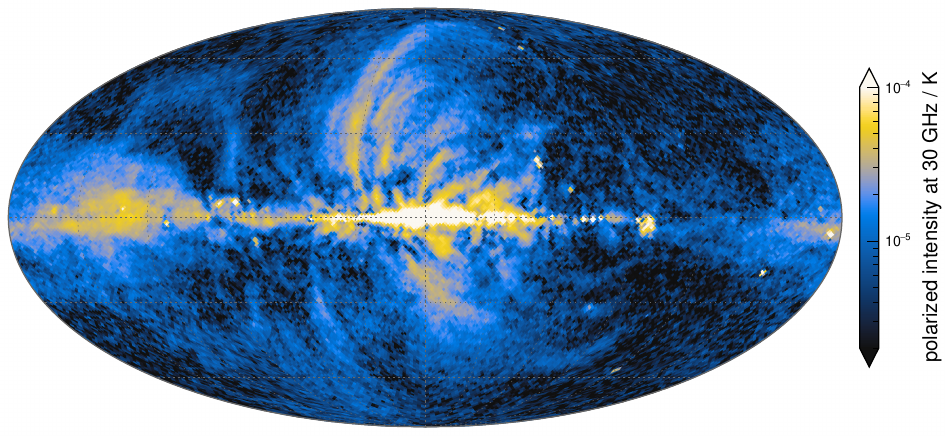}{\columnwidth}{(a) Input map.\hspace*{\labelpad}}}
\gridline{\fig{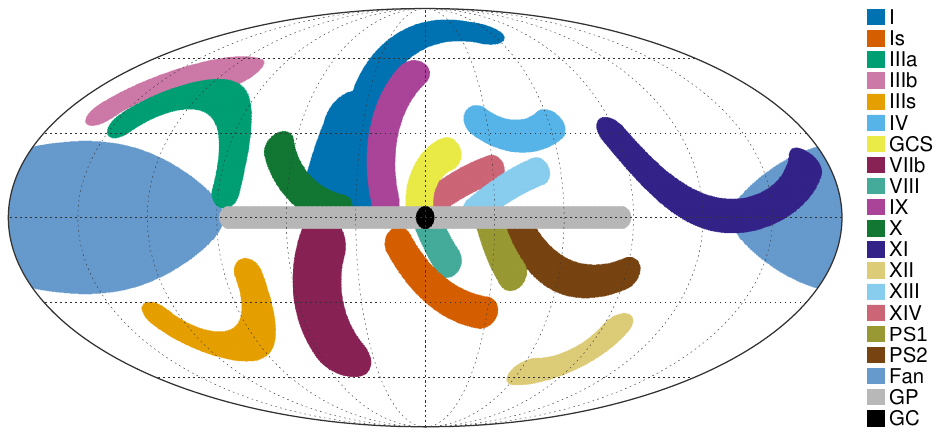}{\columnwidth}{(b) Foreground regions.\hspace*{\labelpad}}}
\gridline{\fig{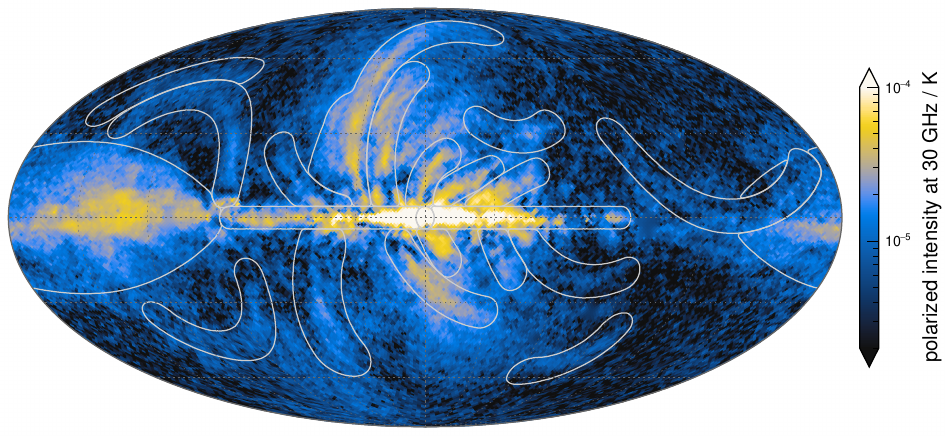}{\columnwidth}{(c) De-sourced map with foreground region contours.\hspace*{\labelpad}}}
\caption{Polarized synchrotron intensity at 30~GHz. Panel (a) shows the
average of the \WMAP and \Planck polarized-intensity maps. Panel (b)
shows the foreground regions used in the analysis. Panel (c) shows the
de-sourced map after applying the source mask and inpainting the masked
pixels for display purposes, with the contours of the foreground regions
from panel (b) superimposed. Polarized intensities in
  (a) and (c) are clipped to the range 2--100~\muK.}
\label{fig:polintensity}
\end{figure}

For the sky template of excess polarized synchrotron intensity
described in Sec.\,\ref{sec:fg}, we use a simple arithmetic
average of the Stokes \Q{} and \U{} sky maps from
\WMAP~\citep{2013ApJS..208...20B}, extrapolated to 30~GHz, and
\Planck~\citep{2020A&A...641A...4P}, as described in
\citepalias{2024ApJ...970...95U}. We average the two \Q{} maps and the
two \U{} maps separately, and then form the polarized intensity,
$\PI=\sqrt{\Q^2+\U^2}$. The resulting sky map is shown in
Fig.\,\ref{fig:polintensity}\,(a). Note that because $\PI$ is
positive definite, it is biased upward in the presence of noise
\citep[e.g.][]{1974ApJ...194..249W}. We estimate this bias from the
sub-pixel variance of \Q{} and \U{} and find that, after averaging to
$N_{\rm side}=16$, the expected polarized intensity in the absence of
a true signal is only $\simeq 0.6\,\muK$.\footnote{We estimate the
  per-component noise variance $\sigma_{QU}^2$ from the distribution
  of sub-pixel \Q{} and \U{} variances within each $N_{\rm side}=16$
  pixel, adopting the mode of its low-variance peak as an estimate of
  the instrumental noise floor. \Q{} and \U{} give similar values. For
  noise-only Stokes parameters of equal variance, the polarized
  intensity is Rayleigh-distributed with $\langle\PI\rangle =
  \sqrt{\pi/2}\,\sigma_{QU}$. We find $\sigma_{QU}=2.5,\,2.2,\,1.8\,\muK$
  for \WMAP, \Planck, and our arithmetic-average map, respectively. For
  comparison, the more sophisticated reprocessings of
  \citet{2024A&A...686A.297W} and \citet{2024A&A...689A.353D} give
  $1.6\,\muK$. Our simple average thus reaches a comparable noise
  floor. Since each final $N_{\rm side}=16$ pixel averages up to 16
  $N_{\rm side}=64$ subpixels, the expected signal-free polarized
  intensity is $\langle\PI\rangle =
  \sqrt{\pi/2}\,\sigma_{QU}/\sqrt{16} \approx 0.6\,\muK$.}


\begin{figure*}[!t]
  \centering
\includegraphics[clip, rviewport=0 0 1 1,width=0.7\linewidth]{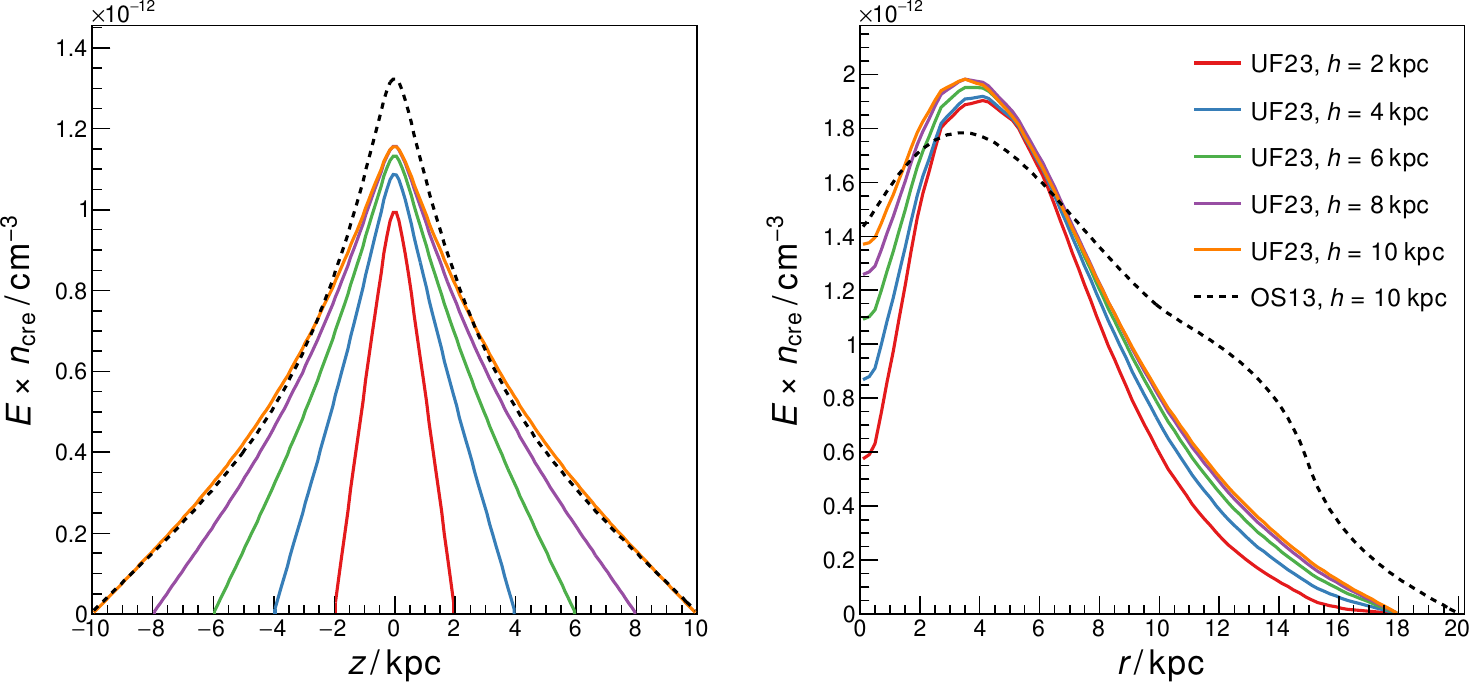}
\caption{Density of 3~GeV
  cosmic-ray electrons as a function of {\itshape Left:} Galactic
  height $z$ at a Galactocentric radius $r=8.2$~kpc and {\itshape
  Right:} Galactocentric radius $r$ at the mid-plane ($z=0$). Different choices of the
  half-height \hDiff of the diffusion volume are shown as lines with
  different colors. The dashed line is the result
  of~\citet{2013MNRAS.436.2127O} for $\hDiff=10$~kpc. }
\label{fig:ncreDens}
\end{figure*}

From both maps, we remove emission from bright sources in two steps,
using a common mask for the intensity and polarized-intensity
maps. First, we apply the 30\,GHz polarized-intensity point-source
mask of \citet{2024A&A...689A.353D}, which covers the brightest
Galactic and extragalactic compact sources. Second, we mask the most
prominent residual extragalactic sources using the maps of
\citet{2015MNRAS.451.4311R}: we compute the temperature difference
$\Delta T$ between their destriped and destriped-and-desourced maps,
i.e.\ the emission attributed to the subtracted sources, average it
within a disk of $1^\circ$ radius around each pixel, and mask all
pixels for which this average exceeds 3.3\,K.  We supplement these
masks by enlarging the regions around the Vela SNR and the LMC, and
adding masks for the SMC, the Cygnus~X region, and four remaining
bright Galactic SNRs, namely Cas~A, HB9, HB21, and the Cygnus~Loop
\citep{2025JApA...46...14G}. The masked regions are shown in
Fig.\,\ref{fig:intensity}\,(b). The mask is applied at a \HEALPix
resolution of $N_{\rm side}=64$, corresponding approximately to the
beam size of the Haslam survey. The mask removes 2.4\% of the
pixels. As can be seen in Fig.\,\ref{fig:intensity}\,(c), the brightest
sources in total intensity are removed, and most of the spurious
low-synchrotron regions in the component-separation map at 408~MHz are
masked as well, since they coincide with bright low-latitude
regions. The corresponding de-sourced polarized-intensity map is shown
in Fig.\,\ref{fig:polintensity}\,(c).

To characterize large-area foreground regions associated with local
structures, such as nearby supernova shells, as well as other
localized emission regions not included in our large-scale model, we
define \nRegions\ sky regions, illustrated in
Fig.\,\ref{fig:polintensity}\,(b). Most are based on the polarized
radio filaments listed by
\citet{2015MNRAS.452..656V,2016A&A...594A..25P} and are labeled by
Roman numerals. We additionally include the \Planck spurs (PS1 and
PS2) and the Galactic Center spur (GCS), as well as extended polarized
emission around the so-called Fan region \citep{1964BAN....17..465B}
(Fan), the Galactic center (GC), and the Galactic plane (GP). The
resulting contours are superimposed on the de-sourced synchrotron maps
in Figs.\,\ref{fig:intensity}\,(c) and \ref{fig:polintensity}\,(c). As
can be seen, many of these features are present in both total and
polarized intensity. See Sec.\,\ref{sec:fg} for further discussion.

The final maps used in this analysis are constructed at a resolution
of $N_{\rm side}=16$. The intensity and polarized-intensity values in
each pixel are obtained as the average over its unmasked $N_{\rm  side}=64$
subpixels, requiring that at least half of the pixel area
remain unmasked, i.e.\ that at least 8 of the 16 subpixels are
unmasked. Only 0.8\% of the $N_{\rm side}=16$ pixels (24 of 3072) are
excluded for failing this criterion, leaving a total of 3048 pixels
for the fit.

\section{Cosmic-Ray Electrons}
\label{sec:ncre}

Predictions for the synchrotron emission require a model for the
three-dimensional density and energy spectrum of Galactic cosmic-ray
electrons and positrons, $\ncre({\bm x},E)$.

We use the same set of cosmic-ray electron models as developed in
\UFCoh. Our baseline model is \modelCreSix, obtained from a
plain-diffusion calculation with the \DRAGON cosmic-ray propagation
code \citep{Evoli:2008dv}, assuming a diffusion-halo half-height of
$\hDiff=6$~kpc.  This is the cosmic-ray electron model used for most
of the \UFCoh magnetic-field models. To estimate the systematic
uncertainty associated with the cosmic-ray electron distribution, we
also use the otherwise identical \modelCreTen model, with
$\hDiff=10$~kpc. In Sec.\,\ref{sec:sysNcre} we additionally explore
$\hDiff=2$, 4, and 8~kpc; note that halo half-heights smaller than
$6$~kpc could not describe the coherent field in \UFCoh and are
therefore not part of the standard \UFCoh ensemble. In these models,
the diffusion coefficient is fixed through
$D/\hDiff=0.046~\mathrm{kpc/Myr}$ at 10~GV, with the rigidity
dependence taken from the PD2 model of \citet{2016ApJ...831...18C}.
The radial distribution of cosmic-ray sources follows the pulsar
distribution of \citet{2006MNRAS.372..777L}, used as a proxy for
supernova remnants, and the electron cooling by synchrotron radiation
is computed using the \citetalias{2012ApJ...761L..11J} coherent and
random magnetic-field model.  The injected proton and lepton spectra
are constrained by Voyager~I measurements of the local interstellar
spectrum \citep{2016ApJ...831...18C} and by AMS-02 measurements inside
the heliosphere \citep{AMS:2014gdf,AMS:2015tnn,AMS:2019rhg}.  As an
independent alternative, we also include the \modelCreOS model,
i.e.\ the \texttt{z10LMPDE} \GALPROP calculation of
\citet{2013MNRAS.436.2127O}.  This model assumes $\hDiff=10$~kpc, a
maximum Galactocentric radius of 20~kpc, the source distribution of
\citet{1998ApJ...509..212S} with a radial cutoff at 15~kpc, and the
magnetic-field model of \citet{2010RAA....10.1287S} for synchrotron
losses. We re-normalize \modelCreOS to the AMS-02
electron-plus-positron flux at 35~GV. For more details and the
rationale for these choices, see \citet{2024ApJ...970...95U}.

In Fig.\,\ref{fig:ncreDens}, we show the models' cosmic-ray electron
densities at an energy of 3\,GeV.\footnote{For an isotropic random
  field of rms strength $b$, the characteristic synchrotron frequency
  is $\nu_{\rm c}\simeq13.1\,{\rm MHz}\,(b/\upmu{\rm G})(E/{\rm
    GeV})^2$ \citep[e.g.][]{2011hea..book.....L}. Thus, for
  $b(z)=b_0\exp(-|z|/z_0)$, with $b_0\simeq4\,\upmu{\rm G}$ and
  $z_0\simeq1\,{\rm kpc}$ as found here, 408-MHz emission corresponds
  to $E_{\rm c}(z)\simeq 2.8\exp(|z|/(2z_0))\,{\rm GeV}$ and is
  produced predominantly by cosmic-ray electrons with energies of
  approximately $2$--$6\,{\rm GeV}$.}  Their vertical profiles closely
resemble tent functions, $n(z)=n_0(1-|z|/\hDiff)$, which are the
steady-state solutions for one-dimensional diffusion with injection at
$z=0$, free escape at $\pm\hDiff$, and no energy losses. This
indicates that \ncre{} in these models at these energies is governed
primarily by escape rather than cooling, as expected for the
plain-diffusion parameters given above.  A more complete analysis
would self-consistently model the magnetic field and cosmic-ray
synchrotron losses, see e.g.\ \citet{Orlando:2026fhk}.  However, as
long as escape is faster than cooling and synchrotron radiation is
only one of several competing loss channels, \ncre{} depends only
weakly on the magnetic field even where energy losses do
matter.\footnote{The mean vertical escape time of a 3\,GeV electron
  injected at the mid-plane, with $D/\hDiff$ from above scaled to that
  energy, is $t_{\rm esc}=\hDiff^2/(2D) \simeq0.12\,\mathrm{Gyr}
  \left({\hDiff}/{6\,\mathrm{kpc}}\right)$. By comparison, its
  synchrotron-loss time is $t_{\rm loss}^{\rm syn} ={E}/{|\dot E_{\rm
      syn}|} \simeq0.17\,\mathrm{Gyr}
  \left({B}/{5\,\upmu\mathrm{G}}\right)^{-2}.$ Since the magnetic
  field strength decreases with Galactic height, the electron samples
  on average a weaker field than the typical mid-plane value used
  here, so synchrotron cooling remains, on average, slower than escape
  at this energy.}  Future systematic studies can quantify the
accuracy of this approximation and, more importantly, should address
the effects of reacceleration and convection
\citep[e.g.][]{2024ApJ...963..111S} as well as the interplay between
the magnetic field and the anisotropic, inhomogeneous diffusion
coefficient of the Galaxy
\citep[e.g.][]{2012A&A...547A.120E,2013JCAP...03..036D}.

\section{Synchrotron Intensity Calculation}
\label{sec:synchro}
To predict the synchrotron intensity in pixel $i$ for a given
magnetic-field configuration, we integrate the synchrotron emissivity
along the corresponding line of sight. At each position, we calculate the
emissivity of a single electron in the coherent and isotropic random field
components following \citet{1965ARA&A...3..297G} and
\citet{1986A&A...164L..16C}, respectively. We then integrate the
single-electron emissivity numerically over the differential energy
spectrum of the cosmic-ray electron
density, $\ncre({\bm x},E)$, predicted by the models described in
Sec.\,\ref{sec:ncre}. This yields the volume emissivity
$\varepsilon_\nu(\boldsymbol{x})$, the power emitted per unit volume
and frequency at the observing frequency $\nu$.  The main dependencies
of the emissivity can be seen most clearly in the special case of a
power-law electron spectrum, $\mathrm{d}N/\mathrm{d}E = \kappa\,
E^{-p}$, for which
\begin{equation}
  \varepsilon_\nu \;\propto\; \kappa\, \nu^{-(p-1)/2}\, B^{(p+1)/2},
  \label{eq:powerlaw}
\end{equation}
where $B$ denotes the perpendicular component $B_\perp$ of the
coherent field or the root-mean-square strength $b$ of the random
field, respectively.  The coherent emission is linearly polarized
perpendicular to $B_\perp$ with the intrinsic polarization fraction
\begin{equation}
  \Pi(p) = \frac{p+1}{p+7/3},
  \label{eq:pimax}
\end{equation}
whereas the emission from the isotropic random field is
unpolarized. Expressed as antenna temperature, $T_\nu = c^2 I_\nu/(2
k_{\rm B} \nu^2)$, the frequency dependence becomes $T_\nu \propto
\nu^{-\beta_S}$ with $\beta_S = (p+3)/2$.

The integration along the line of sight, including the average over
the finite solid angle $\Omega_i$ of the pixel, is performed with a
simplified non-recursive variant of stratified Monte-Carlo sampling.
The predicted pixel-averaged antenna temperature is
the volume integral of $\varepsilon_\nu$ over the cone subtended by
the pixel,
\begin{equation}
  T_i \;=\; \frac{c^2}{2 k_{\rm B} \nu^2}\, \frac{1}{\Omega_i}
        \int_{\Omega_i} {\rm d}\Omega
        \int_0^{R(\nLOS)} {\rm d}r\;
        \frac{\varepsilon_\nu(\boldsymbol{x}_\smallodot + r\,\nLOS)}{4\pi},
  \label{eq:pixelIntegral}
\end{equation}
where $\boldsymbol{x}_\smallodot$ is the position of the observer and
$R(\nLOS)$ is the distance at which the line of
sight leaves the integration volume, taken to be a cylinder of radius 20~kpc and
half-height of 10~kpc. Since ${\rm d}r\,{\rm d}\Omega = {\rm
  d}V/r^2$, the geometric dilution of the flux cancels the
growth of the volume element with distance. Positions drawn uniformly
in distance and  solid angle within
the pixel cone therefore provide an unbiased estimate of
Eq.\,\eqref{eq:pixelIntegral}. The line of sight is divided into
$n_s$ radial shells of length $\Delta R = 500$~pc, and the integral is
estimated as the sum of the per-shell sample means,
\begin{equation}
  \hat{T}_i = \frac{c^2}{2 k_{\rm B} \nu^2}\, \frac{\Delta R}{4\pi}
    \sum_{k=1}^{n_s} \frac{1}{N_k}
    \sum_{j=1}^{N_k} \varepsilon_\nu(\boldsymbol{x}_{kj}),
  \label{eq:mcEstimator}
\end{equation}
where
$\boldsymbol{x}_{kj}
 =\boldsymbol{x}_\smallodot+r_{kj}\hat{\boldsymbol{n}}_{kj}$. The number of points $N_k$
in shell $k$ is set adaptively: the standard deviation $\sigma_k$ of a
single sample's contribution, estimated from an initial exploration
sample, determines the allocation $N_k = \sigma_k \sum_j \sigma_j /
\sigma_{\rm tgt}^2$, which minimizes the total number of points
required for the Monte-Carlo uncertainty $\sigma(\hat{T}_i) = (\sum_k
\sigma_k^2/N_k)^{1/2}$ to reach the target $\sigma_{\rm tgt}$~\citep{press1990}.
We use a $\sigma_{\rm tgt}$ that is $5\%$ of the measurement uncertainty in pixel
$i$, thus introducing a negligible contribution to the uncertainty of the
data--model comparison. In this way, the
integration points concentrate in the volumes in which the emissivity
fluctuates most strongly. The sampled positions are kept fixed during
the model optimization, such that the objective function remains a
deterministic function of the model parameters.  This procedure yields
both the coherent-field intensity templates, $c_i$, evaluated once per
model variation (Sec.\,\ref{sec:coh}), and the random-field
intensities, $r_i(\boldsymbol\eta)$, re-evaluated for every trial
parameter set in the fit (Sec.\,\ref{sec:analysis}).

\begin{figure*}[t]
\def\labelpad{0.8cm}
\gridline{%
  \makebox[\columnwidth][c]{\modelTwist\unskip\hspace*{\labelpad}}%
  \hfill
  \makebox[\columnwidth][c]{\modelXr\unskip\hspace*{\labelpad}}%
}
\gridline{
  \leftfig{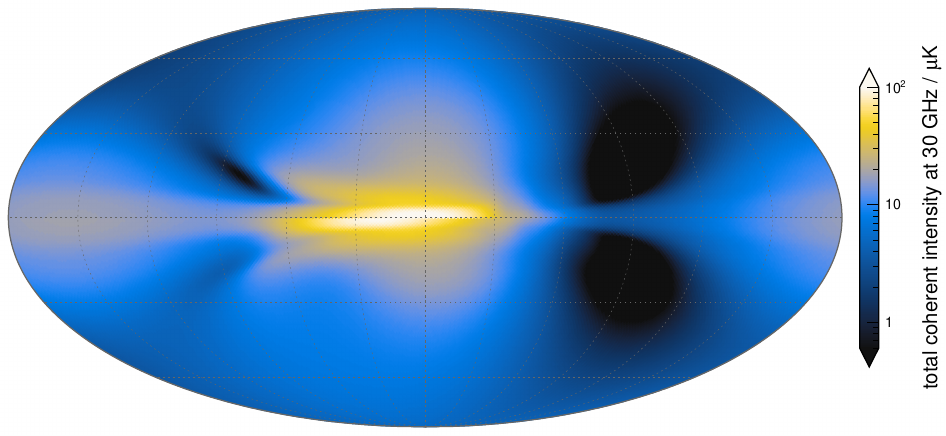}{\columnwidth}{(a)\hspace*{\labelpad}}
  \rightfig{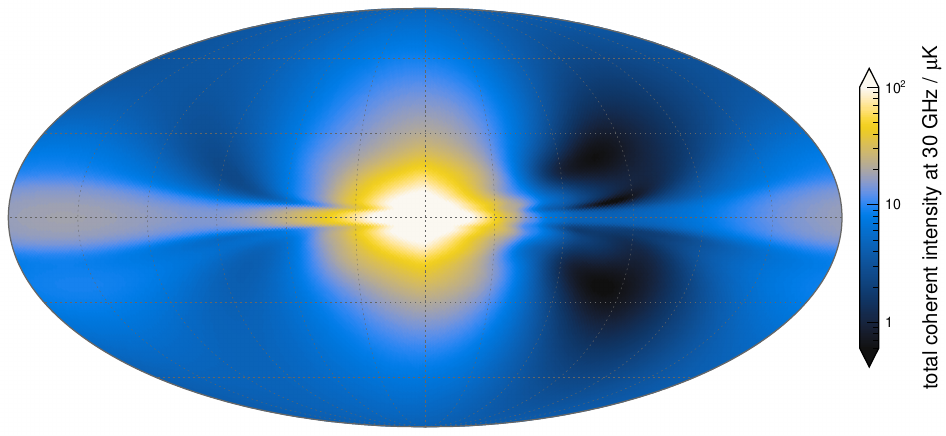}{\columnwidth}{(d)\hspace*{\labelpad}}
}
\gridline{
  \leftfig{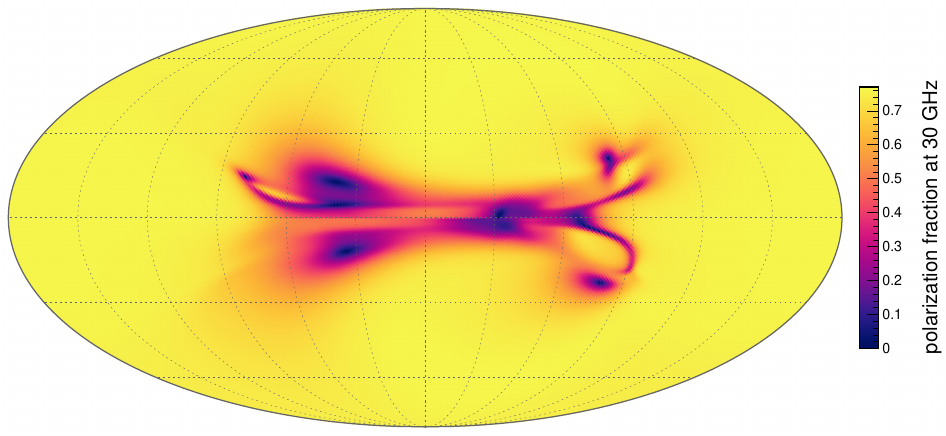}{\columnwidth}{(b)\hspace*{\labelpad}}
  \rightfig{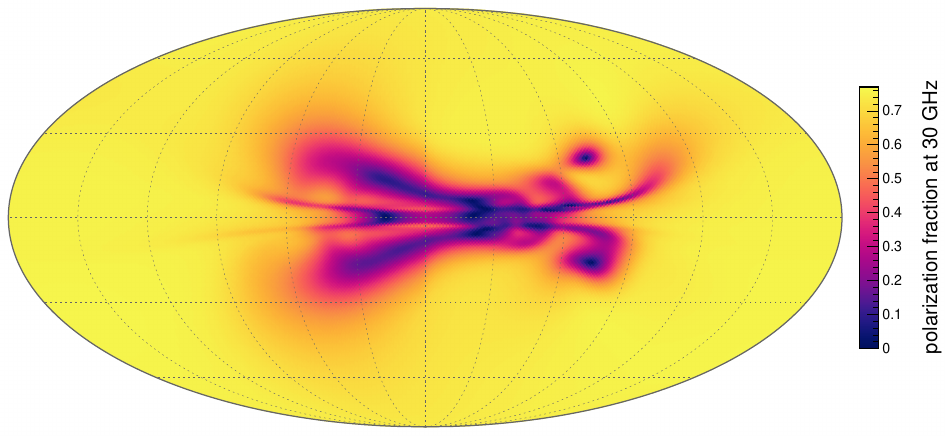}{\columnwidth}{(e)\hspace*{\labelpad}}
}
\gridline{
  \leftfig{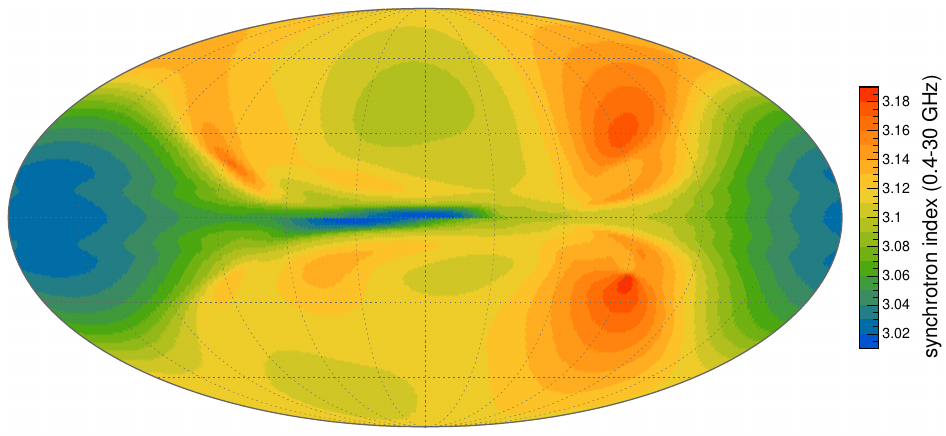}{\columnwidth}{(c)\hspace*{\labelpad}}
  \rightfig{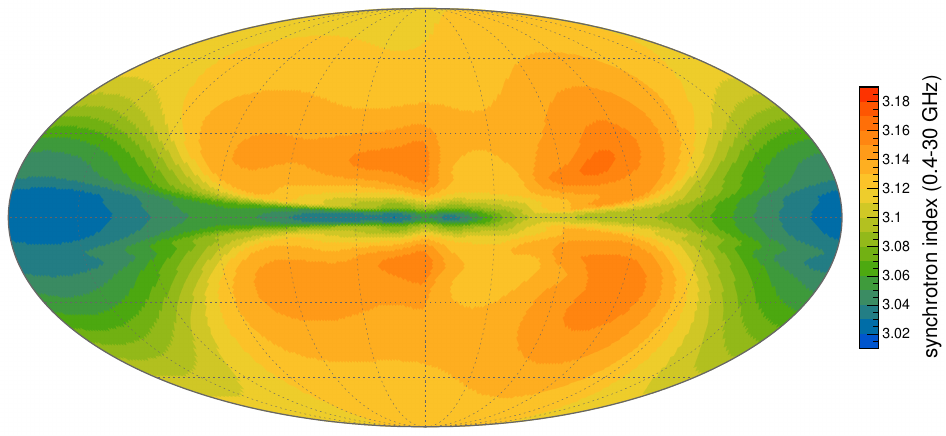}{\columnwidth}{(f)\hspace*{\labelpad}}
}
\caption{\UFCoh model predictions for the synchrotron emission produced by
cosmic-ray electrons in the coherent GMF. For illustration, two model
variations are shown in the left and right columns: \modelTwist and
\modelXr. Top row: total intensity, $I_\text{coh}$, at 30~GHz. Middle
row: polarization fraction at 30~GHz. Bottom row: effective synchrotron
spectral index between 408~MHz and 30~GHz.}
\label{fig:cohModel}
\end{figure*}

\section{Intensity from the Coherent Field}
\label{sec:coh}

Since the aim of this work is to determine the isotropic random
component of the GMF, the contribution of the coherent field to the
total synchrotron intensity (including its striated enhancement) must
be taken into account. We calculate $I_\text{coh}$ in
Eq.\,(\ref{eq:total}) using the suite of coherent field models from
\UFCoh and the calculation described in Sec.\,\ref{sec:synchro}. These
models were adjusted to extragalactic rotation measures and the
polarized synchrotron data at 30~GHz, and we extrapolate their
predictions to 408~MHz. Since the \UFCoh{} fit was performed on only
57.8\% of the synchrotron sky, the predicted intensity in the masked
regions is subject to additional uncertainty. Note that since we use
the same cosmic-ray electron densities, $\ncre$, for the calculation
of both $I_\text{coh}$ and $I_\text{rand}$, the relative magnitude of
the two contributions depends only on the respective field components
and not on differences in the assumed electron distribution.

The model variations considered here are as follows. The fiducial model
is \modelBase. The \modelXr model predicts a much larger polarized
intensity toward the central Galaxy owing to a different (exponential)
extrapolation of the poloidal X~field at small radii. In the
\modelSpur model, there is no coherent field in the spiral disk except
for the local Orion arm. The \modelNe model is a variation in which a
different model of the thermal electron density was used for the
calculation of the rotation measures.  In the \modelTwist model, the
toroidal halo field is replaced by a unified halo field originating
from a dynamical model in which the poloidal field is twisted by the
differential rotation of the Galaxy.  In addition, the \modelKappa model
assumes an anticorrelation between thermal electrons and the magnetic
field, resulting in a larger coherent field strength and no striated
component. The \modelCreTen and \modelCreOS models are re-fits of the
fiducial field configuration using the alternative cosmic-ray electron
models of the same name introduced in Sec.\,\ref{sec:ncre}; whenever we
vary \ncre{} in the following, we therefore use the coherent field that
was fitted with the same electron model.

Two examples of the predicted coherent-field emission are shown in
Fig.\,\ref{fig:cohModel} for the \modelTwist and \modelXr models. As
expected, the \modelXr model exhibits a large intensity in the inner
Galaxy, as can be seen in the top row. The polarization fraction, shown
in the middle row, differs from $\Pi$
(Eq.\,(\ref{eq:pimax})), because the polarization of emission from
regions with different field orientations along the line of sight can
partially cancel, whereas the total intensity is retained. The
polarized intensity therefore provides only a lower bound on the
coherent total intensity, $I_\text{coh}\geq\PI_\text{coh}/\Pi$, and cannot be
used as a proxy for $I_\text{coh}$, which must be accounted for in
full.

The extrapolation of the emission from 30~GHz to 408~MHz follows from
the \DRAGON/\GALPROP calculations without any further freedom, as it
is fully determined by the energy spectrum of the cosmic-ray
electrons. The resulting effective synchrotron spectral index between
$\nu_1 = 408$~MHz and $\nu_2 = 30$~GHz is shown in the bottom row. It
is given by
\begin{equation}
  \beta_S = -\frac{\ln\!\left(T(\nu_2)/T(\nu_1)\right)}
                  {\ln\!\left(\nu_2/\nu_1\right)},
  \label{eq:betaS}
\end{equation}
and coincides with $(p+3)/2$ (Sec.\,\ref{sec:synchro}) only if the
electrons follow a power law in energy with index $p$.  As can be
seen, the predicted effective spectral index of the coherent emission
varies by only $\pm 0.1$ around $\beta_S = 3.1$, but this translates
to appreciable differences in the extrapolated intensity (see the
numerical values in the next section).

\section{Foreground Intensity}
\label{sec:fg}

As noted in Sec.\,\ref{sec:data}, most of the localized features in
\emph{polarized} intensity at 30\,GHz are also clearly visible in the
\emph{total} intensity at 408\,MHz. This correspondence is expected
for two reasons.

Firstly, polarized synchrotron emission is necessarily accompanied by
total emission: since the polarization fraction of synchrotron
radiation from a uniform magnetic field is at most $\Pi(p)$
(Eq.\,(\ref{eq:pimax})), a polarized intensity $\PI$ implies a total
intensity of at least $\PI/\Pi(p)$ at the same frequency.
Extrapolating in brightness temperature with the effective spectral
index of Eq.\,(\ref{eq:betaS}), the minimum contribution of a polarized
feature observed at 30\,GHz to the total intensity at 408\,MHz is
$q(p)\, \PI(30\,\text{GHz})$ with
\begin{equation}
  q(p) = \frac{1}{\Pi(p)}
  \left(\frac{408\,\text{MHz}}{30\,\text{GHz}}\right)^{-(p+3)/2}.
  \label{eq:beta}
\end{equation}
For a typical spectral index of $p = 3$, as expected for an $E^{-2}$
injection spectrum steepened by synchrotron and inverse-Compton
cooling, $\Pi = 0.75$ and $q \approx 5\times10^{5}$. For the value
found for the coherent models in the previous section, $\beta_S = 3.1
\pm 0.1$ (i.e.\ $p = 2\beta_S - 3 = 3.2 \pm 0.2$), the ratio increases
to $q = (8^{+4}_{-3})\times10^{5}$. Individual features may, however,
deviate from this mean value: a young, uncooled electron population,
as might be expected in recently accelerated material of a nearby
supernova shell, would have a harder spectrum closer to the injection
index, $p \simeq 2$--$2.4$, whereas weak shocks inject softer spectra
\citep[e.g.][]{1987PhR...154....1B} and can result, after cooling, in
an index above~3. Since $q$ depends exponentially on $p$, such
regional variations of the spectral index have a large effect on the
expected intensity ratio: relative to the fiducial value at $p = 3.2$,
the ratio $q(p)/q(3.2)$ is only $\approx 0.1$ to $0.2$ for the hard,
uncooled case and reaches a factor of $\approx 2$ for $p \simeq 3.5$.
Note that at 408\,MHz the polarized intensity itself is largely erased
by Faraday rotation and the associated depolarization along the line
of sight and within the beam, but the accompanying total intensity is
unaffected. The polarized emission observed at 30\,GHz, where Faraday
effects are negligible, therefore remains a faithful tracer of a
feature's contribution to the low-frequency total intensity.

Secondly, many of the polarized filaments and loops are thought to
arise from the compression or shearing of an initially isotropic
random field, e.g.\ in old supernova
shells~\citep{1962MNRAS.124..125V}. As shown by
\citet{1980MNRAS.193..439L}, a compressed random field
can produce highly polarized emission,
and this emission is accompanied by an \emph{additional}
total intensity from the same anisotropic random field, spatially
correlated with the polarized intensity. The ratio of the two depends
on the viewing geometry. Defining $\psi$ as the angle between the
compression plane and the line of sight, the polarization fraction
decreases from its maximum value $\Pi(p)$ when the plane contains the
line of sight ($\psi=0$) to zero for a face-on slab
($\psi=90^\circ$), while the total intensity remains finite.
The total intensity associated with a polarized feature can therefore
be written as $\alpha\, \PI/\Pi(p)$ with a geometry factor $\alpha =
\Pi(p)/\Pi_\text{obs}(\psi) \geq 1$. Here $\Pi_\text{obs}$ denotes
the observed polarization fraction of the feature, which in general
varies across it, so that on a per-region basis only the
emission-weighted average $\langle\alpha\rangle$ is accessible.

The same argument applies if the excess polarized intensity originates
from a compression of the coherent field, for example in the bubble model
of \citet{2025A&A...695A.148P}: any compression of the plasma that amplifies
the coherent component inevitably amplifies the pre-existing random
field as well, so regions of enhanced polarized intensity again trace
an enhanced total intensity beyond that of the large-scale field
model.

\citet{2013JCAP...06..041M} modeled the foreground radio loops as
complete spherical shells of compressed field and found that this
geometry is well suited for describing the angular power spectrum of
the 408\,MHz sky. However, it is difficult to apply at the pixel
level, where the observed features are mostly incomplete arcs and
irregular filaments rather than closed circles. We therefore model the
foreground intensity at 408\,MHz as being proportional to the
\emph{excess polarized intensity} $a_i$, defined as the modulus of the
residual polarization vector at 30\,GHz after subtracting the
prediction of the coherent field model (including striation),
\begin{equation}
  a_i = \sqrt{\bigl(\Q_i - \Q_i^{\mathrm{coh}}\bigr)^2
            + \bigl(\U_i - \U_i^{\mathrm{coh}}\bigr)^2},
  \label{eq:excess}
\end{equation}
where the Stokes parameters are averaged to the fit resolution of
$N_{\rm side}=16$. Since the Stokes parameters, unlike the polarized
intensity, are additive along the line of sight, Eq.\,(\ref{eq:excess})
is the appropriate measure of the polarized emission of a localized
structure superimposed on the large-scale field: it captures excess
emission in both amplitude and orientation, i.e.\ also features that
rotate the polarization angle without increasing $\PI$, and it is
non-negative by construction. Note that $a_i$ is also nonzero where
the coherent model mismatches the observed polarization in orientation
rather than amplitude; in the foreground interpretation, this residual
is likewise attributed to an additional emitting structure. Within
each of the $N_R = \nRegions$ regions defined in Sec.\,\ref{sec:data},
we set
\begin{equation}
  I_{\text{fg},i} = f_{R(i)}\, a_i,
  \label{eq:ifg}
\end{equation}
where $R(i)$ denotes the region containing pixel $i$. Pixels outside
all regions carry no foreground term. To avoid a discontinuity of the
model at the region boundaries, the regions are extended beyond their
nominal outlines, with $a_i$ down-weighted by a Gaussian in the
angular distance to the boundary with a standard deviation of $4^\circ$,
comparable to the pixel size at the fit resolution.  The physical
interpretation of the re-weighting factors is
\begin{equation}
  f_j \simeq \langle\alpha\rangle_j\,q(p_j),
\end{equation}
where $j=1,\ldots,N_R$ labels a particular region.  By construction,
this ansatz assigns foreground intensity only where excess polarized
emission is present. It therefore captures the bright limbs of
shell-like structures, where the compression layer is seen close to
edge-on and both $\PI$ and $I$ are enhanced, but not emission from
their projected interiors, where a face-on compressed layer can retain
enhanced total intensity while its net polarization vanishes. The
shorter path through a thin layer suppresses this interior surface
brightness relative to the limb emission, but the corresponding broad,
weakly polarized contribution is not represented by our template and
constitutes a limitation of the foreground model.

Since both the compression geometry and the local electron spectral
index differ from region to region, we do not fix the $f_j$ to their
nominal values but treat them as free parameters that enter the model
prediction linearly and are determined analytically in the fit
(Sec.\,\ref{sec:analysis}). A posteriori, the fitted values can be
confronted with this interpretation: the ratio $f_j/q(p)$ estimates
the emission-weighted geometry factor $\langle\alpha\rangle$ of a
region for the nominal spectral index $p$.  Values below unity can be
interpreted as being due to a younger electron population with a
harder spectrum than the diffuse Galactic spectrum. Conversely, for
regions in which the compression layer is seen mostly edge-on,
i.e.\ $\langle\alpha\rangle \simeq 1$, the fitted $f_j$ directly
constrains the local spectral index between 408\,MHz and 30\,GHz via
Eq.\,\eqref{eq:beta}, and thus the cooling state of the electron
population in the feature. The obtained best-fit $f_j$ values are
listed in Sec.\,\ref{sec:excessPolFactor} in the appendix.

The foreground feature with the largest angular extent is the ``Fan''
region in the outer Galaxy (Fig.\,\ref{fig:polintensity}\,(b)). Here, the
limited flexibility of the data-driven ansatz is potentially most
consequential: the template can be rescaled only by a single factor
$f_j$, although both the compression geometry and the local spectral
index may vary across such an extended region. Any resulting mismatch
would therefore affect many pixels. To assess its impact, we
supplement the polarization-based template with the parametric
total-intensity model described in Sec.\,\ref{sec:outerspur}.

\section{Isotropic and Dipolar Offsets}
\label{sec:offsets}

The zero level of the synchrotron intensity reconstructed by \Planck
(Sec.\,\ref{sec:data}) uses the offset corrections of
\citet{2017A&A...597A.131W}, obtained from linear regressions
(``$T$--$T$ plots'') between the 408~MHz and 1420~MHz surveys
\citep{1986A&AS...63..205R,2001A&A...376..861R}. For the 408~MHz map,
the correction is a monopole of $(8.9\pm1.3)$~K and a dipole of
$(3.2\pm1.5,\,0.7\pm1.4,\,-0.8\pm1.5)$~K in Galactic Cartesian
coordinates. The removed offsets absorb, besides the survey zero level
and the CMB monopole, the uniform part of any emission, be it local,
Galactic, or extragalactic, unless it follows a similar frequency
scaling as the Galactic synchrotron emission and is therefore
degenerate with it.

We therefore additionally allow for a residual large-scale offset
between data and model,
\begin{equation}
  I_\text{off}(\nLOS) = m_0 + \ddip\cdot\nLOS,
  \label{eq:offset}
\end{equation}
with a free monopole $m_0$ and dipole vector \ddip{} determined
analytically for each trial field configuration
(Sec.\,\ref{sec:analysis}). These terms absorb the statistical
uncertainty of the adopted zero levels, as well as any monopole and
dipole of the sky emission itself. The latter is indistinguishable
from the instrumental offsets in the $T$--$T$ determination and hence
removed from the data, but is, in general, present in the prediction
of our parametric model.

In the convention of Eq.\,(\ref{eq:total}), a positive $m_0$
corresponds to isotropic emission present in the data but not
accounted for by the model.  Such an excess intensity may arise from
an instrumental offset or from unmodeled astrophysical
emission. Possible sources of the latter are discussed in
Sec.\,\ref{sec:mono}.

\section{Random Field Model}
\label{sec:rand}
We describe the random magnetic field in terms of its root-mean-square
field strength, $\brms = \sqrt{\langle \mathbf{b}^2\rangle}$. For a
statistically isotropic random field, the synchrotron emissivity
depends on the random field only through \brms{}
(Sec.\,\ref{sec:synchro}), and as long as the correlation length of the
turbulence is small compared to the scales over which \brms{} varies,
each line of sight samples many turbulent cells, so the observed
intensity approaches its ensemble average, independent of the power
spectrum and of the particular realization. Since the rms is thus the
only property of the random field that enters, we write $b \equiv
\brms$ for convenience in the following. For the modeling of the
random field we use a parametric ansatz to describe its large-scale
spatial dependence $b(r,z)$.  In the following we consider a minimal
disk model, an extension with an annular enhancement in the inner
Galaxy, a variant with spiral arms, and, as a check on unmodeled
emission, a phenomenological foreground spur in the outer Galaxy.

\subsection{Disk Model}
Motivated by total-intensity radio observations of
edge-on spiral galaxies
\citep[e.g.][]{2015AJ....150...81W,2018A&A...611A..72K,
  2025A&A...699A.243H}, we model the large-scale vertical
profile of the random field as a disk,
\begin{equation}
  \bdisk(r,z)
  =
  \bdisksun\,
   \Vdisk(z)\, \Sdisk(r)\, \Edisk(r) / \Edisk(\Rsun),
  \label{eq:bdisk}
\end{equation}
with amplitude \bdisksun{} at the solar circle\footnote{\label{fnt:grav}For
  consistency with the \UFCoh coherent field analysis we use the
  \textsc{Gravity} Collaboration's earlier estimate,
  $\Rsun = 8.178$~kpc \citep{2019A&A...625L..10G}, close to their
  subsequent value of $\Rsun = 8.277$~kpc
  \citep{2022A&A...657L..12G}.} and the radial profile
\begin{equation}
  \Edisk(r) = \exp\left(-\frac{\sqrt{r^2+\rcusp^2}}{\lrad}\right),
  \label{eq:edisk}
\end{equation}
with scale length \lrad. The parameter \rcusp{} smooths the cusp of
the exponential profile toward the Galactic center. Our fits yield
only an upper bound, $\rcusp \lesssim 1\,\mathrm{kpc}$, and we fix
$\rcusp = 1\,\mathrm{kpc}$. In the outer Galaxy the field is smoothly
suppressed by the sigmoid $\Sdisk(r) = [1 + e^{(r -
    \Rmax)/\wmax}]^{-1}$, with transition radius \Rmax{} and
transition width \wmax. Note, however, that the field cannot
be constrained beyond the radial extent of \ncre, and
present cosmic-ray electron density models do not extend to large
radii, as seen in the right panel of Fig.\,\ref{fig:ncreDens}.

We truncate the random disk at $\Rmax = 18$~kpc with a transition
width of $\wmax = 2\,\mathrm{kpc}$, the smallest truncation radius at
which the increase in $\chi^2$ over the untruncated fit stays within
$\Delta\chi^2 = 9\;(3~\sigma)$ for all models considered, so that the
model as provided does not carry the field into a region where the
data have no discriminating power.

The vertical profile of the disk is a hyperbolic secant,
\begin{equation}
  \Vdisk(z) = \mathrm{sech}\left(\mathrm{arcosh}(e)\,z/\zdisk\right),
  \label{eq:vdisk}
\end{equation}
motivated by the vertical density profile of a self-gravitating
isothermal gas layer, $\rho \propto \mathrm{sech}^2(z/z_0)$
\citep{1942ApJ....95..329S}: a magnetic field in pressure
equipartition with such a layer would have $b^2$ following this
profile and hence $b \propto \mathrm{sech}$. We allow the height of
the profile to be a free parameter of the fit, not tied to the scale
height of the gas in the Galactic disk. The factor $\mathrm{arcosh}(e)
\approx 1.657$ normalizes the argument such that \zdisk{} is the
characteristic height at which the profile drops to $1/e$ of its
mid-plane value.  Since the data mainly constrain the characteristic
height of the emitting layer and are only weakly sensitive to the
detailed shape of the vertical profile, we also consider a model
variant with a Gaussian disk profile, $\Vdisk(z) =
\exp(-(z/\zdisk)^2)$, and an exponential variant, $\Vdisk(z) =
\exp(-|z|/\zdisk)$, both again satisfying $\Vdisk(\zdisk) =
\Vdisk(0)/e$.

\subsection{Disk+Ring Model}
As a more elaborate random field model, we extend the basic disk and
allow for a region of modified random field in the inner Galaxy which
can arise, e.g., from supernova-driven turbulence tracing the peak of
the star-formation rate at a few kpc, and/or from the stirring of the
interstellar medium by the rotating bar. We model this enhancement as
an axisymmetric annular component,
\begin{equation}
  \bann(r,z)
  =
  \bannpeak\, G(z;\, \zann)\, G(r-\rann;\, \wann),
  \label{eq:bannulus}
\end{equation}
where \bannpeak{} is the peak amplitude,
\begin{equation}
  G(x;\sigma) = \exp\left(-\frac{x^2}{2\sigma^2}\right),
  \label{eq:gauss}
\end{equation}
denotes a Gaussian, \rann{} is the fitted peak radius, \wann{} the
radial width, and \zann{} the vertical Gaussian width. In this model,
the disk is not smoothed at the center ($\rcusp = 0$) but instead is
phased out toward small radii by an additional multiplicative factor
$G(\min(r-\rann,0);\, \wann)$ in Eq.\,\eqref{eq:bdisk} so that for $r <
\rann$, the disk field declines toward the center following the same
Gaussian that forms the inner flank of the annular component. The rms
strengths of the two components add in quadrature, so the total random
field of the ring model is $\sqrt{\bdisk(r,z)^2 + \bann(r,z)^2}$.

\subsection{Spiral Model}
\label{sec:spiral}
Previous models included spiral arms in the random field, and the
synchrotron data can in principle constrain such structure. As pointed
out by \citet{1985A&A...153...17B} and \citet{2010MNRAS.401.1013J},
spiral arms in the magnetic field strength imprint step-like features
on the longitude profiles of the low-latitude synchrotron intensity at
the well-known arm tangents \citep[e.g.][]{2022NewA...9701896V}, see
also Fig.\,\ref{fig:intensity-slices} below.  Similar features in other
line-of-sight--integrated emissions can likewise be interpreted as
tracers of spiral structure
\citep[e.g.][]{2010ApJ...722.1460S,2025CoPhC.31109537K}.

Here we test a spiral random field component with a four-arm
logarithmic spiral geometry plus the Local Arm with the geometry of
\citet{2017ApJ...835...29Y}, adapted to our choice of \Rsun
(footnote~\ref{fnt:grav}) and with the radial evolution of the arm
widths as determined by \citet{2019ApJ...885..131R}. We allow for
different magnetic field strengths in each arm following $b_i
(r/r_\text{ref})^\alpha$ and test $\alpha=0$ and $-1$.  We find that the
arm parameters are strongly degenerate with simpler axisymmetric
structures.  This is also illustrated for the [\ion{N}{2}] emission in
Fig.\,4 of \citet{2025CoPhC.31109537K}: a readjustment of the arm
normalizations creates a longitude-symmetric bump in the inner-Galaxy
longitude profile at the tangents of the Scutum and Norma arms at
$\ell \approx \pm 32\degr$, indistinguishable from that of a ring of
radius $\rann = \sin(32\degr)\,\Rsun \approx 4.3\,\mathrm{kpc}$.

In addition to the anticipated difficulty in distinguishing between
spiral arms and axisymmetric structure in the inner Galaxy, spiral
arms in the outer Galaxy pose a different problem: an arm that
continues beyond its tangent into the outer Galaxy, such as the
Perseus arm, contributes over a wide range of longitudes.  Hence a
localized overdensity in random field and \ncre{} which produces a
sizable region of excess synchrotron intensity can drive the fit to
the strength of the entire arm.  Precisely this happened in previous
modeling efforts, which did not include the foreground intensities as
we do here (see Sec.\,\ref{sec:fg}).  In particular, the large
intensity toward the outer Galaxy drove the exceptionally large random
field assigned to arm~7 in the \citet{2012ApJ...761L..11J} model ($b_7
= 37\,\muG \times (5\,\mathrm{kpc}/r)$). In our test of models with a
spiral component described in Sec.\,\ref{sec:sysComplexity}, we
therefore restrict the azimuthal extent of the arms to $200\degr$,
preventing their outer segments from contributing over a broad range
of outer-Galaxy longitudes. The fitted arm strengths are then driven
primarily by the tangent features in the inner Galaxy by construction.

\subsection{Outer Spur}
\label{sec:outerspur}

We also consider a fit variant that includes a parametric foreground
component in the outer Galaxy, motivated by extended total-intensity
emission that may not be fully captured by our polarization-based
templates. We represent this component as a localized ``spiral spur.''
Importantly, the synchrotron data used here constrain neither its
line-of-sight extent $\Delta s$ nor its distance from us.\footnote{For
  the related polarized emission toward the Galactic anticenter,
  \citet{2017MNRAS.467.4631H} argued that a substantial fraction
  originates at distances $\gtrsim2$~kpc, in or beyond the Perseus
  arm. \citet{2025A&A...693A.284K} likewise modeled it as a
  Galactic-scale feature with a major contribution from the Perseus
  arm, whereas \citet{2021AJ....162...35W} modeled it as a local
  filament about 1~kpc from the Sun.} For concreteness, we place it at
a distance of 1~kpc; this choice only determines the position at which
the unperturbed cosmic-ray electron normalization $n_0$ is
evaluated. If the intensity enhancement originates from a local
perturbation, then the actual normalization $n_\text{os}$ could be
larger, with a ratio $\kappa_\text{os} = n_\text{os} / n_0$; see the
discussion of the density enhancement in adiabatic compression in
Sec.\,\ref{sec:mono}. The observed intensity constrains only the
product \ospurNorm{}. The remaining geometrical parameters of this
phenomenological component are its central longitude,
$\ell_\text{os}$, and its full angular widths in longitude and
latitude, \ospurDeltaL{} and \ospurDeltaB{}, respectively.

\subsection{Summary}

In summary, we consider two main models of the large-scale random
field: the minimal \emph{disk} model with three free parameters
(\bdisksun, \lrad, \zdisk), and the \emph{ring} model, which adds the
annular enhancement and the inner phase-out of the disk, with four
further parameters (\bannpeak, \rann, \wann, \zann), for a total of
seven. We also test a spiral random field component constrained, by
construction, by the intensity in the inner Galaxy.  A final
outer-spur variant supplements the global random field model with a
localized foreground component; it is used to assess the sensitivity
of the results to unmodeled emission in the outer Galaxy. The
comparison of these model variants is presented in
Sec.\,\ref{sec:sysComplexity}.

\section{Analytic Insights}
\label{sec:synpar}

Before turning to the fit of the data, we derive analytic
approximations for the synchrotron intensity in two simplified limits.
These reveal which features of the sky constrain the random-field disk
parameters. For this purpose, we approximate the random-field
emissivity as $\varepsilon_{\rm rand}\propto\ncre\,b^2$, corresponding
to $(p+1)/2\simeq2$ in Eq.\,\eqref{eq:powerlaw}, and omit the coherent
field and foreground contributions.  In this section we denote the
magnitude of the Galactic latitude as $\beta$.

In the plane-parallel slab approximation, an observer in the mid-plane of a
horizontally homogeneous emission layer measures
\begin{align}
I(\ell,\beta)
& = m_0+\int_0^\infty
\varepsilon\left(s\sin\beta\right)\,\mathrm{d}s \\
& = m_0+\frac{1}{\sin \beta}
\int_0^\infty\varepsilon(z)\,\mathrm{d}z.
\label{eq:coseclaw}
\end{align}
Equation~(\ref{eq:coseclaw}) is the classical cosecant law,
independent of the shape of the vertical profile. If \ncre{} and $b$
decline exponentially with scale heights $h_\text{cre}$ and \zdisk,
the emissivity scale height is
$h_\varepsilon=\left(2/\zdisk+1/h_\text{cre}\right)^{-1}$.  For
$h_\text{cre}\gg\zdisk$ the electron density is nearly constant across
the emitting layer and the vertical column reduces to
\begin{equation}
\int_0^\infty\varepsilon(z)\,\mathrm{d}z
\propto
n_0\,
b_0^2\,\zdisk\,c_V,
\label{eq:vintegral}
\end{equation}
where $n_0$ and $b_0$ are the mid-plane cosmic-ray density and
magnetic field strength, and $c_V$ is the vertically averaged squared
field profile
\begin{equation}
 c_V\equiv
\frac{1}{\zdisk}\int_0^\infty\Vdisk^2(z)\,\mathrm{d}z~.
\label{eq:verticalColumn}
\end{equation}
 For the sech, Gaussian, and exponential profiles,
 $c_V=1/\mathrm{arcosh}(e)\approx 0.60$, $\sqrt{\pi/8} \approx 0.63$,
 and $1/2$, respectively. The latitude dependence of the intensity
 \eqref{eq:coseclaw} robustly fixes the monopole $m_0$ and the product
 $n_0\,b_0^2\,\zdisk\,c_V$. The near-equality of the coefficients then
 explains why the three profiles yield similar values of $b_0$ and
 \zdisk{} in the numerical fits.

The individual parameters $b_0$ and $\zdisk$ can be constrained
separately only through departures from the plane-parallel limit. In
our fits this is done numerically, but the origin of this sensitivity
can be understood from a tangent-point approximation. For lines of
sight toward the inner Galaxy, $|\ell|<90\degr$, an inward increase in
the emissivity weights the intensity toward the segment closest to the
Galactic center. The projected distance to this tangent point is
$s_{\rm t}\cos\beta=\Rsun\cos\ell$, and its height above the mid-plane
is therefore
\begin{equation}
z_{\rm t}=\Rsun\cos\ell\tan\beta.
\end{equation}
The latitude profile thus traces the vertical emissivity profile near
the tangent point, while the longitude dependence of its width carries
information on \zdisk. In this approximation, the intensity falls to
$1/e$ of its in-plane value when $z_{\rm t}$ reaches the emissivity
scale height $h_\varepsilon\simeq\zdisk/2$, giving
\begin{equation}
\label{eq:bethalf}
\tan\beta_{1/e}
\approx
\frac{\zdisk}{2\,\Rsun\cos\ell}.
\end{equation}

In summary, assuming \ncre{} is known, the sensitivity of the
synchrotron sky to the magnetic-field parameters can be understood as
follows. In the plane-parallel approximation, the latitude dependence
fixes the column $b_0^2\,\zdisk\,c_V$ and the global offset $m_0$. The
in-plane longitude profile constrains the radial emissivity gradient
and hence the radial scale length \lrad{} of the random field. Finally,
the latitude widths of inner-Galaxy sightlines constrain \zdisk{}
through Eq.\,\eqref{eq:bethalf}, providing the additional information
needed to determine $b_0$ separately.

\section{Data Analysis}
\label{sec:analysis}
We model the measured intensity $I_i$ at pixel $i$ as
\begin{equation}
  I_i = \mu_i + \epsilon_i,
\end{equation}
with
\begin{equation}
  \mu_i = r_i(\boldsymbol\eta) + k\,c_i + m_0
        + \ddip\cdot{\nLOS}_i + f_{R(i)}\,a_i,
  \label{eq:model}
\end{equation}
where $r_i$ is the synchrotron intensity produced by the random field
with model parameters $\boldsymbol\eta$ (Secs.\,\ref{sec:synchro}
and \ref{sec:rand}), $c_i$ is the intensity from
the coherent field scaled by a global normalization $k$ associated
with the extrapolation between \Planck and Haslam frequencies, $m_0$ is
a global monopole (constant offset), \ddip{} is a global dipole vector
contracted with the unit direction vector ${\nLOS}_i$ of the pixel,
and $a_i$ is the excess polarized intensity scaled by a factor $f_R$
for each region $R(i)\in\{1,\ldots,N_R\}$; for pixels outside the
foreground regions this term is absent. The residuals $\epsilon_i$
are assigned the variance $\mathrm{Var}(\epsilon_i) =
\sigma_{I,i}^{2} + f_{\mathrm{ref}}^{2}\,\sigma_{a,i}^{2}$, where
$\sigma_{I,i}^{2}$ and $\sigma_{a,i}^{2}$ are the pixel variances of
$I_i$ and $a_i$, and the second term is the propagated contribution of
$a_i$, evaluated at the fixed reference value $f_{\mathrm{ref}} =
q(3.2)$ (see Eq.\,(\ref{eq:beta})). This choice keeps the fit weights
$w_i = 1/\mathrm{Var}(\epsilon_i)$ parameter-independent, so that the
minimization of the $\chi^2$ remains a linear least-squares problem in
the nuisance parameters.  As in our previous analysis of the \UFCoh
coherent field, we use the observed variance of the unmasked
high-resolution subpixels within each $N_{\rm side}=16$ pixel to
define the fit weights. We interpret this variance as an empirical
uncertainty of the binned datum, not as the statistical variance of
the mean, since the subpixels are correlated by the finite angular
resolution of the maps and by coherent Galactic variance.

The full parameter vector of the synchrotron fit splits into two
groups,
\begin{equation}
  \boldsymbol\Theta = (\!\!\!\!\underbrace{\boldsymbol\eta}_{\text{GMF}}\!\!\!\!,\;\underbrace{k,\, m_0,\, \ddip,\, \bm{f}}_{\text{nuisance}}\,),
\end{equation}
with $\bm{f} = (f_1,\ldots,f_{N_R})^{\!\top}$. The parameters
$\boldsymbol\eta$ of the random field model affect $r_i$ non-linearly
and are optimized numerically using \MINUIT~\citep{1975CoPhC..10..343J}.
The nuisance parameters $\boldsymbol\theta = (k, m_0, \ddip, \bm{f})$,
on the other hand, enter Eq.\,\eqref{eq:model} linearly. Their best-fit
values at given $\boldsymbol\eta$ therefore follow from the
stationarity condition of the weighted sum of squared residuals,
\begin{equation}
  \chi^{2} = \sum_i w_i\,(I_i - \mu_i)^{2},
  \qquad
  \frac{\partial\chi^{2}}{\partial\boldsymbol\theta} = \bm{0},
  \label{eq:chi2}
\end{equation}
a linear system of $5+N_R$ equations that we solve exactly in each
\MINUIT{} iteration (Appendix~\ref{sec:opti}). The uncertainties and
correlations of the nuisance parameters are given by the inverse of
the coefficient matrix of this system.

The rescaling factors $f_R$ quantify additional emission and are
therefore physically non-negative. We enforce this constraint
iteratively: regions for which the unconstrained solution of
Eq.\,\eqref{eq:chi2} yields a negative factor are removed from the
parameter vector (i.e.\ their factors are set to zero) and the fit is
repeated until only non-negative factors remain, making the
optimization a non-negative linear least-squares fit.

As a final step, we guard against features not captured by our
foreground model that could bias the inferred global parameters. From
the initial best-fit solution, we compute the pull of each pixel,
$\Delta_i=\sqrt{w_i}\,(I_i-\mu_i)$, and refit after excluding pixels
with $|\Delta_i|\geq3$. For the $N=3048$ pixels entering the initial fit,
Gaussian fluctuations alone would yield about eight pixels above this
threshold, $N\,\mathrm{erfc}(3/\sqrt{2})\simeq8$. In practice, 78
pixels (2.6\%) are excluded. Their spatial distribution is mixed (see
Fig.\,\ref{fig:intensityFit}): most excluded pixels are scattered
across the sky in groups of one or two, while the two largest coherent
clusters (9 and 18 pixels) coincide with intensity enhancements near
the upper end of the North Polar Spur at
$(\ell,b)\simeq(-60^\circ,+60^\circ)$ and with the brightest part of a
broader emission swath near $(150^\circ,-45^\circ)$, plausibly
associated with Radio Loop~II. The emission in the regions thus
excluded is undoubtedly produced by localized phenomena which should
not be allowed to influence the fit to the large-scale random field;
the excluded pixels inflated the $\chi^2$ of the initial fit by about
1000.

\section{Impact of Modeling Choices}
\label{sec:modelingChoices}

As will become clear below, we find that two basic models are
sufficient---a pure disk and a disk+ring---with model variants being
subsumed into uncertainties in the basic models' parameters.  In this
section we investigate the sensitivity of the fit to details of
various modeling choices. For this purpose, we repeat the fit with
alternative assumptions on the coherent magnetic field, the cosmic-ray
electron density, the pixel mask, and the shape and symmetry of the
vertical profile of the random field. The corresponding best-fit
parameters of these model variants are collected in
Tables~\ref{tab:Icoh} to \ref{tab:heightNS} in the Appendix.  Model
parameters mentioned below are defined in Sec.\,\ref{sec:rand}.

\subsection{Coherent Intensity}
\label{sec:sysIcoh}

First we study the dependence of the results on the assumed
synchrotron intensity of the coherent GMF, $I_\text{coh}$ in
Eq.\,\eqref{eq:total}. We repeat the fit for five variants of the
coherent-field model: \modelBase, \modelXr, \modelNe, \modelKappa, and
\modelTwist{} (fits a1--a5 in Table~\ref{tab:Icoh}). The random field
parameters are stable under this variation: \bdisksun{} changes by
at most $0.2~\muG$ and \zdisk{} by at most $0.02$~kpc, i.e.,
within one to two statistical standard deviations. Larger,
anticorrelated variations occur for the radial scale-length \lrad{}
and $k$, the scaling to 408~MHz.  The \modelXr{} variant, which
exhibits the largest coherent intensity toward the inner Galaxy,
prefers a flat radial profile with $\lrad = 21$~kpc and $k = 0.77$,
whereas the other variants give $\lrad = 12$--15~kpc and $k =
1.05$--1.27. The best description of the data is obtained with
\modelTwist{} (a5), which is better by $\Delta\chi^2 \geq 145$ than
the other variants; we therefore adopt a5 as the reference fit in the
following.

\subsection{Cosmic-Ray Electrons}
\label{sec:sysNcre}

The synchrotron emissivity is proportional to the product of the
cosmic-ray electron density and a power of the transverse
magnetic-field strength, so the fitted field parameters depend on the
assumed electron distribution.  To judge the sensitivity of the
inferred random field to this degeneracy, we vary the half-height
\hDiff{} of the CR-electron diffusion halo between 2 and 10~kpc (fits
b1--b4 in Table~\ref{tab:ncre}) and also use the electron density
model of \citet[][OS13]{2013MNRAS.436.2127O} (fit b5). For each
electron model, we adopt the corresponding \UFCoh{} coherent
intensities, rederived with that model; fit a1 therefore serves as the
comparison here. Note that small halo heights, $\hDiff \leq 4$~kpc,
are problematic in the coherent-field analysis: the corresponding
\UFCoh fits yield only a lower limit on the height of the toroidal
halo. These solutions are therefore not included in the standard
\UFCoh ensemble.

The parameters of the random field are remarkably stable against the
assumed halo height: for $\hDiff \geq 4$~kpc, all parameters agree
within about one statistical standard deviation, since for
$\hDiff \gg \zdisk$ the electron density in the emission region
becomes insensitive to the halo height. Only for the extreme case of
$\hDiff = 2$~kpc does the fit compensate the vertically concentrated
electrons with $\zdisk = 0.69$~kpc and an increased \lrad{} and $k$.
The OS13 model illustrates the additional dependence on the spatial
form of \ncre{} at fixed halo height: compared to fit b4, with the same
$\hDiff = 10$~kpc, it gives $\zdisk = 1.23$~kpc and
$\bdisksun = 3.8~\muG$.

The quality of the fit to the synchrotron emission depends only weakly
on the electron model ($\chi^2/n_\mathrm{df} = 1.38$--1.42). Excluding
the disfavored $\hDiff = 2$~kpc fit, the systematic uncertainties are
dominated by the choice of the electron model itself, about
$({+0.33}/{-0.04})$~kpc on \zdisk{} and $({+0.2}/{-0.7})~\muG$ on
\bdisksun, well above the statistical uncertainties.

These variations are, however, strongly anticorrelated, as anticipated
by the analytic discussion of Sec.\,\ref{sec:synpar}: the latitude
dependence of the intensity fixes the emission column
$n_0\,\bdisksun^2\,\zdisk\,c_V$ (Eq.\,\ref{eq:vintegral}), while the
local cosmic-ray electron density $n_0$ is normalized to the same
Voyager\,I and AMS-02 data in all models considered. The choice of electron
model therefore mainly redistributes the field between amplitude and
thickness rather than changing the emission column that the data
actually determine.  Accordingly, the vertical column $\int_0^\infty
\bdisk^2(\Rsun,z)\,{\rm d}z$ of Table~\ref{tab:parSummary} below is
determined to $9\%$, much better than one would infer by propagating
the uncertainties of \bdisksun{} and \zdisk{} independently.

\subsection{Pixel Mask}
\label{sec:sysMask}

The default mask excludes pixels with an absolute pull $|\Delta| \geq
3$ and retains 97\% of the sky. To check whether the results are
driven by the Galactic plane, where the model is expected to be least
reliable, we redo the fit excluding, in addition to the default mask,
iso-latitude rings of \HEALPix cells around the Galactic plane (fits
c1--c3 in Table~\ref{tab:mask}), retaining 95\%, 91\%, and 87\% of the
sky.  Excluding the Galactic plane changes all parameters by less than
about one standard deviation, showing that the results for the random
field model are not driven by the Galactic plane; the mild increase of
$\chi^2/n_\mathrm{df}$ from 1.35 to 1.39 shows that the residuals are
not concentrated at low latitudes. Tightening the pull cut on
individual pixels to $|\Delta| < 2$ (fit c4) yields
$\chi^2/n_\mathrm{df} = 0.85$, close to the value of 0.77 expected for
Gaussian residuals after a $2\sigma$ pull cut (0.97 for the default
cut at $3\sigma$). The fitted random-GMF parameters \bdisksun, \zdisk,
and \lrad{} change by less than about one standard deviation. This
stability shows that the default $3\sigma$ mask is adequate for
determining the random GMF, and that the fit is both robust and not
biased by remaining outliers.

\subsection{Height Dependence}
\label{sec:sysHeight}

Replacing the default sech vertical profile by an exponential or
Gaussian profile (fits d1 and d2 in Table~\ref{tab:heightNS}) changes
the fit by $\Delta\chi^2=-11$ and $+3$, respectively. Thus, the
exponential is modestly preferred, whereas the Gaussian and sech
profiles are essentially indistinguishable. Given the overall excess
of $\chi^2/n_{\mathrm{df}}$ above unity and the presence of structured
residuals, we do not regard this modest difference as robust evidence
for a particular functional form and retain sech as the default
choice. The corresponding parameter ranges are
$\zdisk=0.91$--$0.99$~kpc and $\bdisksun=4.3$--$4.7\,\muG$.

\begin{figure*}[t]
  \centering
  \includegraphics[width=0.9\textwidth]{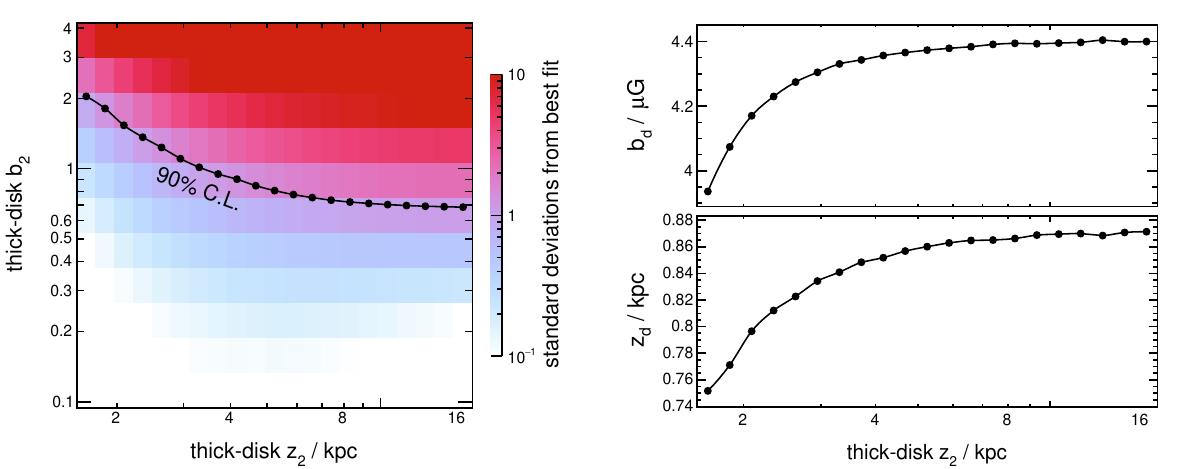}
  \caption{Profile-likelihood scan of the thick-disk parameters,
    $(b_2, z_2)$.  \emph{Left:} Deviation from the single-disk best fit, in
    standard deviations, as a function of the thick-disk scale height
    $z_2$ and field strength $b_2$, profiled over the remaining model
    parameters. The line indicates the 90\% C.L.\ upper limit contour
    on $b_2$. \emph{Right:} Values of the refit standard-disk
    parameters along this contour: field strength $\bdisksun$ (top)
    and scale height $\zdisk$ (bottom).}
  \label{fig:deltaSigma}
\end{figure*}

\subsection{North-South Symmetry}
\label{sec:sysNS}
In fit d3 of Table~\ref{tab:heightNS} we allow for independent field
strengths, scale heights, and radial scale lengths above and below the
Galactic plane. The field strengths of the two hemispheres agree
within 7\% ($\bdisksun^\mathrm{N} = 4.2~\muG$ versus
$\bdisksun^\mathrm{S} = 4.5~\muG$) and the radial scale lengths within
10\% ($\lrad^\mathrm{N} = 13$~kpc versus $\lrad^\mathrm{S} = 12$~kpc),
while a moderately thicker northern disk, $\zdisk^\mathrm{N} =
1.10$~kpc versus $\zdisk^\mathrm{S} = 0.83$~kpc, is slightly preferred
with all other parameters essentially unchanged. The improvement of
$\Delta\chi^2 = 20$ for three additional degrees of freedom is
nominally significant, but small compared to the overall excess of
$\chi^2/n_\mathrm{df}$ above unity. We therefore retain the symmetric
profile as our default and include the N--S difference as an
indicative systematic uncertainty of the scale height.

It is reassuring that the two hemispheres yield such overall
consistent parameters, given that the northern and southern skies contain
very different local foreground structures. Strong residual foreground
contamination that is asymmetric between north and south would tend to drive the fitted
parameters of the two hemispheres apart. Their approximate equality
therefore provides a useful consistency check: if the random field of
the Galaxy is intrinsically symmetric about the plane, it implies that
our foreground treatment removes most spurious structures that would
otherwise bias the fitted field parameters.

\subsection{Dipolar Offset and Outer Spur}
\label{sec:dipoSpur}

The fit parameters for the \ModelD{} configuration without and with an
outer spur are given in Tables~\ref{tab:Icoh} and \ref{tab:DF},
respectively; those for the \ModelDR{} configuration are given in
Tables~\ref{tab:DR} and \ref{tab:DRF}. When included, the spur is
typically centered at $\ospurL\sim170^\circ$, with broad full angular
widths of $\ospurDeltaL\sim70^\circ$ and
$\ospurDeltaB\sim60^\circ$. Its central longitude is close to the
direction of the fitted dipole in models without a spur, typically
$\ell\sim150^\circ$ with $|b|\lesssim20^\circ$. The two components
therefore absorb much of the same broad feature on the sky and are
strongly degenerate.

For example, the best-fit dipole vectors of the \ModelDR{} and
\ModelDRF{} models are $(-2.3\pm0.1,\,0.7\pm0.1,\,0.7\pm0.1)$~K and
$(-0.2\pm0.1,\,1.2\pm0.1,\,0.6\pm0.1)$~K, respectively. Adding the
spur thus reduces the dipole amplitude from approximately $2.5$ to
$1.4$~K and shifts its direction from
$(\ell,b)\simeq(163^\circ,16^\circ)$ to $(100^\circ,27^\circ)$. This
degeneracy should be considered together with the substantial
uncertainty in the dipolar zero level. The dipole correction of
\citet{2017A&A...597A.131W}, used in the \Planck reconstruction, has
uncertainties of about 1.5~K in each component (see
Sec.\,\ref{sec:offsets}), comparable to the magnitudes of the
components inferred here. Residual errors in the dipole correction
applied to the Haslam survey can therefore plausibly account for at
least part of the fitted signal. We consequently regard the dipole and
outer-spur terms primarily as alternative phenomenological
descriptions of the same broad residual, and do not assign a robust
astrophysical interpretation to either the fitted dipole or the
detailed spur parameters.

\subsection{Thick Disk}
\label{sec:thickdisk}

The synchrotron data constrain the random field only within the layer
populated by cosmic-ray electrons. To quantify how much additional
turbulent field could be present at larger heights, we superimpose a
second, thicker disk on the fiducial disk. It has the same radial
scale length, \lrad{}, and a vertical profile
$\mathrm{sech}\!\left[\mathrm{arcosh}(e)\,z/z_2\right]$, where $z_2$
is the $1/e$ scale height and $b_2$ is the mid-plane field strength of
the thick disk at the solar circle. We fix $(b_2,z_2)$ on a grid and
refit all remaining parameters at each point. For
$z_2\lesssim2\,\zdisk$, the two vertical components become degenerate
and only their combined contribution is constrained. We therefore
regard only $z_2\gtrsim2$~kpc as a physically distinct thick disk,
although we begin the scan at $z_2>1$~kpc to display the onset of this
degeneracy.

The fit results are shown in Fig.\,\ref{fig:deltaSigma}. In the left panel,
we display the deviation from the best fit in units of standard
deviations, $n_\sigma =
\sqrt{\alpha\left[\chi^2(b_2,z_2)-\chi^2_0\right]}, $ where $\chi^2_0$
is the minimum obtained without the thick-disk component and
$\alpha=n_{\mathrm{df}}/\chi^2_0$ is the error-inflation factor
recommended by \citet{Rosenfeld:1975fy}, which accounts for the
larger-than-unity reduced $\chi^2$ of the single-disk fit. The
one-sided upper limit at 90\% confidence level is shown as a line at
$n_\sigma=1.28$. The two panels on the right show how the
parameters of the fiducial disk, $\bdisksun$ and $\zdisk$
adjust themselves in response to the presence of a thick disk of given thickness. At small
$z_2$, the fit reduces both parameters to compensate for the
contribution of the additional component. As $z_2$ approaches the
cosmic-ray diffusion-halo height, $\hDiff=6$~kpc in this model, the
upper limit on $b_2$ approaches $0.7~\upmu\mathrm{G}$. We conclude
that while the data cannot exclude the presence of a random field at
large Galactic heights, any such component is tightly constrained
through the mid-plane field strength; as far as our analysis can tell,
there is no evidence for a thick disk.

\begin{table}[t]
  \centering\small
\caption{Fit quality of models of increasing complexity. The number of
  free and fixed parameters refers to the random-field model; the
  normalization and zero-level parameters common to all fits are not
  counted.}
  \begin{tabular}{l|ccc}
    \multirow{2}{*}{name} & \multicolumn{2}{c}{parameters} &  \multirow{2}{*}{$\chi^2/n_\mathrm{df}$} \\
    & free & fixed & \\\hline
    \ModelD  & 3 & 3 & $3982/2946 = 1.35$ \\
    \ModelDR & 7 & 2 &  $3890/2942 = 1.32$ \\
    \ModelDF & 7 & 3 & $3683/2942 = 1.25$ \\
    \ModelDRF & 11 & 3 & $3499/2938 = 1.19$ \\
    \ModelDRS & 14 & 22 & $3879/2935 = 1.32$ \\
\end{tabular}
\label{tab:models}
\end{table}

\subsection{Model Complexity}
\label{sec:sysComplexity}

Finally, we compare the fit quality of models of increasing
complexity, as listed in Table~\ref{tab:models}. Adding a ring
component to the disk-only model gives ``\ModelDR{}'', having four
additional free parameters; this improves the fit by $\Delta\chi^2 =
92$. Although nominally significant, the improvement in reduced
$\chi^2$ is small.

Adding an outer spur (\ModelDRF) yields a large improvement of
$\Delta\chi^2 = 391$ with four more free parameters, reaching
$\chi^2/n_\mathrm{df} = 1.19$ due to a significantly better
description of the intensity toward the outer Galaxy. In fits without
the spur component, part of that emission is absorbed by the dipole,
so the dipole changes significantly when the spur is added, whereas
the random-field parameters remain stable. Since the outer spur is an
ad hoc component, we use both \ModelDR{} and \ModelDRF{} results to
define the range of possible parameter values.

In the \ModelDRS{} model, we add to \ModelDR{} a full spiral pattern
with seven additional free parameters (field strength in five arms and
one common scale height and width factor). The fixed geometry of the
4+1 spiral arms is described by 22 fixed parameters, see
Sec.\,\ref{sec:spiral}. The resulting improvement of $\Delta\chi^2 =
11$ with respect to \ModelDR{} is not significant, and we therefore do
not include spiral arms in the random field in this analysis.

\begin{table}[t]
 \caption{Parameters of the two representative random-field models.
For \modelRing{}, only a lower limit on \lrad{} is obtained; the adopted
value is given in parentheses. The derived rows give the one-sided
vertical integral of $b^2$, relevant for synchrotron intensity
(Sec.\,\ref{sec:synpar}) and UHECR deflections
(Sec.\,\ref{sec:uhecrDeflections}). The final three rows are fixed
parameters.\label{tab:parSummary}}
 \centering
\begin{tabular}{clr@{\,${}\pm{}$\,}lr@{\,${}\pm{}$\,}ll}
  \toprule
  & parameter &\multicolumn{2}{c}{\modelDisk} & \multicolumn{2}{c}{\modelRing} \\
  \midrule
  \multirow{7}{*}{\rotatebox[origin=c]{90}{\footnotesize GMF parameters}}
  & \bdisksun{}                & $4.4$ & $0.6$ & $4.1$ & $0.6$                       & \muG   \\
  & \zdisk{}                   & $1.0$ & $0.2$ & $0.9$ & $0.1$                       & kpc    \\
  & \lrad{}                    & $15$  & $6$   & \multicolumn{2}{c}{$\geq 30\,(100)$}& kpc    \\
  & \bannpeak{}                & \multicolumn{2}{c}{--} & $4.1$ & $0.7$              & \muG   \\
  & \rann{}                    & \multicolumn{2}{c}{--} & $4.5$ & $0.6$              & kpc    \\
  & \wann{}                    & \multicolumn{2}{c}{--} & $2.0$ & $0.2$              & kpc    \\
  & \zann{}                    & \multicolumn{2}{c}{--} & $1.9$ & $0.2$              & kpc    \\ \cmidrule{2-7}
  \multirow{5}{*}{\rotatebox[origin=c]{90}{\footnotesize nuisance}}
  & $k$                         & $0.9$ & $0.4$ & $0.7$ & $0.4$  &     \\
  & $m_0$                       & $5.2$ & $0.9$ & $6.0$ & $1.0$  & K   \\
  & $\ell_d$                    & $127$ & $45$  & $118$ & $46$   & deg.\\
  & $b_d$                       & $20$  & $9$   & $19$  & $9$    & deg.\\
  & $|\mathbf{d}|$              & $2.2$ & $0.7$ & $2.1$ & $0.7$  & K   \\
  \cmidrule{2-7}
  \multirow{2}{*}{\rotatebox[origin=c]{90}{\footnotesize derived}}
  & $\int_0^\infty \bdisk^2(\Rsun,z)\,{\rm d}z$ & $11$ & $1$ & $10$ & $1$              & \muG$^2\,$kpc\\
  & $\int_0^\infty \bann^2(\rann,z)\,{\rm d}z$ & \multicolumn{2}{c}{--} & $30$ & $9$ & \muG$^2\,$kpc\\
  \cmidrule{2-7}
  \multirow{3}{*}{\rotatebox[origin=c]{90}{\footnotesize fixed}}
  & \rcusp & \multicolumn{2}{c}{1}  & \multicolumn{2}{c}{--} & kpc \\
  & \Rmax  & \multicolumn{2}{c}{18} & \multicolumn{2}{c}{18} & kpc \\
  & \wmax  & \multicolumn{2}{c}{2}  & \multicolumn{2}{c}{2}  & kpc \\
  \bottomrule
\end{tabular}

\end{table}

\section{Results}
\label{sec:results}

\def\fitmodel{twistX_corr}
\newcommand{\fitpic}[1]{fitPics/\fitmodel_#1.pdf}

\begin{figure*}[t]
\def\labelpad{1.8cm}
\gridline{\fig{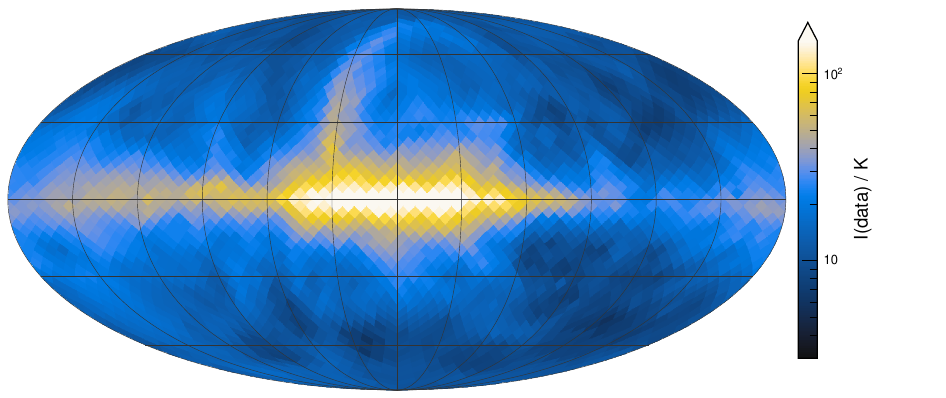}{1.37\columnwidth}{(a) Data.\hspace*{\labelpad}}}
\gridline{\fig{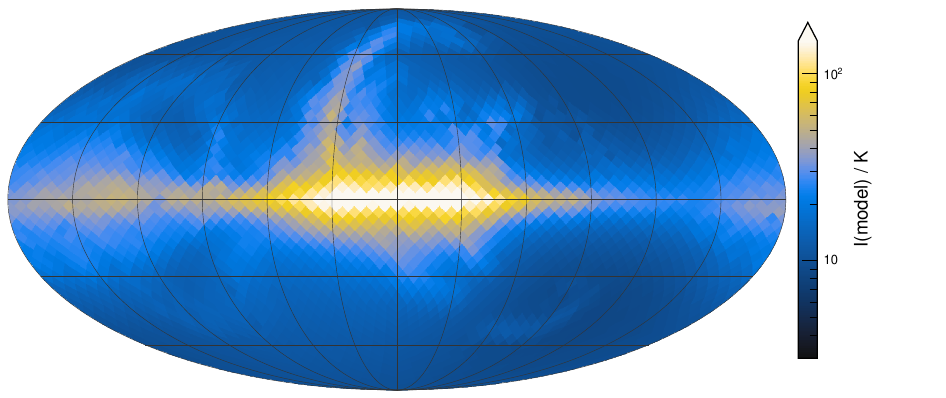}{1.37\columnwidth}{(b) Model.\hspace*{\labelpad}}}
\gridline{\fig{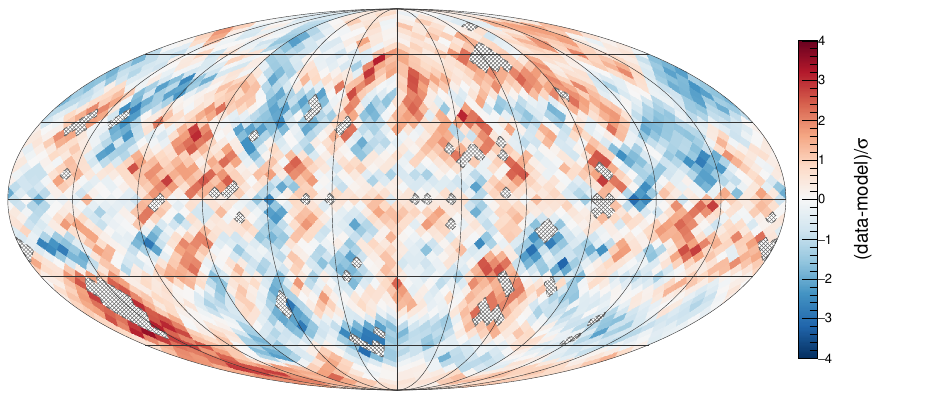}{1.37\columnwidth}{(c) Pull.\hspace*{\labelpad}}}
\caption{Intensity fit for the \ModelDRF model ($\chi^2/n_\mathrm{df}
  = 3499/2938 = 1.19$). Panel (a) shows the synchrotron intensity data at
  408~MHz, panel (b) the total model intensity $I_\text{tot}$, and
  panel (c) the pull, i.e.\ the difference between data and model in
  units of the pixel uncertainty. Masked pixels are shown as hatched
  regions.}
\label{fig:intensityFit}
\end{figure*}

\begin{figure*}[t]
\def\labelpad{1.2cm}
\begingroup
\setkeys{Gin}{trim=0mm 0mm 10mm 0mm,clip}
\gridline{\fig{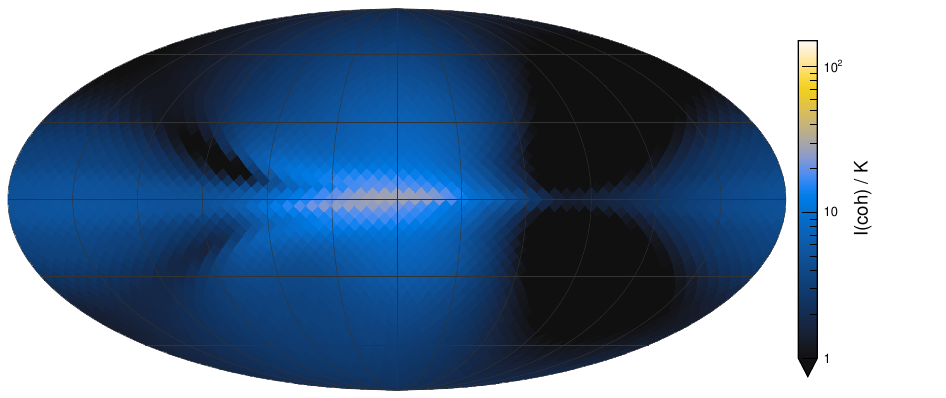}{0.49\textwidth}{(a) $I_\text{coh}$.\hspace*{\labelpad}}
          \fig{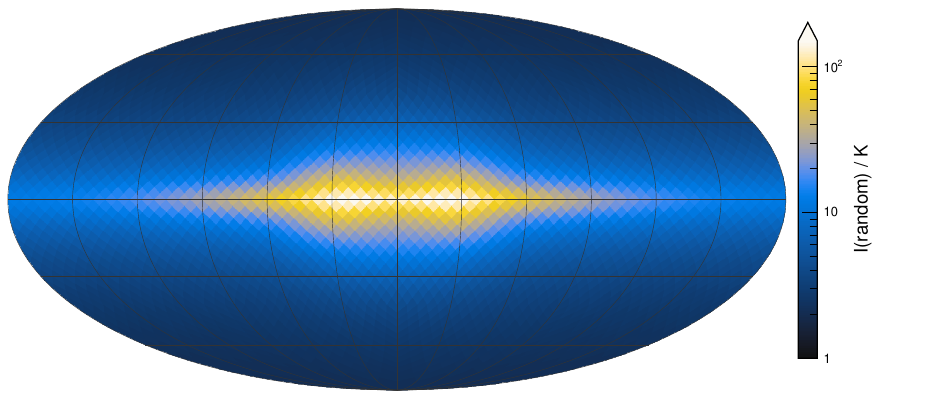}{0.49\textwidth}{(b) $I_\text{rand}$.\hspace*{\labelpad}}}
\gridline{\fig{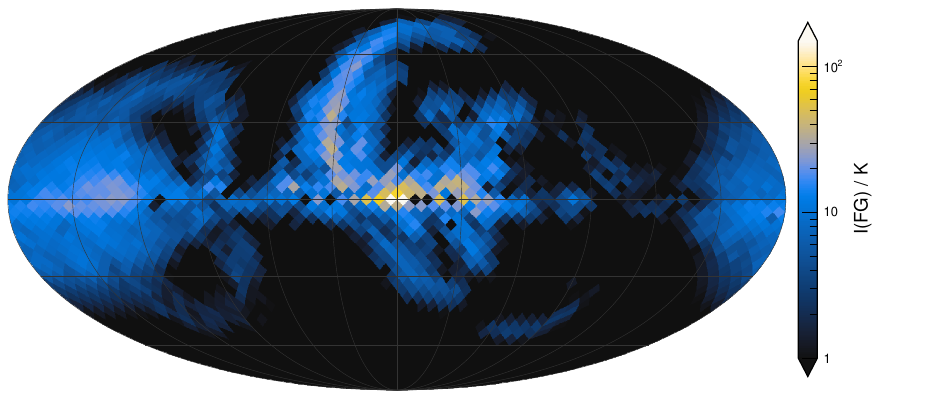}{0.49\textwidth}{(c) $I_\text{fg}$.\hspace*{\labelpad}}
          \fig{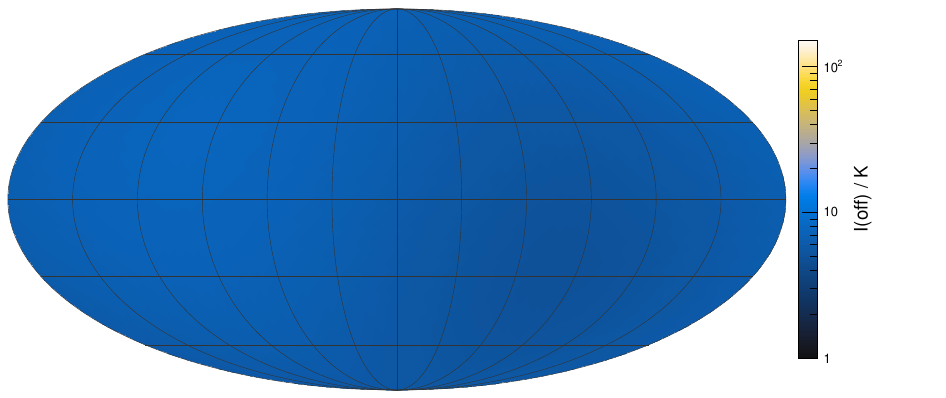}{0.49\textwidth}{(d) $I_\text{off}$.\hspace*{\labelpad}}}
\endgroup
\caption{Components of the model intensity, $I_\text{tot} =
  I_\text{coh} + I_\text{rand} + I_\text{fg} + I_\text{off}$, for the
  \ModelDRF model shown in Fig.\,\ref{fig:intensityFit}: (a)
  synchrotron emission from the coherent field, (b) from the random
  field, (c) the foreground excess of the local regions plus the outer
  spur intensity, and (d) the offset $I_\text{off}$, i.e.\ the fitted
  monopole and dipole.}
\label{fig:intensityFitComponents}
\end{figure*}

\begin{figure*}[t]
  \centering
  \def\figh{0.505}
    \includegraphics[clip,rviewport=0.02 0 1 1,height=\figh\textheight]
      {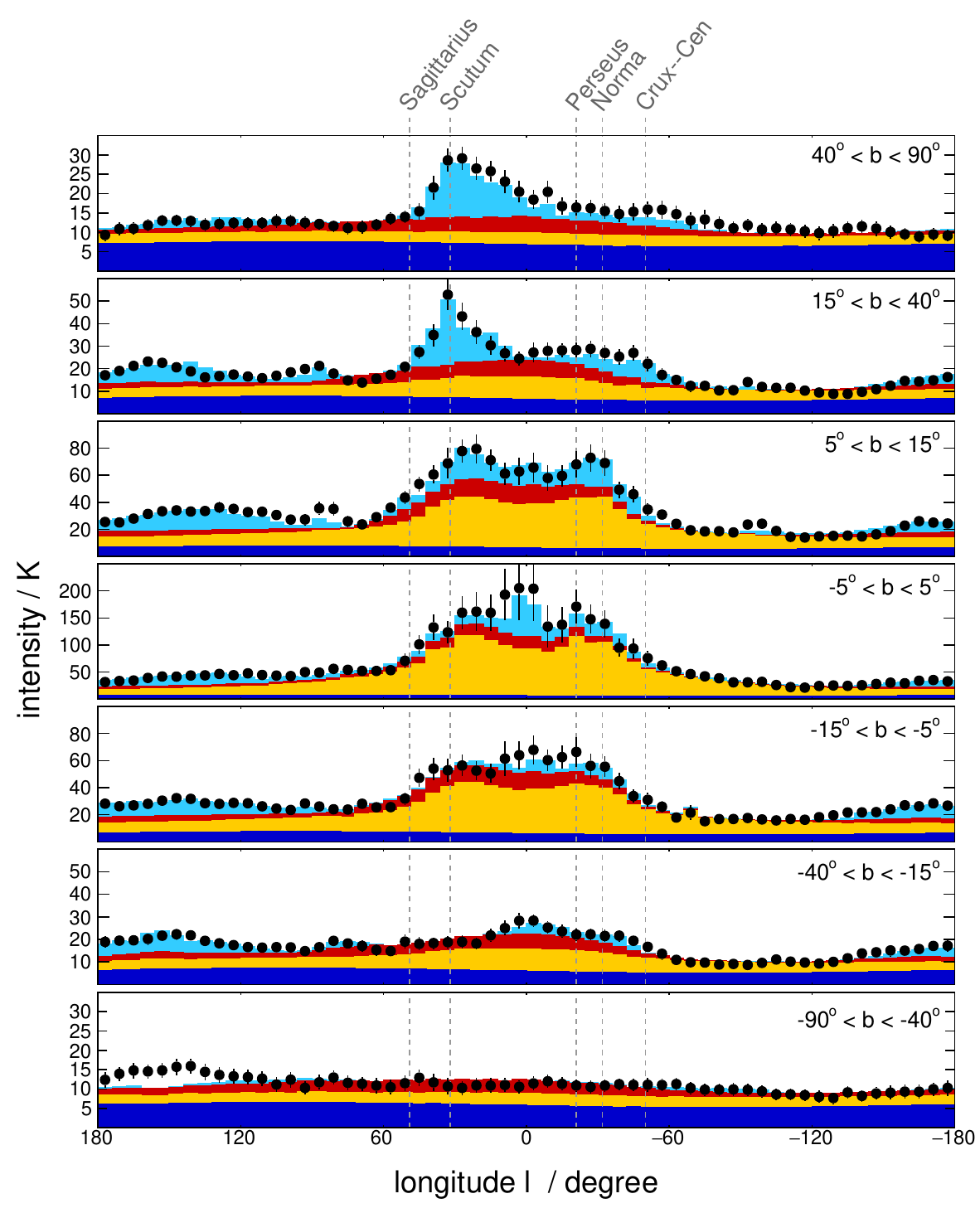}\includegraphics[clip,rviewport=0.06 0 1 1,height=\figh\textheight]
      {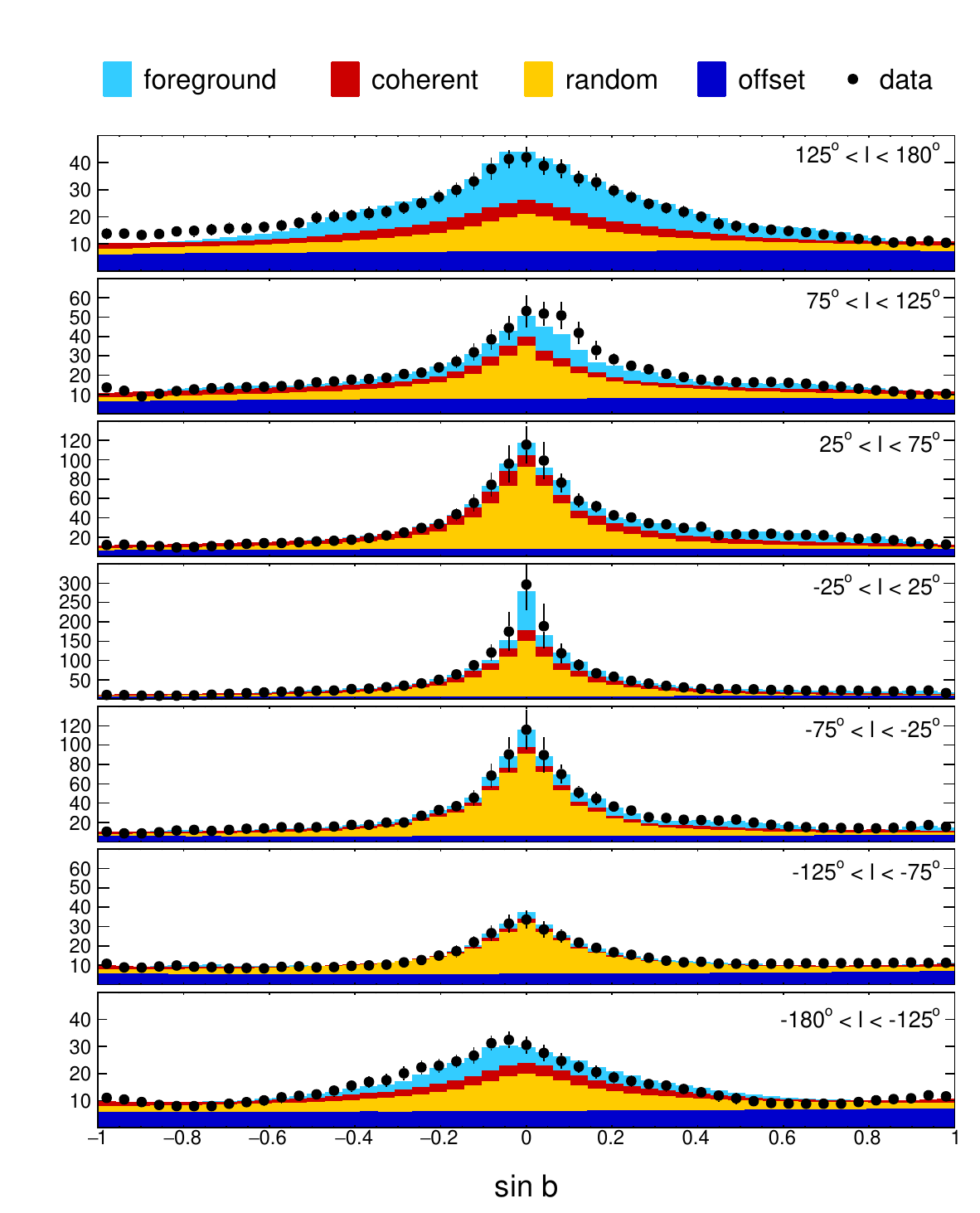}
  \caption{Comparison of the \Planck 408-MHz synchrotron intensity and
    the fitted intensity model (\ModelDRF model). The model intensity
    is shown cumulatively for the offset, random-field,
    coherent-field, and foreground contributions, adding up to the
    total model intensity. Note the different scales in the different
    plots. {\itshape Left:} Longitude profiles in seven
    Galactic-latitude intervals.  {\itshape Right:} Latitude profiles
    in seven Galactic-longitude intervals. Intensities are the
    arithmetic mean over the unmasked pixels in each bin. The error
    bars show the mean per-pixel uncertainty. The vertical dashed
    lines mark the 408-MHz tangent directions of
    \citet{1985A&A...153...17B} associated with the Crux--Centaurus,
    Norma, Perseus, Scutum, and Sagittarius arms, as compiled by
    \citet{2022NewA...9701896V}.}
  \label{fig:intensity-slices}
\end{figure*}

Of the variants discussed in the previous section, the \ModelDRF{}
model provides the best description of the data, with
$\chi^2/n_{\mathrm{df}}=3499/2938=1.19$. Its best-fit total-intensity
sky is compared with the data in Fig.\,\ref{fig:intensityFit}. The
individual model components are shown in
Fig.\,\ref{fig:intensityFitComponents}.

Latitude and longitude profiles of the data and model are displayed in
Fig.\,\ref{fig:intensity-slices}.  These plots illustrate how the
different components combine to reproduce the data.  The offset
component provides a nearly uniform baseline, the coherent field and
the polarization-based foreground templates account for substantial
and spatially varying fractions of the emission, and the random field
describes only the remaining large-scale component.  Their relative
contributions vary strongly with both longitude and latitude, but
their sum closely follows the data in nearly all angular slices.  This
decomposition is essential: an analysis based on the random-field
component alone would necessarily absorb coherent emission, local
foregrounds, and the offset into the inferred random-field
distribution, thereby biasing both its strength and spatial scales.
In the low-latitude longitude profiles, the ring component reproduces
the broad enhancement across the inner Galaxy.  Once this component
and the foregrounds are included, no systematic residual enhancements
or changes of slope remain at the displayed spiral-arm tangent
longitudes that would call for an additional spiral-arm modulation of
the random field.

Overall, the model reproduces the observed sky well, although some
systematic differences remain visible in the pull map in
Fig.\,\ref{fig:intensityFit}\,(c).  These residuals do not necessarily
indicate deficiencies of the large-scale model: the prediction
represents the ensemble-averaged sky, whereas the observed sky is a
particular realization of the random field.  Consequently, some of the
large-pull regions may be compatible with nearby fluctuations in $b$,
depending on the coherence length of the random field.  We defer a
quantitative assessment of this possibility to future work. Other
residuals, such as the extended region of positive pull surrounding
the masked area near $(\ell,b)=(150^\circ,-45^\circ)$, are clearly
associated with localized structures not captured by the model. In
either case, the fit using the more aggressive $2\sigma$ pull mask
demonstrates that these regions do not bias the inferred
parameters. We therefore conclude that our model provides a good
overall description of the data. To our knowledge, no previous model
has reproduced the 408-MHz sky with comparable fidelity.

We use the variations explored in the previous section to define two
representative models. Both are built on the sech disk of
Eq.\,(\ref{eq:bdisk}) and differ only in their treatment of the inner
Galaxy. In the \modelDisk{} model, which summarizes \ModelD{} and
\ModelDF{}, the exponential radial profile continues inward to the
Galactic center, where it is smoothed on the scale \rcusp{}. In the
\modelRing{} model, summarizing \ModelDR{} and \ModelDRF{}, the disk
is instead phased out below \rann{} and the field of the inner Galaxy
is carried by the annular component of Eq.\,(\ref{eq:bannulus}).  For
each model parameter, we determine the minimum and maximum best-fit
values among the corresponding variants and adopt their midpoint as
the central value and half their difference as an estimate of the
modeling uncertainty.  This deliberately simple procedure is
appropriate because the spread among the variants reflects systematic
modeling uncertainty rather than a statistical one.  Fewer model
variants were explored for \modelRing{} than for \modelDisk{}, so the
spread of its \bdisksun{} values alone would imply a spuriously small
uncertainty; we therefore conservatively assign \bdisksun{} the same
relative uncertainty in both models.  To take into account parameter
correlations, we evaluate the one-sided vertical integrals of $b^2$
separately for every model variant: $\bdisksun^2\,\zdisk\,c_V$ for the
disk, where $c_V$ is defined in Eq.~\eqref{eq:verticalColumn}, and
$\bannpeak^2\,\zann\,\sqrt{\pi}/2$ for the Gaussian annular component.
We quote the midpoint and half-width of their resulting ranges rather
than constructing them from the independently summarized parameters.

The resulting parameters of the two models are listed in
Table~\ref{tab:parSummary}. We infer a local rms random-field strength
of about 4~\muG, consistent with estimates obtained from pulsar
rotation and dispersion measures, e.g.\ 4--6~\muG{} estimated by
\citet{1993MNRAS.262..953O}.  The scale height of the random-field
disk is about 1~kpc, substantially smaller than values inferred or
adopted in previous random-field models, see Sec.\,\ref{sec:thickness}
and the comparison to previous models in
Sec.\,\ref{sec:previousModels} for further discussion. Toward the
inner Galaxy, the disk field of the \modelDisk{} model increases with
radial scale length \lrad{}. In the \modelRing{} model, only lower
limits on \lrad{} are obtained: the disk field is nearly constant with
radius, while an annular component comparable in strength to the disk
centered at a Galactocentric radius of $4.5\pm0.6$~kpc accounts for
the inner-Galaxy enhancement of synchrotron intensity.

Among the nuisance parameters fitted jointly with the magnetic-field
parameters, the normalization $k$ of the coherent-field contribution
$I_\mathrm{coh}$ at 408~MHz shows a large variation of $\pm 0.4$ among
the variants.  The best fits of \ModelDF{} (e5) and \ModelDRF{} (g5)
yield $k=1.0$ and $0.7$, respectively, i.e.\ a good agreement with the
nominal frequency extrapolation of the coherent field intensity
predictions.  Since $I_\mathrm{coh}$ is extrapolated over nearly two
decades in frequency, a rescaling by $k$ is equivalent to a shift of
the effective spectral index of Eq.\,(\ref{eq:betaS}).  The value
$k=0.7$ therefore corresponds to $\Delta\beta_S=-0.08$, well inside
the $\pm0.1$ range spanned by the coherent-model predictions across
the sky (bottom row of Fig.\,\ref{fig:cohModel}); the full spread of
$k$ among all variants corresponds to $|\Delta\beta_S|\lesssim0.24$.
The inferred monopole lies between 4 and 7~K and is discussed further
in Sec.\,\ref{sec:mono}. The analysis also favors a sizable residual
dipole. Its components are, however, compatible with the uncertainties
in the dipolar zero-level correction applied to the Haslam map. As
discussed in Sec.\,\ref{sec:dipoSpur}, this dipole therefore may well
reflect residual zero-level systematics rather than astrophysical
emission. The region-dependent re-weighting factors for the excess
polarized emission are reported in Sec.\,\ref{sec:excessPolFactor}.

\begin{figure*}[!tp]
  \centering
\def\figw{0.48}
\def\padding{1.3cm}
\includegraphics[
  width=\figw\textwidth,
  rviewport=0 0 0.84 1,
  clip
]{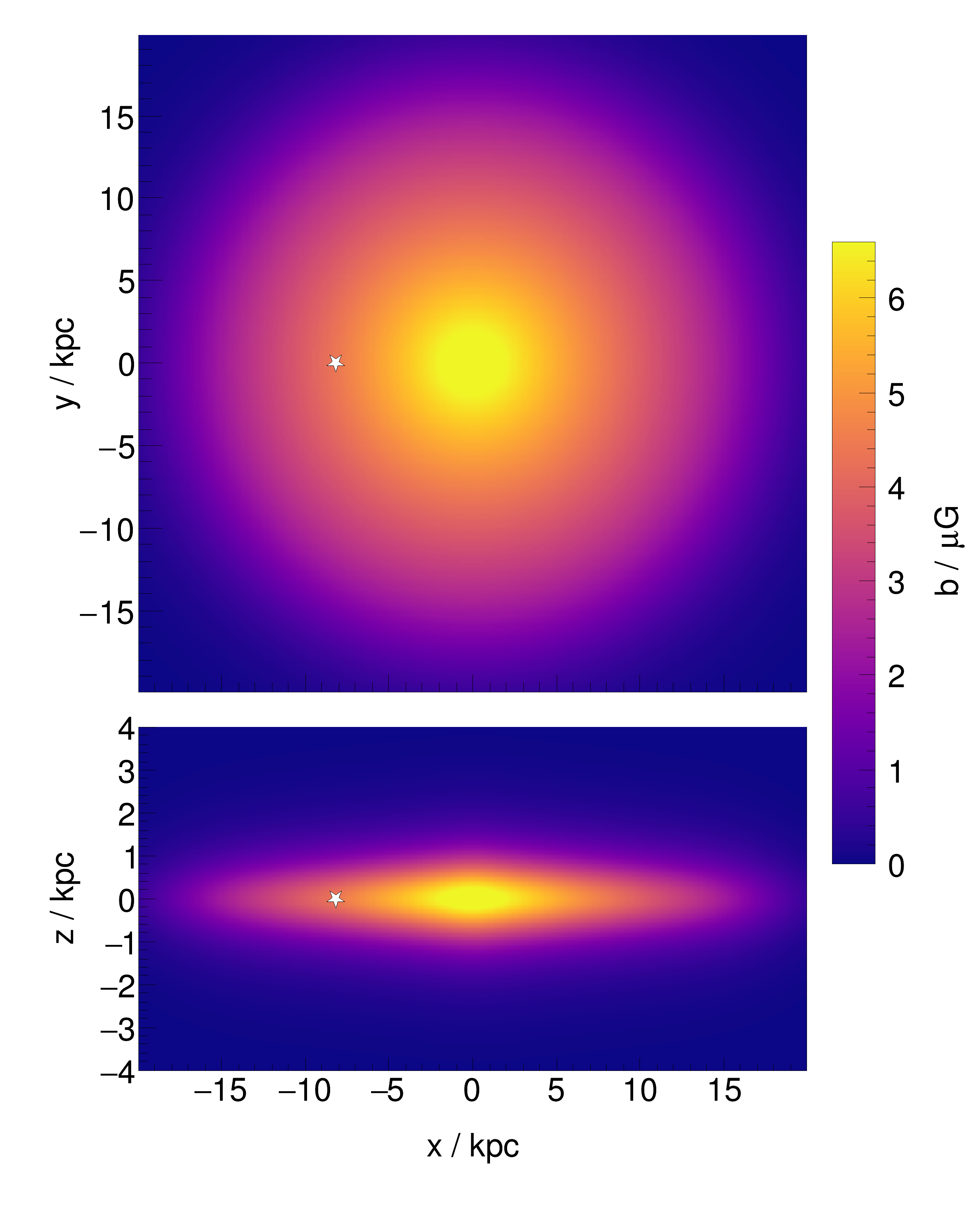}~\includegraphics[
  rviewport=0.14 0 0.98 1, clip,
  width=\figw\textwidth,
]{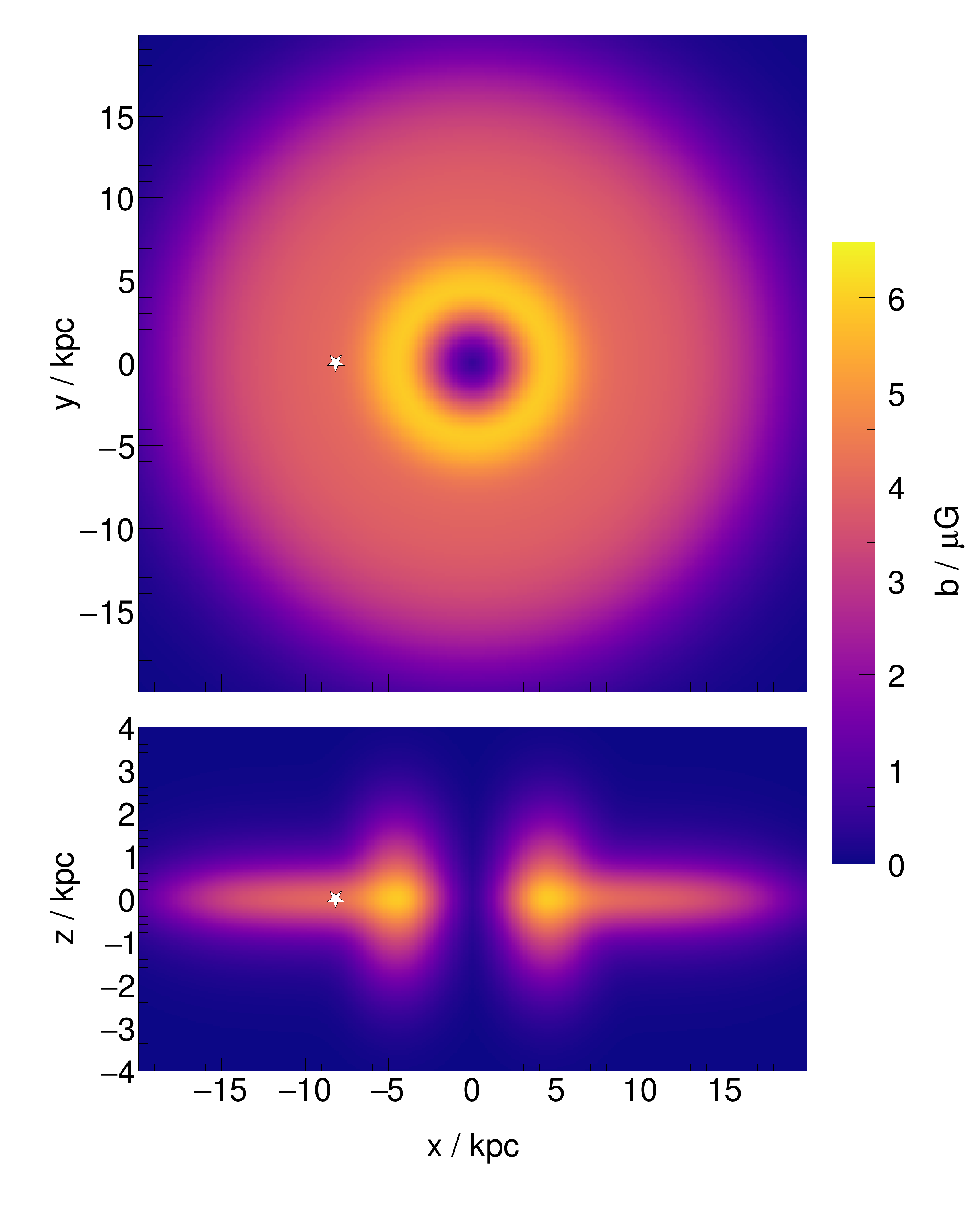}

\makebox[\figw\textwidth]{\hspace*{\padding}(a) Disk field.}\makebox[\figw\textwidth]{(b) Ring field.\hspace*{\padding}}
\caption{Random magnetic field of the \modelDisk{} (a) and
  \modelRing{} (b) models. In each panel, the top
  plot shows the face-on view of the Galactic plane at $z=0$ and the
  bottom plot the edge-on view in the plane containing the Sun and the
  Galactic center. The star marks the position of the Sun.}
\label{fig:randomfield}
\end{figure*}

Face-on and edge-on maps of the random-field strength, $b(x,y,0)$ and
$b(x,0,z)$, are shown in Fig.\,\ref{fig:randomfield}. The maps
highlight the different behavior of the two models near the inner
Galaxy. Interestingly, the data appear insensitive to whether a large
random field exists at $r=0$.  This likely reflects the low cosmic-ray
electron density at small Galactocentric radii, visible in the right
panel of Fig.\,\ref{fig:ncreDens}, which suppresses the synchrotron
emissivity and weakens the constraints on the random field there. The
two models thus illustrate the range of central random-field
configurations compatible with the present data.

\section{Discussion}
\label{sec:discussion}

\subsection{Isotropic Offset}
\label{sec:mono}
Across the main model variants, we find a sizable residual monopole,
$m_0\simeq4$--$7$~K, in addition to the $(8.9\pm1.3)$~K
zero-level correction of \citet{2017A&A...597A.131W} already applied
to the 408~MHz map (Sec.\,\ref{sec:offsets}). This additional isotropic
component is not accounted for by any of the large-scale disk or halo
components in our models.

A similar difficulty has long been encountered in Galactic
synchrotron modeling. To reproduce the nearly latitude-independent
high-latitude intensity, \citet{2008A&A...477..573S} introduced an ad
hoc, spherical enhancement of the cosmic-ray electron density
extending to 1~kpc around the Sun. Our residual monopole may also be
related to the unexplained radio synchrotron background reported by
\citet{2011ApJ...734....5F}, whose recently revised amplitude is about
15~K at 408~MHz \citep{2026NatAs.tmp..122M}.

A natural candidate for a nearby astrophysical contribution is the
Local Bubble, a cavity of hot, tenuous plasma of radius
$\sim100$--300~pc created by supernovae over the past $\sim10$--15~Myr
and bounded by a shell of swept-up gas and dust
\citep[e.g.][]{2024ApJ...973..136O}. Dust polarization indicates that
the magnetic field is compressed within, and predominantly tangential
to, the shell
\citep{2018A&A...611L...5A,2020A&A...636A..17P,2025ApJ...988..191O},
so the wall should emit synchrotron radiation
\citep{2025A&A...693A.284K,2025A&A...695A.148P}. Since the observer
lies inside the cavity, this contribution is nearly isotropic and could
enter the fitted $m_0$.\footnote{The polarized part should already be
captured by our excess-polarization template. We nevertheless repeated
the fits with the Local Bubble contribution to $I_\text{coh}$ from the
preliminary \modelBubBest{} and \modelBubDust{} variants of
\citet{Pelgrims:2025nbj}, which include a realistic shell distance and
thickness. All parameters, including $m_0$, remain within the range
spanned by our model variants.}

To estimate the unpolarized contribution, we consider an initially
uniform sphere of radius $R$ whose isotropic random field is swept
into a shell of thickness $d$.  We denote its brightness before
compression by $T_{\rm swept}$ and define the volume compression
factor as
\begin{equation}
C\equiv \frac{R^3}{R^3-(R-d)^3}
    \simeq\frac{R}{3d}.
\end{equation}
For a cosmic-ray electron spectrum with $p=3$,
Eqs.~\eqref{eq:powerlaw} and \eqref{eq:pixelIntegral} then give the
shell brightness seen by an observer at the center as
\begin{equation}
\begin{aligned}
T_{\rm LB}
&=T_{\rm swept}
\begin{cases}
C,       & \text{$b$ only},\\
C^{8/3}, & \text{$b$ and CRE},
\end{cases}
\label{eq:LBcompression}
\end{aligned}
\end{equation}
corresponding to compression of the magnetic field alone and to joint
compression of the magnetic field and cosmic-ray electrons (CREs),
respectively.  The first case follows from flux conservation%
\footnote{For a fluid element with initial radius $r_0$, uniform
  volume compression by $C$ corresponds to $r_0^3=C\,(r^3-a^3)$, where
  $a=R-d$.  For the initially isotropic random field $b_0$,
  $b_{\perp,0}^2=2/3\, b_0^2$ denotes the mean-square field
  perpendicular to the radial line of sight. A small material surface
  perpendicular to either tangential field component has area
  proportional to $r_0\,{\rm d}r_0$ before compression and to $r\,{\rm
    d}r$ afterward. Flux conservation therefore gives the
  amplification of the transverse rms field as $ C_b(r)\equiv
  {b_\perp(r)}/{b_{\perp,0}} = (r_0\,{\rm d}r_0)/(r\,{\rm d}r)
  =C\,{r}/{r_0}.$ For an observer at the center and $p=3$, the
  synchrotron emissivity is proportional to $b_\perp^2$. Hence $T_{\rm
    swept}\propto b_{\perp,0}^2R$, whereas $T_{\rm LB}\propto
  b_{\perp,0}^2 \int_a^R C_b^2(r)\,{\rm d}r$.  Evaluating this
  integral, one finds that the intensity amplification factor is
  simply equal to the volume compression factor, $ C_I\equiv {T_{\rm
      LB}}/ {T_{\rm swept}} = C.$ Although the \textit{linear}
  displacement field adopted by \citet{2025A&A...695A.148P} produces a
  different radial profile of $b(r)$, it has the same thin-shell
  limit, $C_I\simeq R/(3d)$. The radially uniform prescription of
  \citet[Eq.\,8]{2025A&A...693A.284K} conserves the same flux, and
  hence the same shell-averaged amplification, $\langle
  C_b\rangle\simeq R/(2d)$, but its square gives only $C_I\simeq
  R/(4d)$, since $\langle b_\perp\rangle^2\le\langle
  b_\perp^2\rangle$.}  and the second case applies if the cosmic-ray
electrons remain confined while being compressed adiabatically with
the gas.  Their power-law normalization is then enhanced by
$C^{(p+2)/3}=C^{5/3}$ \citep[Eq.\,11]{2001A&A...366...26E}.  Numerical
simulations are needed to determine where between these limits the
actual enhancement would lie today.  A simple argument would be that
flux freezing leaves the radial field approximately unchanged, so for
a $D$ typical of the Galactic disk at the energies that dominate the
408~MHz emission, the diffusive escape time across the wall,
$d^2/(2D)$, is only $10^3$ years, four orders of magnitude below the
lifetime of the Bubble. Flux freezing also compresses the radial
coherence length, but standard scalings of the diffusion
coefficient~\citep[e.g.][]{2023arXiv230900298E} would increase the
escape time by only a factor of a few. However, transport in the wall
becomes highly anisotropic, self-generated turbulence could further
suppress radial diffusion, and cosmic-ray electrons may also be
accelerated in the shell, so this simple estimate need not apply. The
adiabatic case can therefore be regarded as an upper limit.

With the local 408~MHz emissivity of $\varepsilon_{408,\odot} =
4.3~\mathrm{K}\,\mathrm{kpc}^{-1}$ in the \modelDisk{} model, a bubble
with $R=200$~pc and $d=30$~pc has $C=2.6$ and $T_{\rm swept}=0.9$~K,
so that
\begin{equation}
\Delta T_{\rm LB}=T_{\rm LB}-T_{\rm swept}\simeq
\begin{cases}
1.4~\mathrm{K},  & \text{$b$ only},\\
10~\mathrm{K},   & \text{$b$ and CRE}.
\end{cases}
\label{eq:LBexcess}
\end{equation}
Magnetic compression alone therefore yields an excess of order a
Kelvin for $d\simeq30$~pc, a non-negligible fraction of our fitted
$m_0$. The contribution increases for thinner bubble walls and could account
for most of the monopole for $d\simeq10$~pc, although such a thin wall
lies below current Local Bubble estimates. For the nominal geometry,
accounting for the full monopole would instead require additional
compression of the cosmic-ray electrons and hence the confinement
discussed above. A contribution toward the lower end of this range is
consistent with \citet{2021MNRAS.502.2807K}, who modeled the Bubble
as a homogeneous cavity with the locally measured cosmic-ray electron
spectrum and an equipartition-limited field.

The observed monopole could instead originate on much larger scales.
A component associated with the Galactic halo or the circumgalactic
medium \citep[e.g.][]{2023ARA&A..61..131F} is a natural candidate for a
monopole: seen from within, the emission from a quasi-spherical distribution
 is isotropic to first order.  \citet{2025PhRvD.111f3057B}
invoked an extended turbulent Galactic halo, possibly reaching beyond
200~kpc, to explain the low polarized synchrotron intensity fraction
suggested by preliminary C-BASS results. More recently,
\citet{2026arXiv260623317L} showed that Galactic cosmic-ray transport
in an extended halo can describe the observed secondary-to-primary
ratios of cosmic-ray nuclei. Whether cosmic-ray electrons can populate
such an extended halo and thereby contribute to the near-isotropic
synchrotron emission is less clear.  Even if synchrotron losses are
negligible in a sub-$\upmu$G circumgalactic field, cosmic-ray
electrons still suffer substantial energy losses through
inverse-Compton scattering on the cosmic microwave background. This
may prevent the electrons from reaching large distances unless they
are continuously supplied or reaccelerated.

The considerations above do not identify a unique origin for $m_0$.
Depending on the shell geometry and transport of cosmic-ray
electrons, the Local Bubble could make a modest or substantial
contribution. An extended Galactic or circumgalactic component also
remains possible, but depends on a poorly constrained population of
cosmic-ray electrons. We therefore retain $m_0$ as an empirical offset
rather than assigning it to a specific physical component.

\begin{figure*}[t]
  \centering
  \def\figh{0.18}
\includegraphics[clip,rviewport=0 0 0.9 1,height=\figh\textheight]{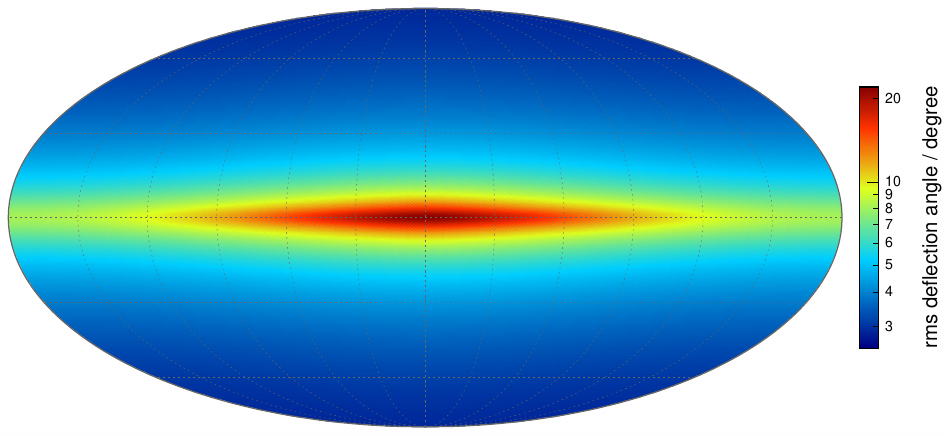}\hfill\includegraphics[height=\figh\textheight]{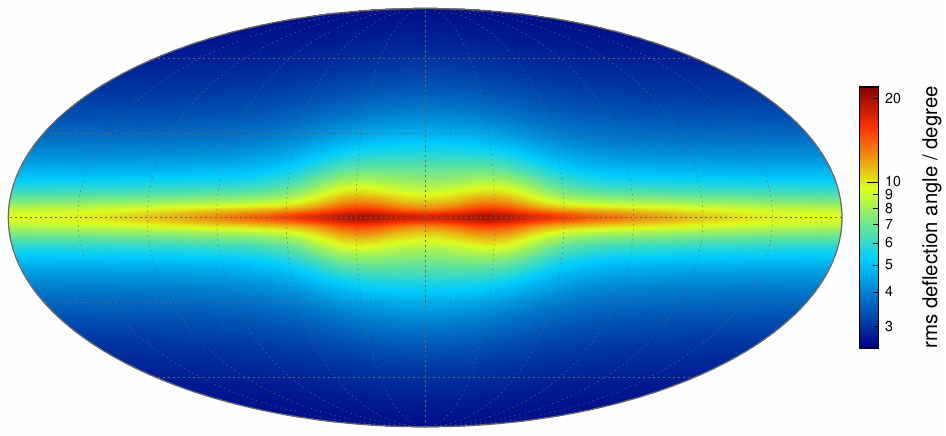}
\caption{Root-mean-square UHECR deflection angle for \modelDisk{} (left) and \modelRing{} (right) for
a particle rigidity of $\mathcal{R} = 10$~EV and a coherence length of $\lambda_{\mathrm{c}} = 50$~pc.}
\label{fig:deflections}
\end{figure*}

\subsection{Disk Thickness}
\label{sec:thickness}
The thickness of our random-field disk, $\zdisk=1.0\pm0.2$~kpc, is
substantially smaller than the scale heights of the extended
random-field distributions inferred or adopted in previous studies.
The equivalent $1/e$ height of the Gaussian halo reconstructed by
\citet{2012ApJ...761L..11J} is $4.0\pm1.8$~kpc and likewise 4~kpc for
the single exponential component of
\citet{2013MNRAS.436.2127O}. \citet{2013JCAP...03..036D} argued that
exponential scale heights $\leq2$~kpc are excluded at $3\sigma$.

We attribute these differences to our explicit separation of emission
that does not trace the smooth random-field disk.  Local foregrounds
are described using polarized-intensity templates, while the fitted
monopole $m_0$ accounts phenomenologically for any nearly isotropic
residual.  Without these
components, high-latitude emission is instead absorbed into the
global random-field model, biasing its inferred scale height upward.
Supernova-driven shocks and shells produce magnetized structures with
characteristic sizes of order 100~pc.  Nearby structures subtend large
angles and can therefore contribute prominently at high latitudes, as
is evident from the $\PI$ map in Fig.\,\ref{fig:polintensity}.  If such
emission is attributed to a smooth Galactic component, its large
angular extent is interpreted as a large vertical extent, further
increasing the inferred scale height.


The scale height $\zdisk$ describes the random field strength, whereas
radio observations of external galaxies measure the vertical extent of
the synchrotron emission.  Since the modeled CR electron density
varies only mildly across the vertical extent of our random-field
disk, its synchrotron emissivity is approximately proportional to the
squared random field strength.  For the exponential and sech profiles
considered here, this gives an effective emissivity scale height of
$h_\varepsilon\simeq(0.5$--$0.65)\,\zdisk$, which is approximately
$(0.6\pm0.1)$~kpc in the \modelDisk{} model.

A comparison with the thin and thick radio disks measured in nearby
edge-on galaxies by CHANG-ES \citep{2025A&A...699A.243H} would help
assess whether the vertical structure we have inferred for the Milky Way is typical of spiral
galaxies. However, the CHANG-ES measurements describe the total radio
continuum at 3~GHz, including both synchrotron and thermal free--free
emission, whereas the value derived here describes only the
synchrotron contribution of the random field at 408~MHz. The different
observing frequencies probe different CR electron energies and hence
potentially different vertical emissivity profiles, and free--free
emission contributes mainly to the thin disk. Most importantly, the Galactic
coherent field contributes an extended synchrotron component even when the random field is confined to a
relatively thin disk, see \UFCoh{}. A quantitative comparison therefore requires
modeling the combined synchrotron emission of the random and coherent
fields, together with the thermal contribution, at the observing
frequency and viewing geometry of the external galaxies, which we
leave to future work.

\subsection{Implications for Cosmic-Ray Deflections}
\label{sec:uhecrDeflections}

At sufficiently high rigidity, $\mathcal{R}=E/(Ze)$, ultrahigh-energy
cosmic rays (UHECRs) propagate approximately rectilinearly through the
random Galactic magnetic field. Following \citet{2002JHEP...03..045H},
and assuming a spatially constant coherence length $\lambda_{\mathrm{c}}$,
their root-mean-square deflection angle is given by
\begin{equation}
  \delta_{\mathrm{rms}}
  \simeq
  2.7^\circ
  \left(\frac{10\,\mathrm{EV}}{\mathcal{R}}\right)
  \left(\frac{\lambda_{\mathrm{c}}}{50\,\mathrm{pc}}\right)^{1/2}
  \left(
    \frac{\mathcal{I}_b}
         {10\,\muG^{2}\,\mathrm{kpc}}
  \right)^{1/2},
  \label{eq:deflection}
\end{equation}
where
\begin{equation}
  \mathcal{I}_b
  \equiv
  \int_0^\infty \mathrm{d}s\,b^2(s).
  \label{eq:deflectionIntegral}
\end{equation}

For the height profile $\Vdisk(z)$ used here (see
Eq.\,(\ref{eq:vdisk})), the plane-parallel approximation close to the
Galactic poles gives
\begin{equation}
  \mathcal{I}_b = \frac{1}{\sin\beta}\int_0^\infty \bdisk^2(\Rsun,z)\,{\rm d}z,
\end{equation}
where $\beta$ denotes the absolute value of the Galactic latitude. The rms
random-field contribution to the arrival-direction smearing of UHECRs
incident close to the poles can therefore be read off directly from
Table~\ref{tab:parSummary}, with a modeling uncertainty below $5\%$
since $\delta_{\mathrm{rms}}\propto\sqrt{\mathcal{I}_b}$. The
\modelDisk{} model yields a moderate deflection of
$\delta\simeq3^\circ$ perpendicular to the Galactic plane for
$\mathcal{R}=10$~EV and $\lambda_{\mathrm{c}}=50$~pc. An additional
thick random-field disk compatible with the limits from
Sec.\,\ref{sec:thickdisk}, $b_2^2 z_2\leq3~\upmu{\rm G}^2\,{\rm kpc}$
for $z_2=3$~kpc, would contribute at most $1.1^\circ$ in quadrature,
increasing the total deflection to $\delta^\prime\simeq3.2^\circ$.

To study the directional dependence, we numerically integrate
Eq.\,(\ref{eq:deflectionIntegral}) along different lines of sight for
our two representative models. For $\mathcal{R}=10$~EV and
$\lambda_{\mathrm{c}}=50$~pc, the median deflections are
$4.1^\circ$ and $3.9^\circ$ for \modelDisk{} and \modelRing{},
respectively. The corresponding sky maps are shown in
Fig.\,\ref{fig:deflections}. As expected, the deflections are largest
in the Galactic plane, where they reach up to $20^\circ$. Toward the
Galactic poles, we recover the analytical estimate given above.

The Pierre Auger Observatory reported UHECR anisotropies above $E_0
\sim40$~EeV that are driven by an overdensity of events in the
Centaurus region with a fitted Gaussian angular scale of
$\sigma=15^{+8}_{-4}\degr$ \citep{2022ApJ...935..170A}. As an
application of our random-field constraints, let us consider the
possibility that a point source produces the excess and ask what the
typical charge of these CRs must be, for deflections in the random
field to produce the observed spreading. In the direction of the
Centaurus excess, both models yield
$\mathcal{I}_b(\mathrm{Cen})=(45\pm7)~\upmu\mathrm{G}^2\,\mathrm{kpc}$.
Since the angular variance scales as $E^{-2}$, we use the effective
spectrum-weighted energy above $E_0$,
$E_{\mathrm{eff}}=1/\sqrt{\langle E^{-2}\rangle}\simeq50$~EeV. Setting
$\delta_{\mathrm{rms}}=\sqrt{2}\,\sigma$ for a two-dimensional
Gaussian, we obtain
\begin{equation}
  Z(\mathrm{Cen})
  =
  \left(19^{+10}_{-5}\pm2\pm3\right)
  \left(\frac{50~\mathrm{pc}}{\lambda_{\mathrm{c}}}\right)^{1/2}.
\end{equation}
The uncertainties arise, respectively, from the fitted angular scale,
the range of $\mathcal{I}_b(\mathrm{Cen})$, and the Auger energy scale.
The inferred charge lies between those of silicon and iron and is
consistent with recent interpretations of the mass composition at
these energies \citep{Mollerach:2025fmx, PierreAuger:2026qbt}.

We also compare our results with two older random-field models widely
used in the literature: the original \citetalias{2012ApJ...761L..11J}
model and its ``JF12b'' retuning by the \citet{2016A&A...596A.103P}.
Both older models predict larger deflections than those of \modelDisk{}
and \modelRing{}, with ratios reaching 2.4 in some directions. The largest
sky-median ratio is 1.7, for JF12b relative to
\modelRing{}. See Sec.\,\ref{sec:previousModels} in the appendix for
more details.

Finally, an astrophysical origin of the fitted monopole could imply an
additional UHECR deflection not included above. For $p=3$, synchrotron
intensity and deflection variance depend on the same magnetic column,
$\mathcal{I}_b=\int b^2\,{\rm d}s$. If the normalization $\ncre$ of the
electron spectrum is spatially constant in the emitting region, the
column implied by the intensity $m_0$ follows from
Eq.~\eqref{eq:powerlaw}, and Eq.~\eqref{eq:deflection} gives
\begin{equation}
 \delta_{m_0}\simeq
 3.7^\circ
 \left(\frac{m_0}{5\,{\rm K}}\right)^{1/2}
 \left(\frac{\ncre}{n_\smallodot}\right)^{-1/2}
 \left(\frac{10\,{\rm EV}}{\mathcal R}\right)
 \left(\frac{\lambda_{\rm c}}{50\,{\rm pc}}\right)^{1/2},
 \label{eq:deflectionMonopole}
\end{equation}
where $n_\smallodot$ is the local value of the electron density.
If the full $m_0=4.3$--$7.0$~K originated in an extended turbulent halo
with $\ncre=0.1\,n_\smallodot$, it would therefore produce an rms
deflection of $11^\circ$--$14^\circ$ at $\mathcal R=10$~EV, much
larger than the approximately $3^\circ$ polar disk contribution.

The effect of the Local Bubble is much smaller because its wall
replaces, rather than supplements, the smooth local field. Before
compression, the swept-up region contributes $\delta_{\rm
  swept}\simeq1.5^\circ$ at the reference values above.  If the
turbulence is flux-frozen, its magnetic column is amplified by
$C\simeq R/(3d)$ while its radial coherence length is reduced. If
$\lambda_{\rm c,LB}/\lambda_{\rm c}\simeq d/R$, then the ratio
$(\delta_{\rm LB}/\delta_{\rm swept})^2 = {C\,\frac{\lambda_{\rm
      c,LB}}{\lambda_{\rm c}}} \simeq 1/3$, i.e.\ in this case the
local rms deflection is \textit{reduced} to $1/\sqrt{3}$ of its
uncompressed value.

\section{Conclusions and Outlook}
\label{sec:conclusions}

In this paper we presented a new analysis of the random field of the
Galaxy based on the total synchrotron intensity at 408\,MHz.  We find
that, once localized foreground emission is separated from the
Galaxy-wide signal, the large-scale isotropic random field is
considerably simpler and more vertically compact than suggested by
previous models. Across the analysis variations, the dominant
component of the random field is consistently a disk with a local rms
strength of about $4~\upmu\mathrm{G}$ and a $1/e$ scale height of
approximately $1~\mathrm{kpc}$. An annular enhancement in the inner
Galaxy improves the description of the data, but does not alter the
properties of the main disk substantially. We find no evidence for
either a large-scale spiral pattern or an additional thick
random-field disk.

Our new determination of the thickness of the Galactic random field
disk, which we find to be significantly thinner than in previous
models, illustrates the importance of distinguishing nearby
synchrotron structures from the large-scale random field. Because
nearby structures subtend large angles on the sky, failing to identify
them as local leads to incorrect random field inference, with
excessively strong and extended components.  The polarization-based
foreground template used here avoids this, while retaining 97\% of the
sky for the fit. We provide two representative models that encompass
the principal modeling uncertainties and can be used in applications
requiring a global random-field model.

A direct next step is to repeat the analysis at other frequencies. Of
particular interest is the forthcoming final C-BASS intensity
map at 5~GHz \citep{2018MNRAS.480.3224J}. Together with the 408~MHz
sky, it will provide an independent test of the inferred spatial
structure and constrain spatial variations of the synchrotron
spectrum, thereby probing the magnetic field with a higher-energy
population of cosmic-ray electrons. Broadband intensity maps from the
1.3--1.8~GHz GMIMS high-band surveys
\citep{2021AJ....162...35W,2025A&A...694A.169S} will provide
complementary information. A joint analysis will help distinguish
residual zero-level systematics from an astrophysical monopole and
constrain the spectral index of the foreground contributions.

The dependence of the inferred random field on the assumed cosmic-ray
electron distribution is an important systematic uncertainty in the
field strength and scale height. A more complete treatment should
determine the coherent field, random field, and electron distribution
in a mutually consistent framework. The magnetic fields affect the
electron distribution through radiative cooling and, likely more
importantly for its spatial distribution at the energies relevant
here, through cosmic-ray diffusion.

Finally, the absence of significant spiral structure in the present
analysis should not be interpreted as evidence that the Galactic
random-field disk is intrinsically smooth or axisymmetric. Total
synchrotron intensity integrates the emissivity along the entire line
of sight, making spiral arms difficult to distinguish from
axisymmetric enhancements such as the ring considered here.
Distance-resolved information is needed to break these degeneracies.
Low-frequency absorption by \ion{H}{2} regions at known distances can
provide distance-resolved constraints on the synchrotron emissivity
\citep{2020A&A...636A...2P}, while the fluctuations and spatial
correlations of pulsar rotation measures in different distance ranges
can constrain the strength and organization of the random field,
provided that fluctuations of the thermal electron density can be determined
simultaneously.

Further progress will come from new multi-frequency maps of total and
polarized synchrotron emission and their Faraday depolarization,
denser pulsar and extragalactic rotation-measure grids, and the
distance-resolved dispersion of starlight polarization angles in the
nearby interstellar medium \citep[see][and references
  therein]{2026arXiv260602149H}.

\begin{acknowledgments}
We thank Rainer Beck, Michael Kachelriess and Foteini Oikonomou for fruitful
discussions.  MU acknowledges the hospitality of CCPP/NYU, where part
of this work was done. This research of GRF has been supported by
National Science Foundation grant NSF-PHY-2413153.
\end{acknowledgments}


\bibliography{main}{}

\software {The results of this paper have been derived using the
  software packages \HEALPix~\citep{2005ApJ...622..759G},
  \ROOT~\citep{Antcheva:2009zz}, as well as basic functionality of
  the \Offline framework of the Pierre Auger
  Observatory~\citep{Argiro:2007qg}.}

\bibliographystyle{aasjournalMU}

\appendix

\section{Analytic Optimization of Linear Parameters}
\label{sec:opti}
At fixed random-field parameters $\boldsymbol\eta$, we split the
nuisance parameters into a global part and the regional amplitudes,
$\boldsymbol\theta = (\boldsymbol\theta_{\rm G}, \bm{f})$, and collect
the pixel templates multiplying the global parameters into a vector,
\begin{equation}
\boldsymbol{\theta}_{\rm G}=(k,m_0,d_x,d_y,d_z)^{\top},
\qquad
\bm{t}_i=(c_i,1,n_{x,i},n_{y,i},n_{z,i})^{\top},
\end{equation}
such that the model of Eq.\,\eqref{eq:model} reads
$\mu_i=r_i+\boldsymbol{\theta}_{\rm G}^{\top}\bm{t}_i+f_{R(i)}a_i$ and
the $\chi^2$ of Eq.\,\eqref{eq:chi2} becomes
\begin{equation}
\chi^2
=\sum_i w_i
 \left[\Delta_i-\boldsymbol{\theta}_{\rm G}^{\top}\bm{t}_i
                   -f_{R(i)}a_i\right]^2,
\qquad
\Delta_i\equiv I_i-r_i.
\end{equation}
Since each $f_R$ affects only the pixels of its region, the
corresponding normal equations decouple from one another. Defining
\begin{gather}
\mathbf{A}_{\rm G}=\sum_i w_i\,\bm{t}_i\bm{t}_i^{\top},
\qquad
\bm{y}_{\rm G}=\sum_i w_i\,\bm{t}_i\Delta_i,
\\
\bm{b}_R=\sum_{i\in R}w_i a_i\bm{t}_i,
\qquad
D_R=\sum_{i\in R}w_i a_i^2,
\qquad
y_R=\sum_{i\in R}w_i a_i\Delta_i,
\end{gather}
the regional amplitudes can be eliminated analytically, leaving the
five-dimensional system
\begin{equation}
\left(
\mathbf{A}_{\rm G}
-\sum_R\frac{\bm{b}_R\bm{b}_R^{\top}}{D_R}
\right)
\boldsymbol{\theta}_{\rm G}
=
\bm{y}_{\rm G}
-\sum_R\frac{y_R}{D_R}\bm{b}_R .
\label{eq:linear}
\end{equation}
After solving Eq.\,\eqref{eq:linear}, the regional factors follow
from
\begin{equation}
f_R
=\frac{y_R-\bm{b}_R^{\top}\boldsymbol{\theta}_{\rm G}}{D_R}.
\end{equation}

\begin{figure}[t]
  \centering
  \def\mapheight{0.114\textwidth}
  \def\righttrim{2.7cm}
  \def\off{0.3cm}
  \def\offf{0.4cm}

  \newcommand{\maphead}[2][0pt]{%
    \hspace*{-#1}#2\hspace*{#1}%
  }
  \setlength{\tabcolsep}{0pt}
  \caption{Foreground and offset components and their sum. The direction
    of the fitted dipolar offset, $\bm{d}$, is marked with a white cross.}
  \begin{tabular}{@{}c@{\hspace{0.25em}}c@{\hspace{0.25em}}
      c@{\hspace{0.25em}}c@{\hspace{0.25em}}
      c@{\hspace{0.25em}}c@{\hspace{0.25em}}c@{}}
    \maphead{\small excess pol.}
    & \(+\) &
    \maphead{\small offset ($m_0$+$\bm{d}$)}
    &  &
    & \(=\) &
    \maphead[\offf]{\small total foreground}
    \\[0.3ex]

    \includegraphics[
      height=\mapheight,
      trim=0 0 \righttrim{} 0,clip
    ]{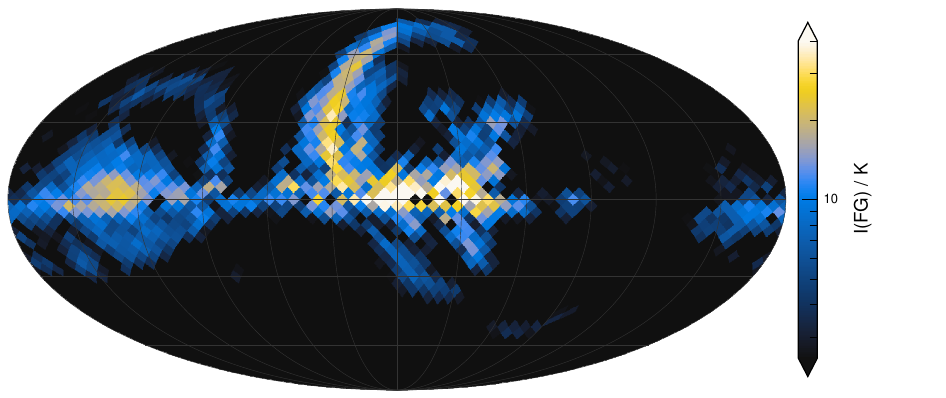}
    & &
    \includegraphics[
      height=\mapheight,
      trim=0 0 \righttrim{} 0,clip
    ]{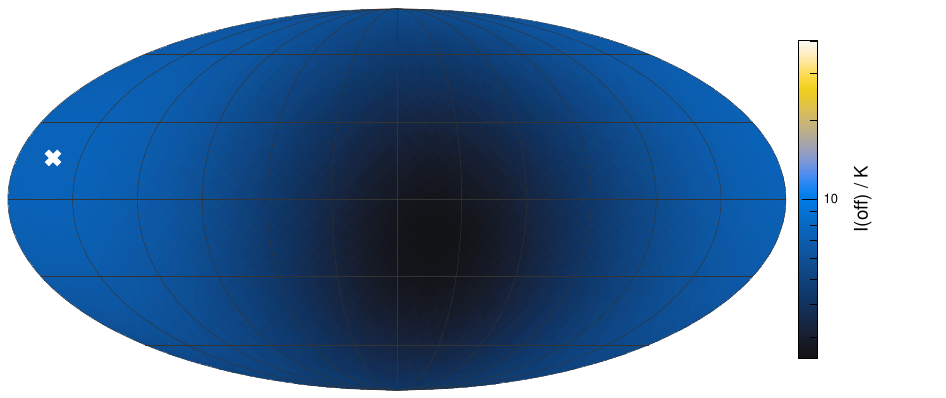}
    & &
    {}
    & &
    \includegraphics[
      height=\mapheight
    ]{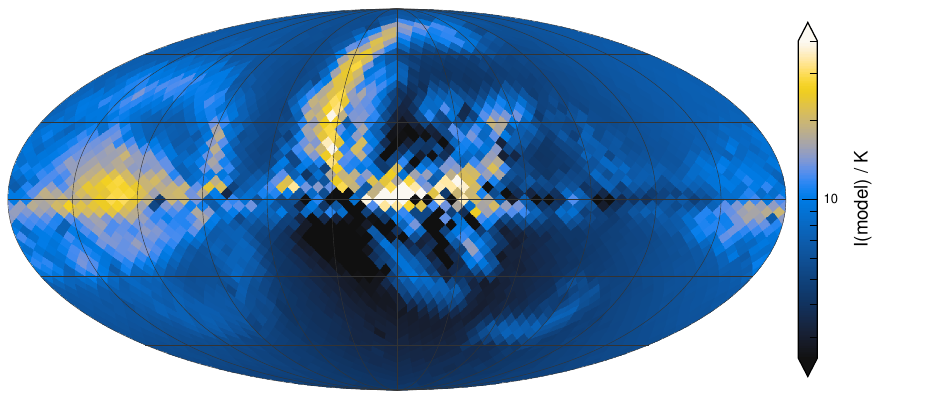}
    \\[-0.3ex]

    \multicolumn{7}{c}{\small (a) \ModelDR}
    \\[2.5ex]

    \maphead{\small excess pol.}
    & \(+\) &
    \maphead{\small offset ($m_0$+$\bm{d}$)}
    & \(+\) &
    \maphead{\small outer spur}
    & \(=\) &
    \maphead[\offf]{\small total foreground}
    \\[0.3ex]
    \includegraphics[
      height=\mapheight,
      trim=0 0 \righttrim{} 0,clip
    ]{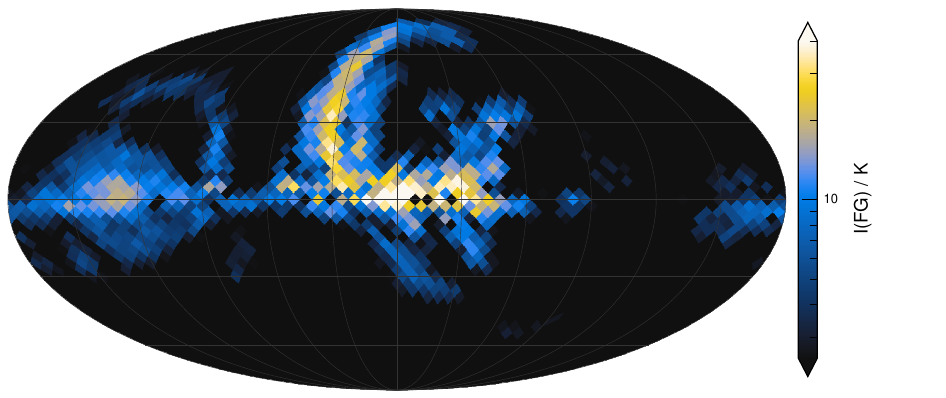}
    & &
    \includegraphics[
      height=\mapheight,
      trim=0 0 \righttrim{} 0,clip
                    ]{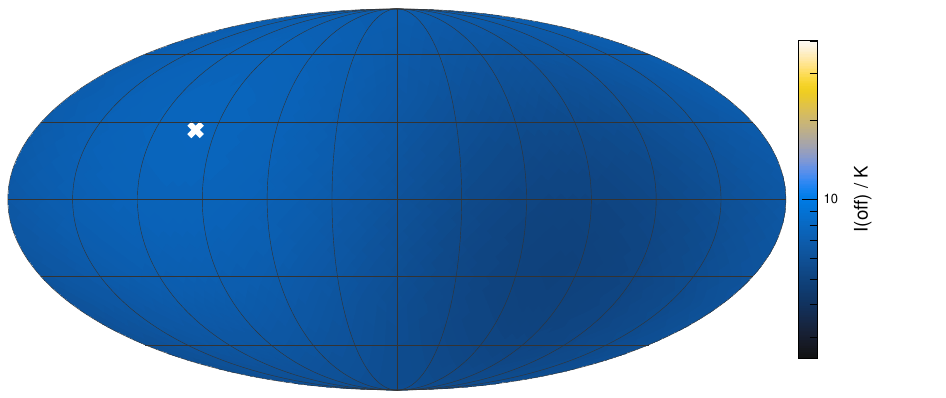}
    & &
    \includegraphics[
      height=\mapheight,
      trim=0 0 \righttrim{} 0,clip
    ]{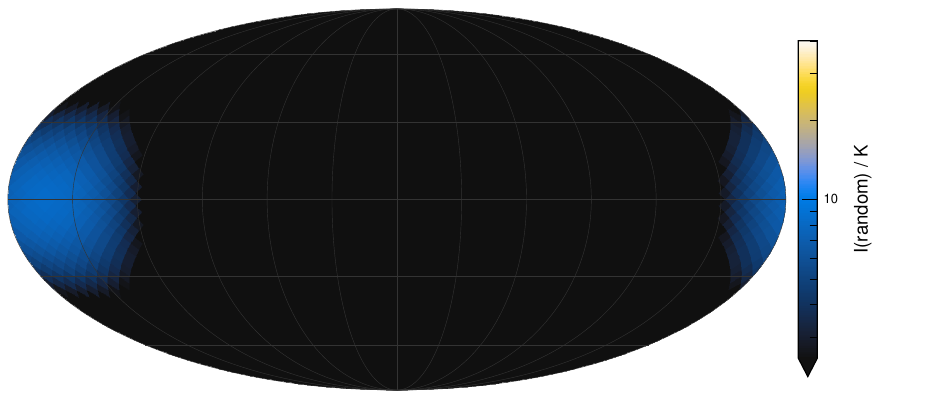}
    & &
    \includegraphics[
      height=\mapheight
    ]{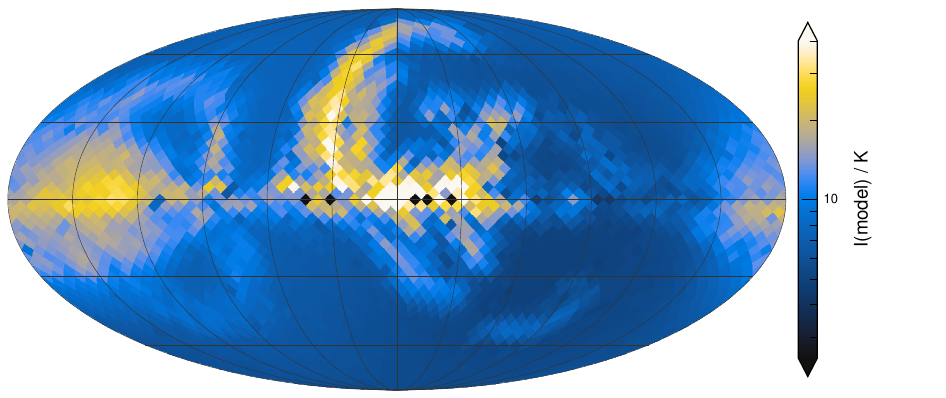}
    \\[-0.3ex]

    \multicolumn{7}{c}{\small (b) \ModelDRF}
  \end{tabular}
  \label{fig:foregroundComponents}
\end{figure}

\section{Comparison of Foreground and Offset Components}
\label{sec:foregroundmaps}
The individual foreground and offset contributions in the best-fit \ModelDR{} and
\ModelDRF{} models (variants f5 and g5) are shown in
Fig.\,\ref{fig:foregroundComponents}. Without an outer spur, the excess
intensity near the Galactic anticenter is largely absorbed by the
dipole component (panel (a)). Adding an outer spur reduces the dipole
amplitude by nearly a factor of two. The remaining differences between
the combined foreground and offset contributions in the rightmost column are
compensated by the normalization of $I_\mathrm{coh}$ (not shown), with
fitted $k$-factors of approximately 1.1 and 0.7, respectively; see
Tables~\ref{tab:DR} and \ref{tab:DRF} below.

\section{Individual Fit Results}
\label{sec:fitresults}
The
individual fit results discussed in Sec.\,\ref{sec:modelingChoices},
from which the parameters of \modelDisk{} and \modelRing{} in
Table~\ref{tab:parSummary} are derived, are listed in Tables~\ref{tab:Icoh}--\ref{tab:DRF}.

\begin{table}[!htb]
  \centering\small
\caption{Best-fit parameters of the \ModelD{} model for different choices of the coherent GMF model and hence $I_\mathrm{coh}$. The average statistical parameter uncertainty is given in the last row.}
\begin{tabular}{ll|cccccccc}
\multirow{2}{*}{id} & \multirow{2}{*}{$I_\mathrm{coh}$} & \bdisksun{} & \zdisk{} & \lrad{} & \multirow{2}{*}{$k$} & $m_0$ & $(\ell, b)_d$ & $|\ddip|$ & \multirow{2}{*}{$\chi^2/n_\mathrm{df}$} \\
 & & [\muG] & [kpc] & [kpc] & & [K] & [deg.] & [K] & \\
\hline\hline
a1 & \modelBase & $4.5$ & $0.90$ & $13$ & $1.05$ & $4.8$ & $(148, 11)$ & $2.5$ & $4140/2946 = 1.41$ \\
a2 & \modelXr & $4.6$ & $0.93$ & $21$ & $0.77$ & $5.5$ & $(133, 11)$ & $2.5$ & $4423/2946 = 1.50$ \\
a3 &\modelNe & $4.5$ & $0.91$ & $15$ & $1.14$ & $4.8$ & $(154, 12)$ & $2.5$ & $4136/2946 = 1.40$ \\
a4 & \modelKappa & $4.5$ & $0.89$ & $13$ & $1.06$ & $4.8$ & $(147, 14)$ & $2.5$ & $4127/2946 = 1.40$ \\
a5 & \modelTwist & $4.3$ & $0.93$ & $12$ & $1.27$ & $4.4$ & $(171, 15)$ & $2.7$ & $3982/2946 = 1.35$ \\
\hline
\multicolumn{2}{r}{$\langle\sigma\rangle$}& $0.1$ & $0.04$ & $\hphantom{0}2$ & $0.02$ & $0.1$ & $(\hphantom{00}2, \hphantom{0}2)$ & $0.1$ & \\
\end{tabular}
  \label{tab:Icoh}
\end{table}

\begin{table}[!htb]
  \centering\small
  \caption{Best-fit parameters of the \ModelD{} model using the specified \ncre{} model for the given \hDiff{} (half-height of the CR electron diffusion halo). Model a1 from Table~\ref{tab:Icoh} is repeated for comparison.}
\begin{tabular}{lcc|cccccccc}
\multirow{2}{*}{id} & \ncre & \hDiff & \bdisksun{} & \zdisk{} & \lrad{} & \multirow{2}{*}{$k$} & $m_0$ & $(\ell, b)_d$ & $|\ddip|$ & \multirow{2}{*}{$\chi^2/n_\mathrm{df}$} \\
 & model & [kpc] & [\muG] & [kpc] & [kpc] & & [K] & [deg.] & [K] & \\
\hline\hline
b1 & UF23 & 2  & $5.0$ & $0.69$ & $24$ & $1.41$ & $5.9$ & $(128, 19)$ & $2.1$ & $4056/2946 = 1.38$ \\
b2 & UF23 &4 & $4.7$ & $0.86$ & $15$ & $1.13$ & $4.8$ & $(147, 12)$ & $2.5$ & $4071/2946 = 1.38$ \\
a1 & UF23 &6 & $4.5$ & $0.90$ & $13$ & $1.05$ & $4.8$ & $(148, 11)$ & $2.5$ & $4140/2946 = 1.41$ \\
b3 & UF23 &8 & $4.4$ & $0.92$ & $13$ & $1.03$ & $4.8$ & $(147, 11)$ & $2.4$ & $4162/2946 = 1.41$ \\
b4 & UF23 &10 & $4.4$ & $0.93$ & $12$ & $1.04$ & $4.8$ & $(146, 11)$ & $2.4$ & $4167/2946 = 1.41$ \\
b5 & OS13 &10 & $3.8$ & $1.23$ & $10$ & $0.90$ & $4.7$ & $(147, 11)$ & $2.5$ & $4181/2946 = 1.42$ \\
\hline
\multicolumn{3}{r}{$\langle\sigma\rangle$}& $0.1$ & $0.05$ & $\hphantom{0}2$ & $0.03$ & $0.1$ & $(\hphantom{00}2, \hphantom{0}2)$ & $0.1$ & \\
\end{tabular}
  \label{tab:ncre}
\end{table}

\begin{table}[!htb]
  \centering\small
\caption{Best-fit parameters of the \ModelD{} model for different
  choices of the pixel mask. Model a5 from Table~\ref{tab:Icoh} is
  repeated for comparison. The latitude cuts successively exclude
  iso-latitude pixel rings of the $N_{\mathrm{side}}=16$ \HEALPix grid:
  the ring centered at $b=0$ for c1, additionally the rings at
  $|b|=2.4^\circ$ for c2, and additionally those at
  $|b|=4.8^\circ$ for c3. The quoted cut angles are the pixel-center
  latitudes of the lowest retained rings.}
\begin{tabular}{lclc|cccccccc}
\multirow{2}{*}{id} & \multirow{2}{*}{pull cut} & \multirow{2}{*}{latitude cut} & sky fraction & \bdisksun{} & \zdisk{} & \lrad{} & \multirow{2}{*}{$k$} & $m_0$ & $(\ell, b)_d$ & $|\ddip|$ & \multirow{2}{*}{$\chi^2/n_\mathrm{df}$} \\
 & &  & [\%] & [\muG] & [kpc] & [kpc] & & [K] & [deg.] & [K] & \\
\hline\hline
a5 & $|\Delta|<3$ & none           & 97 & $4.3$ & $0.9$ & $12$ & $1.27$ & $4.4$ & $(171, 15)$ & $2.7$ & $3982/2946 = 1.35$ \\
c1 & $|\Delta|<3$ & $|b|\geq 2.4^\circ$ & 95 & $4.4$ & $0.9$ & $12$ & $1.26$ & $4.4$ & $(171, 16)$ & $2.7$ & $3923/2890 = 1.36$ \\
c2 & $|\Delta|<3$ & $|b|\geq 4.8^\circ$ & 91 & $4.2$ & $1.0$ & $12$ & $1.25$ & $4.4$ & $(171, 16)$ & $2.7$ & $3827/2768 = 1.38$ \\
c3 & $|\Delta|<3$ & $|b|\geq 7.2^\circ$ & 87 & $4.6$ & $0.8$ & $\hphantom{0}9$ & $1.23$ & $4.5$ & $(170, 16)$ & $2.7$ & $3684/2643 = 1.39$ \\
c4 & $|\Delta|<2$ & none & 88 & $4.3$ & $1.0$ & $14$ & $1.40$ & $3.8$ & $(181, 19)$ & $3.0$ & $2271/2673 = 0.85$ \\
\hline
\multicolumn{4}{r}{$\langle\sigma\rangle$}& $0.4$ & $0.1$ & $\hphantom{0}2$ & $0.03$ & $0.1$ & $(\hphantom{00}2, \hphantom{0}2)$ & $0.1$ &
\end{tabular}
  \label{tab:mask}
\end{table}

\begin{table}[!htb]
  \centering\small
\caption{Best-fit parameters of the \ModelD{} model under different
  assumptions about the functional form (sech, exp or gauss) and
  symmetry of the random-GMF height profile, with independent magnetic
  field and radial and vertical scale lengths above (N) and below (S)
  the Galactic plane in model d3. Model a5 from Table~\ref{tab:Icoh}
  is repeated for comparison and the average statistical parameter
  uncertainties are given in the last row.}
\begin{tabular}{ll|cccccccc}
\multirow{2}{*}{id} & \multirow{2}{*}{\Vdisk{}} & \bdisksun{} & \zdisk{} & \lrad{} & \multirow{2}{*}{$k$} & $m_0$ & $(\ell, b)_d$ & $|\ddip|$ & \multirow{2}{*}{$\chi^2/n_\mathrm{df}$} \\
 & & [\muG] & [kpc] & [kpc] & & [K] & [deg.] & [K] & \\
\hline\hline
a5 & sech & $4.3$ & $0.93$ & $12$ & $1.27$ & $4.4$ & $(171, 15)$ & $2.7$ & $3982/2946 = 1.35$ \\
d1 & exp  & $4.7$ & $0.99$ & $11$ & $1.23$ & $4.3$ & $(171, 14)$ & $2.9$ & $3971/2946 = 1.35$ \\
d2 & gauss & $4.3$ & $0.91$ & $13$ & $1.28$ & $4.4$ & $(172, 16)$ & $2.7$ & $3985/2946 = 1.35$ \\
d3 & sech & ${}^{\text{N:}\,4.2}_{\,\text{S:}\,4.5}$ & ${}^{\text{N:}\,1.10}_{\,\text{S:}\,0.83}$ & ${}^{\text{N:}\,13}_{\,\text{S:}\,12}$ & $1.27$ & $4.3$ & $(172, 11)$ & $2.7$ & $3962/2943 = 1.35$\\
\hline
\multicolumn{2}{r}{$\langle\sigma\rangle$} & $0.1$ & $0.06$ & $\hphantom{0}2$ & $0.03$ & $0.1$ & $(\hphantom{00}2, \hphantom{0}1)$ & $0.1$ &
\end{tabular}
  \label{tab:heightNS}
\end{table}

\begin{table}[!htb]
  \centering\small
  \caption{Best-fit parameters of the \ModelDF{} model for different choices of $I_\mathrm{coh}$.}
\begin{tabular}{ll|cccccccccccc}
\multirow{2}{*}{id}& \multirow{2}{*}{$I_\mathrm{coh}$}& \bdisksun{} & \zdisk{} & \lrad{} & $\ospurNorm$ & $\ell_\text{os}$ & \ospurDeltaL{} & \ospurDeltaB{} & \multirow{2}{*}{$k$} & $m_0$ & $(\ell, b)_d$ & $|\ddip|$& \multirow{2}{*}{$\chi^2/n_\mathrm{df}$} \\
 & & [\muG] & [kpc] & [kpc] & [\muG$^2\,$kpc] & [deg.] & [deg.] & [deg.] &  & [K] & [deg.] & [K] & \\
 \hline\hline
e1 & \modelBase & $4.7$ & $0.83$ & $11$ & $32$ & $168$ & $75$ & $59$ & $0.79$ & $5.5$ & $(102, 19)$ & $1.6$ & $3736/2942 = 1.27$ \\
e2 & \modelXr & $4.7$ & $0.84$ & $13$ & $39$ & $170$ & $73$ & $58$ & $0.58$ & $6.0$   & $(\hphantom{0}83, 17)$ & $1.9$ & $3777/2942 = 1.28$ \\
e3 & \modelNe & $4.7$ & $0.84$ & $11$ & $32$ & $167$ & $74$ & $59$ & $0.85$ & $5.2$ & $(108, 21)$ & $1.5$ & $3742/2942 = 1.27$ \\
e4 & \modelKappa & $4.7$ & $0.82$ & $11$ & $32$ & $168$ & $75$ & $59$ & $0.80$ & $5.5$ & $(100, 22)$ & $1.6$ & $3715/2942 = 1.26$ \\
e5 & \modelTwist & $4.5$ & $0.86$ & $11$ & $26$ & $163$ & $64$ & $65$ & $1.00$ & $5.0$ & $(148, 29)$ & $1.4$ & $3683/2942 = 1.25$ \\
\hline
\multicolumn{2}{r}{$\langle\sigma\rangle$} & $0.2$ & $0.05$ & $\hphantom{0}1$ & $\hphantom{0}4$ & $\hphantom{00}3$ & $\hphantom{0}9$ & $\hphantom{0}3$ & $0.02$ & $0.1$ & $(\hphantom{00}4, \hphantom{0}3)$ & $0.09$ & \\
\end{tabular}
  \label{tab:DF}
\end{table}

\begin{table}[!htb]
 \centering\small
\caption{Best-fit parameters of the \ModelDR{} model for different
  model choices of $I_\mathrm{coh}$. In all variants \lrad{} reaches
  the upper bound of the fit range, $120$~kpc, i.e.\ the data are
  consistent with a radially constant disk field.}
 \begin{tabular}{ll|cccccccccccc}
\multirow{2}{*}{id}& \multirow{2}{*}{$I_\mathrm{coh}$}& \bdisksun{} & \zdisk{} & \lrad{} & \bannpeak{} & \rann{} & \wann{} & \zann{} & \multirow{2}{*}{$k$} & $m_0$ & $(\ell, b)_d$ & $|\ddip|$ &\multirow{2}{*}{$\chi^2/n_\mathrm{df}$} \\
 & & [\muG] & [kpc] & [kpc] & [\muG] & [kpc] & [kpc] & [kpc] &  & [K] & [deg.] & [K] & \\
 \hline\hline
f1 & \modelBase & $4.11$ & $0.98$ & $120$ & $\hphantom{0}4.0$ & $3.9$ & $2.2$ & $\hphantom{0}1.8$ & $0.88$ & $5.5$ & $(139, 12)$ & $2.4$ & $4061/2942 = 1.38$ \\
f2 & \modelXr & $4.25$ & $0.85$ & $120$ & $\hphantom{0}3.4$ & $5.0$ & $2.1$ & $\hphantom{0}2.0$ & $0.62$ & $6.1$ & $(136, 10)$ & $2.8$ & $4295/2942 = 1.46$ \\
f3 & \modelNe & $4.13$ & $1.00$ & $120$ & $\hphantom{0}3.7$ & $4.0$ & $2.1$ & $\hphantom{0}1.8$ & $0.98$ & $5.0$ & $(146, 12)$ & $2.4$ & $4073/2942 = 1.38$ \\
f4 & \modelKappa & $4.11$ & $0.97$ & $120$ & $\hphantom{0}4.1$ & $3.9$ & $2.2$ & $\hphantom{0}1.7$ & $0.89$ & $5.5$ & $(138, 14)$ & $2.4$ & $4049/2942 = 1.38$ \\
f5 & \modelTwist & $3.97$ & $0.99$ & $120$ & $\hphantom{0}4.0$ & $4.0$ & $2.1$ & $\hphantom{0}2.0$ & $1.09$ & $5.1$ & $(163, 16)$ & $2.5$ & $3890/2942 = 1.32$ \\
\hline
\multicolumn{2}{r}{$\langle\sigma\rangle$} & $0.07$ & $0.05$ & $\hphantom{0}50$ & $\hphantom{0}0.1$ & $0.7$ & $0.7$ & $\hphantom{0}0.8$ & $0.03$ & $0.1$ & $(\hphantom{00}2, \hphantom{0}2)$ & $0.1$ & \\
\end{tabular}
  \label{tab:DR}
\end{table}

\begin{table}[!htb]
 \centering\small
\caption{Best-fit parameters of the \ModelDRF{} model for different
  choices of $I_\mathrm{coh}$. In all variants \lrad{} reaches the
  upper bound of the fit range, $120$~kpc, i.e.\ the data are
  consistent with a radially constant disk field. The obtained outer
  spur parameters are very similar to the ones reported in
  Table~\ref{tab:DF} and are omitted for clarity.}
 \begin{tabular}{ll|cccccccccccc}
\multirow{2}{*}{id}& \multirow{2}{*}{$I_\mathrm{coh}$}& \bdisksun{} & \zdisk{} & \lrad{} & \bannpeak{} & \rann{} & \wann{} & \zann{} & \multirow{2}{*}{$k$} & $m_0$ & $(\ell, b)_d$ & $|\ddip|$ &\multirow{2}{*}{$\chi^2/n_\mathrm{df}$} \\
 & & [\muG] & [kpc] & [kpc] & [\muG] & [kpc] & [kpc] & [kpc] &  & [K] & [deg.] & [K] & \\
 \hline\hline
g1 & \modelBase & $4.3$ & $0.85$ & $120$ & $4.7$ & $4.1$ & $2.0$ & $2.0$ & $0.45$ & $6.9$ & $(\hphantom{0}75, 15)$ & $2.0$ & $3574/2938 = 1.22$ \\
g2& \modelXr & $4.3$ & $0.83$ & $120$ & $4.6$ & $4.6$ & $1.8$ & $1.7$ & $0.36$ & $7.1$ & $(\hphantom{0}72, 14)$ & $2.2$ & $3611/2938 = 1.23$ \\
g3&\modelNe & $4.3$ & $0.85$ & $120$ & $4.6$ & $4.1$ & $1.9$ & $2.1$ & $0.43$ & $6.9$ & $(\hphantom{0}75, 15)$ & $2.0$ & $3566/2938 = 1.21$ \\
g4&\modelKappa & $4.3$ & $0.85$ & $120$ & $4.7$ & $4.1$ & $2.0$ & $2.0$ & $0.46$ & $6.8$ & $(\hphantom{0}74, 16)$ & $2.0$ & $3554/2938 = 1.21$ \\
g5&\modelTwist & $4.2$ & $0.85$ & $120$ & $4.6$ & $4.1$ & $1.9$ & $2.0$ & $0.66$ & $6.4$ & $(100, 27)$ & $1.4$ & $3499/2938 = 1.19$ \\
\hline
\multicolumn{2}{r}{$\langle\sigma\rangle$} & $0.1$ & $0.06$ & $\hphantom{0}90$ & $0.3$ & $0.2$ & $0.2$ & $0.2$ & $0.02$ & $0.1$ & $(\hphantom{00}3, \hphantom{0}2)$ & $0.1$ & \\
\end{tabular}
  \label{tab:DRF}
\end{table}

\begin{table}[h]
  \centering
  \small
  \caption{Fitted rescaling factors for the polarized-excess regions,
    expressed relative to the extrapolation expected for an electron
    spectral index $p=3.2$, and the corresponding effective spectral
    indices $p_{\rm eff}$. Entries are means over the \ModelDRF{}
    variants in Table~\ref{tab:DRF}; uncertainties denote their
    standard deviations.}
  \begin{tabular}[t]{lcc}
\toprule
region & $\langle q/q(3.2)\rangle$ & $p(\langle q\rangle)$ \\
\midrule
I & $1.21 \pm 0.04$ & $3.29 \pm 0.02$ \\
Is & $0.43 \pm 0.01$ & $2.79 \pm 0.01$ \\
IIIa & $0.88 \pm 0.04$ & $3.14 \pm 0.02$ \\
IIIb & $0.57 \pm 0.04$ & $2.93 \pm 0.04$ \\
IIIs & $0.48 \pm 0.04$ & $2.85 \pm 0.04$ \\
IV & $1.37 \pm 0.07$ & $3.35 \pm 0.02$ \\
VIII & $0.40 \pm 0.04$ & $2.76 \pm 0.05$ \\
IX & $0.76 \pm 0.04$ & $3.07 \pm 0.02$ \\
\bottomrule
\end{tabular}
\begin{tabular}[t]{lcc}
\toprule
region & $\langle q/q(3.2)\rangle$ & $p(\langle q\rangle)$ \\
\midrule
XI & $0.55 \pm 0.04$ & $2.92 \pm 0.04$ \\
XII & $0.39 \pm 0.03$ & $2.74 \pm 0.04$ \\
XIII & $0.81 \pm 0.05$ & $3.10 \pm 0.03$ \\
XIV & $0.78 \pm 0.10$ & $3.08 \pm 0.06$ \\
PS1 & $0.74 \pm 0.04$ & $3.05 \pm 0.03$ \\
Fan & $0.64 \pm 0.01$ & $2.98 \pm 0.01$ \\
GP & $0.49 \pm 0.01$ & $2.86 \pm 0.01$ \\
GC & $0.88 \pm 0.10$ & $3.14 \pm 0.06$ \\
\bottomrule
\end{tabular}

  \label{tab:regions}
\end{table}

\section{Excess Polarized Intensity Regions}
\label{sec:excessPolFactor}

For all regions except GCS, VIIb, X, and PS2, the fit yields a
non-negative rescaling factor. The unconstrained best-fit factors for
these four exceptions are negative but statistically compatible with
zero. These regions are therefore removed by the iterative procedure
described in Sec.\,\ref{sec:analysis}. Table~\ref{tab:regions} lists
the factors for the remaining \nFitRegions{} regions, averaged over
the \ModelDRF{} variants in Table~\ref{tab:DRF}. Other model variants
give similar results.

The fitted extrapolation factors from 30~GHz to 408~MHz range from
about 0.4 to 1.4 times the values expected for an electron spectrum
with $p=3.2$, corresponding to a synchrotron brightness-temperature
spectral index $\beta_S=(p+3)/2=3.1$.  If attributed entirely to
variations in the electron spectrum, the factors correspond to
effective indices $p_{\rm eff}=2.7$--$3.4$. Even the smooth Galactic
component exhibits variations of $\beta_S=3.1\pm0.1$, or equivalently
$p=3.2\pm0.2$ (bottom row of Fig.\,\ref{fig:cohModel}), so a somewhat
broader range for localized structures with different electron ages
and acceleration histories is not unexpected.

These values should nevertheless be regarded only as effective
spectral indices. As discussed in Sec.\,\ref{sec:fg}, the conversion
also contains a geometry factor $\langle\alpha\rangle\geq1$. Overall,
the fitted rescaling factors are therefore of a plausible
magnitude. Allowing them to vary retains the substantial contribution
of these regions to the 408~MHz intensity while profiling over the
uncertain conversion from 30~GHz to 408~MHz in the determination of the global
magnetic-field parameters.

\section{Comparison with Previous Models}
\label{sec:previousModels}

\begin{figure}[!thb]
\def\padding{0cm}
  \makebox[0.25\textwidth]{\large\hspace*{0.03\textwidth}JF12}\makebox[0.25\textwidth]{\large\Planck JF12b}\makebox[0.25\textwidth]{\large\modelDisk\unskip\hspace*{0.03\textwidth}}\makebox[0.25\textwidth]{\large\modelRing\unskip\hspace*{0.08\textwidth}}\\
\def\figh{0.319}
\includegraphics[
  height=\figh\textheight,
  rviewport=0.05 0 0.84 1,
  clip
]{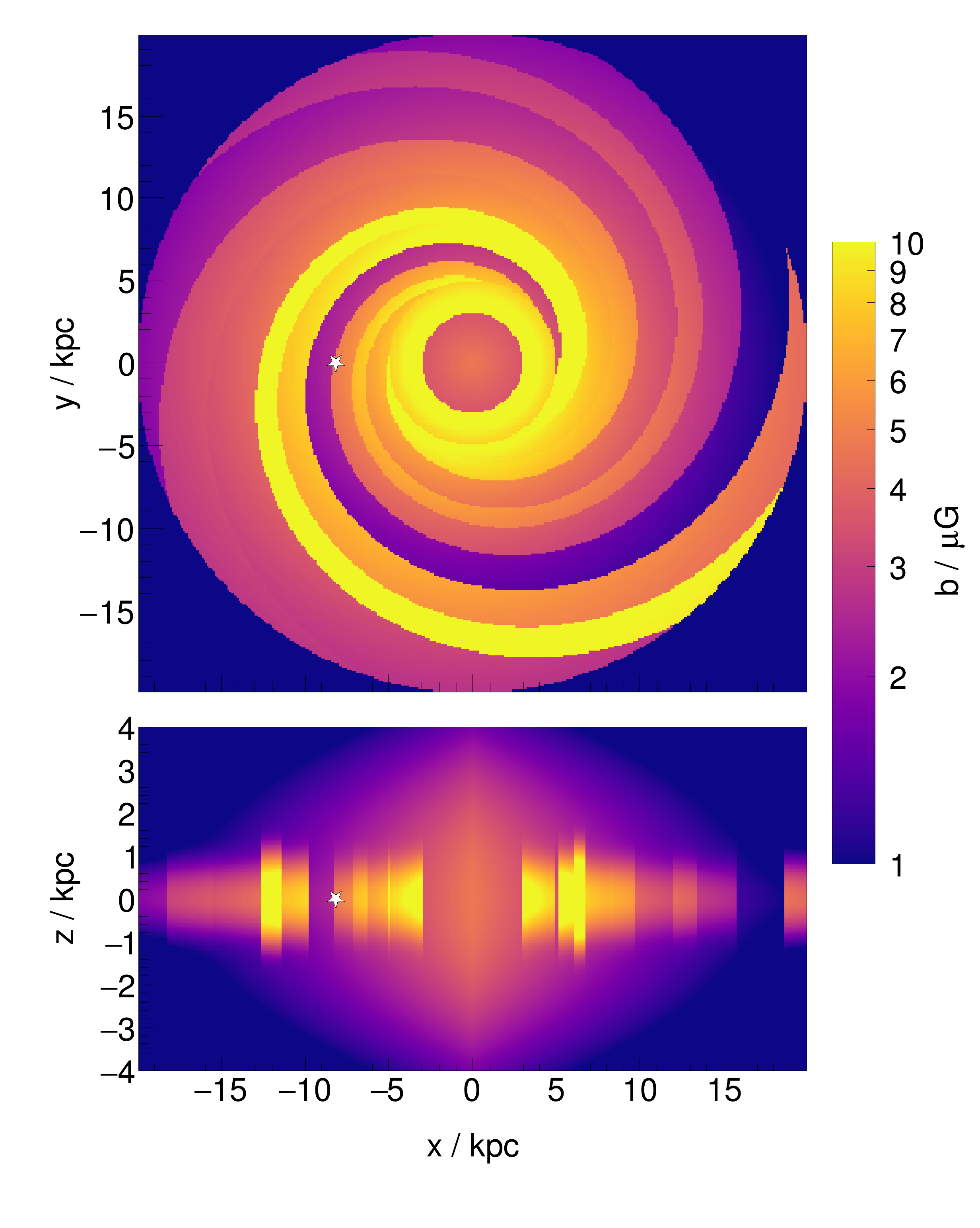}~\includegraphics[
  height=\figh\textheight,
  rviewport=0.14 0 0.84 1,
  clip
]{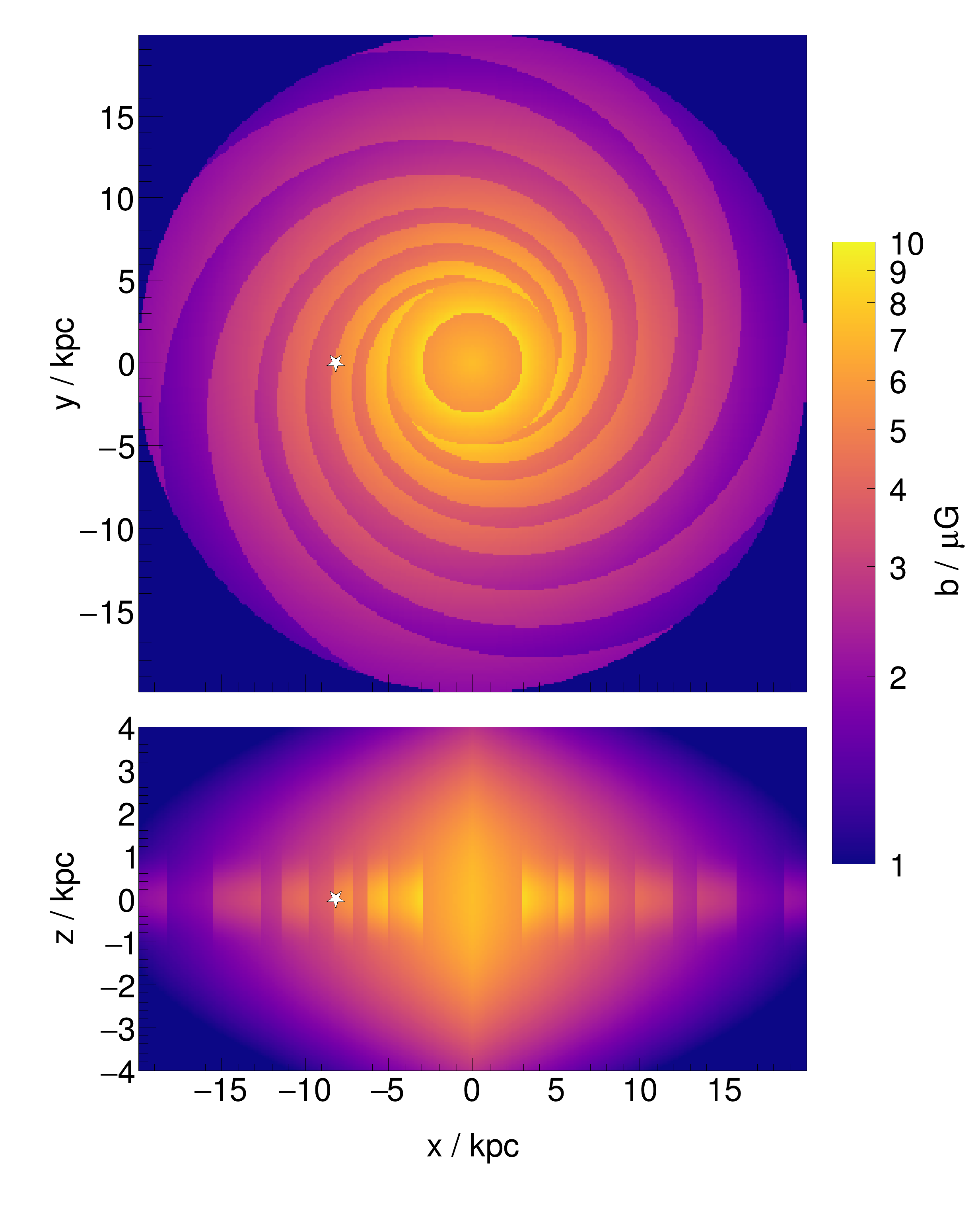}~\includegraphics[
  height=\figh\textheight,
  rviewport=0.14 0 0.84 1,
  clip
]{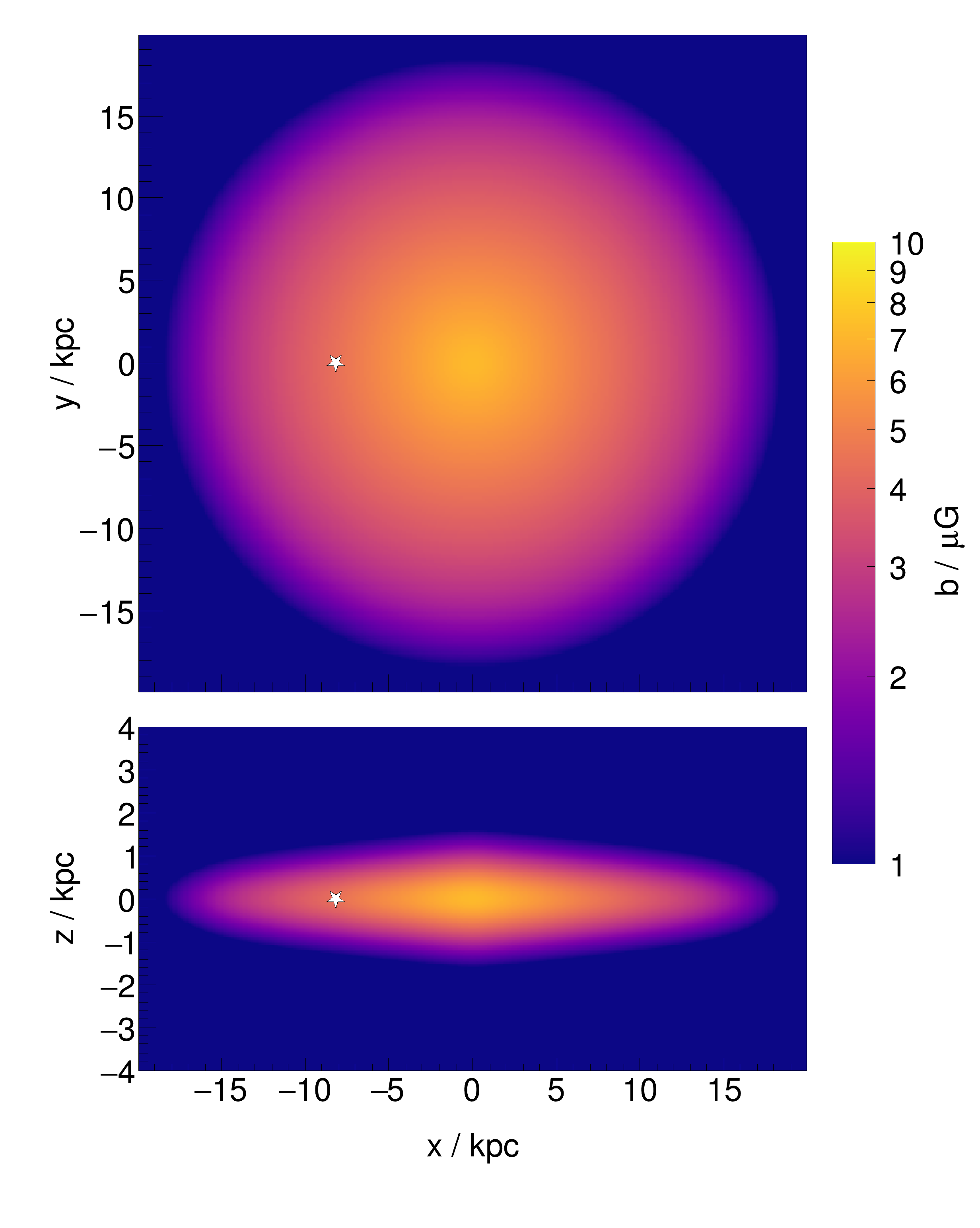}~\includegraphics[
  rviewport=0.14 0 0.98 1, clip,
  height=\figh\textheight,
]{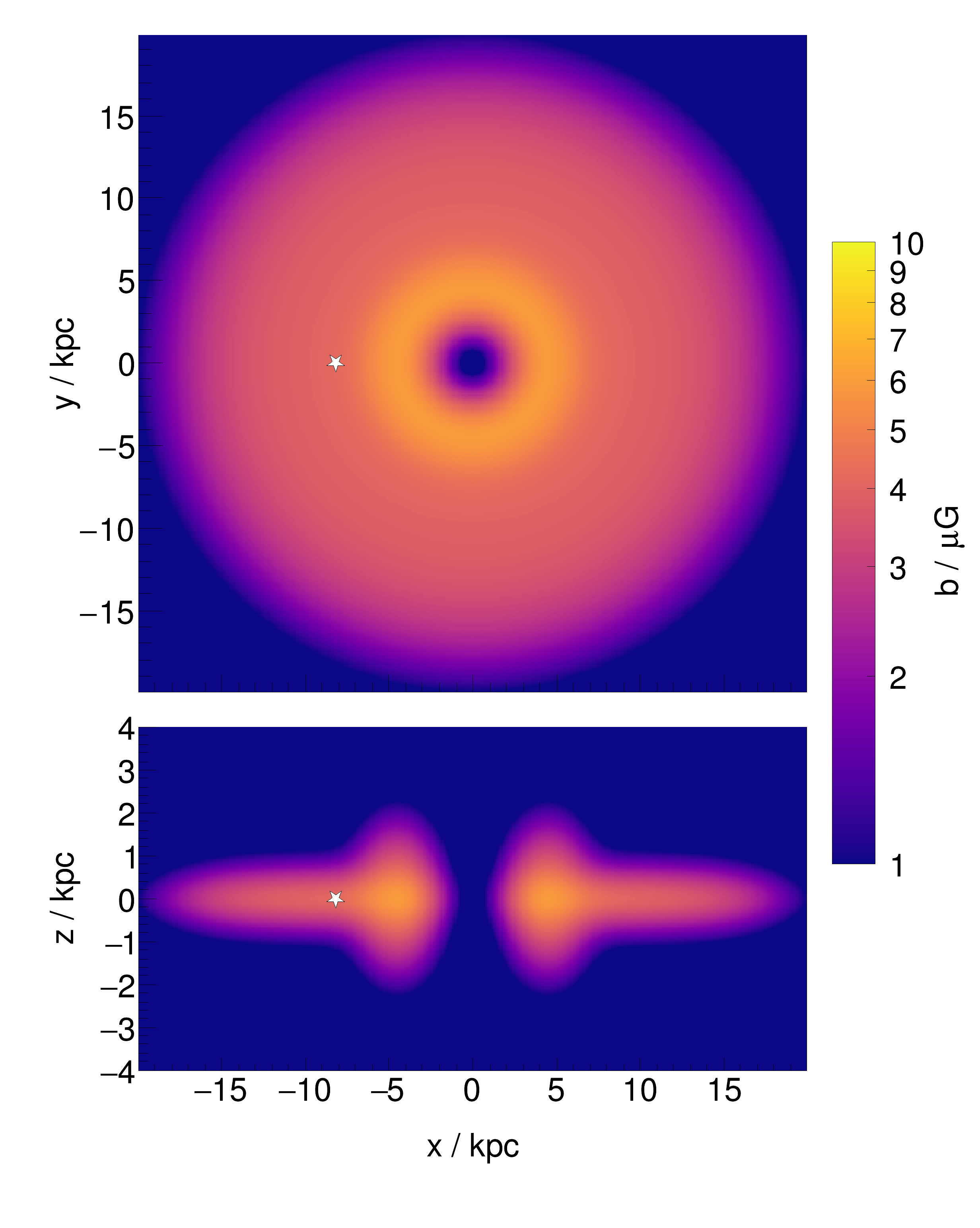}
\caption{Comparison of the previous JF12 and Planck JF12b random-field
models with the \modelDisk{} and \modelRing{} models presented
here. The two
right-hand panels show the same views as Fig.\,\ref{fig:randomfield},
but with adjusted color scales. The star marks the position of the
Sun.}
\label{fig:previousModels}
\end{figure}

\begin{figure}[!htb]
  \centering
  \def\figw{0.48}
\includegraphics[clip,rviewport=0 0 1 1,width=\figw\textwidth]{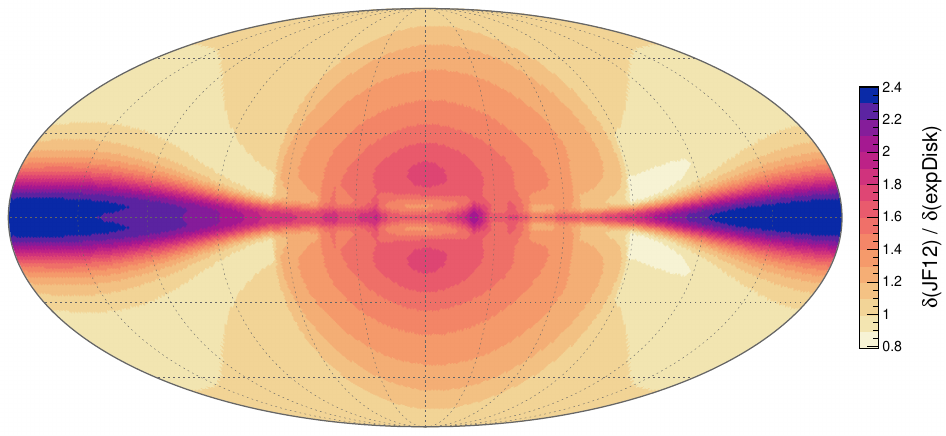} \includegraphics[width=\figw\textwidth]{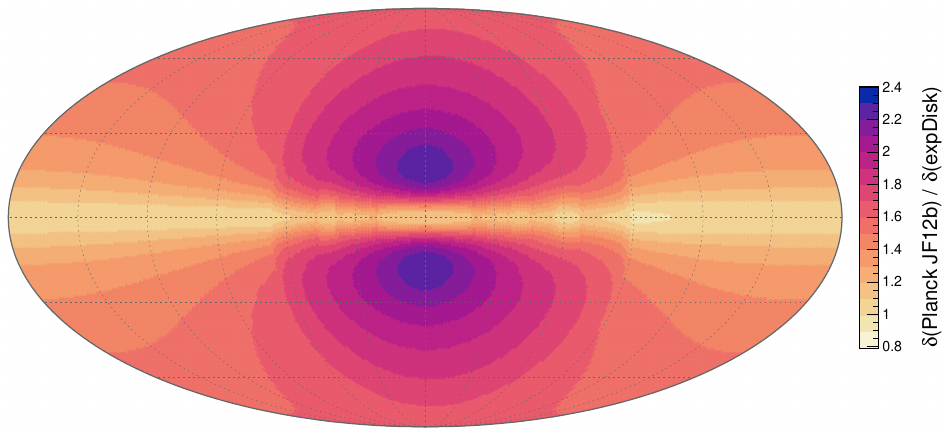}
\caption{Ratio of the rms deflection angle $\delta$ for JF12
  (left) and for JF12b (right) to that of \modelDisk{}.}
\label{fig:defratio}
\end{figure}

The two representative random-field models developed in this work,
\modelDisk{} and \modelRing{}, are compared in
Fig.\,\ref{fig:previousModels} with the
JF12 model of \citet{2012ApJ...761L..11J} and its modification by the
\citet{2016A&A...596A.103P}, usually denoted as JF12b. JF12 features
strong magnetic spiral arms, probably driven by the bright Fan-region
emission in the outer Galaxy. The arm strengths are reduced in JF12b,
which instead has a stronger halo field with a pronounced cusp in the
inner Galaxy. Owing to their larger vertical scale heights, both older
models also predict stronger random fields at large $|z|$ than the
models presented here. These differences translate directly into their
predictions for the rms UHECR deflection, $\delta$.
Figure~\ref{fig:defratio} shows the ratios of the JF12 and JF12b
deflections to those predicted by \modelDisk{}. Relative to
\modelDisk{}, JF12
predicts substantially stronger angular smearing in the outer Galaxy,
whereas the excess for JF12b is largest toward the Galactic center,
reaching a factor of 2.4 in both cases.

\end{document}